\documentclass[iop,revtex4]{emulateapj}

\let\pwiflocal=\iffalse \let\pwifjournal=\iffalse

\usepackage{amsmath,amssymb}
\usepackage[breaklinks,colorlinks,urlcolor=blue,citecolor=blue,linkcolor=blue]{hyperref}
\usepackage{epsfig}    
\usepackage{graphicx}    
\usepackage{lineno}
\usepackage{natbib}
\usepackage{bigints}
\usepackage[outdir=./]{epstopdf}

\usepackage[T1]{fontenc}
\pwifjournal\else
  \usepackage{microtype}
\fi

\pwifjournal\else
  \makeatletter
  \renewcommand\plotone[1]{%
    \centering \leavevmode \setlength{\plot@width}{0.95\linewidth}
    \includegraphics[width={\eps@scaling\plot@width}]{#1}%
  }%
  \makeatother
\fi

\makeatletter

\newcommand\@simpfx{http://simbad.u-strasbg.fr/simbad/sim-id?Ident=}

\newcommand\MakeObj[4][\@empty]{
  \pwifjournal%
    \expandafter\newcommand\csname pkgwobj@c@#2\endcsname[1]{\protect\object[#4]{##1}}%
  \else%
    \expandafter\newcommand\csname pkgwobj@c@#2\endcsname[1]{\href{\@simpfx #3}{##1}}%
  \fi%
  \expandafter\newcommand\csname pkgwobj@f#2\endcsname{#4}%
  \ifx\@empty#1%
    \expandafter\newcommand\csname pkgwobj@s#2\endcsname{#4}%
  \else%
    \expandafter\newcommand\csname pkgwobj@s#2\endcsname{#1}%
  \fi}%

\newcommand\MakeTrunc[2]{
  \expandafter\newcommand\csname pkgwobj@t#1\endcsname{#2}}%

\newcommand{\obj}[1]{%
  \expandafter\ifx\csname pkgwobj@c@#1\endcsname\relax%
    \textbf{[unknown object!]}%
  \else%
    \csname pkgwobj@c@#1\endcsname{\csname pkgwobj@s#1\endcsname}%
  \fi}
\newcommand{\objf}[1]{%
  \expandafter\ifx\csname pkgwobj@c@#1\endcsname\relax%
    \textbf{[unknown object!]}%
  \else%
    \csname pkgwobj@c@#1\endcsname{\csname pkgwobj@f#1\endcsname}%
  \fi}
\newcommand{\objt}[1]{%
  \expandafter\ifx\csname pkgwobj@c@#1\endcsname\relax%
    \textbf{[unknown object!]}%
  \else%
    \csname pkgwobj@c@#1\endcsname{\csname pkgwobj@t#1\endcsname}%
  \fi}

\makeatother

\pwifjournal\else
  \usepackage{etoolbox}
  \makeatletter
  \patchcmd{\NAT@citex}
    {\@citea\NAT@hyper@{%
       \NAT@nmfmt{\NAT@nm}%
       \hyper@natlinkbreak{\NAT@aysep\NAT@spacechar}{\@citeb\@extra@b@citeb}%
       \NAT@date}}
    {\@citea\NAT@nmfmt{\NAT@nm}%
     \NAT@aysep\NAT@spacechar\NAT@hyper@{\NAT@date}}{}{}
  \patchcmd{\NAT@citex}
    {\@citea\NAT@hyper@{%
       \NAT@nmfmt{\NAT@nm}%
       \hyper@natlinkbreak{\NAT@spacechar\NAT@@open\if*#1*\else#1\NAT@spacechar\fi}%
         {\@citeb\@extra@b@citeb}%
       \NAT@date}}
    {\@citea\NAT@nmfmt{\NAT@nm}%
     \NAT@spacechar\NAT@@open\if*#1*\else#1\NAT@spacechar\fi\NAT@hyper@{\NAT@date}}
    {}{}
  \makeatother
\fi

\MakeObj{n33370}{NLTT\%2033370}{NLTT~33370\,AB}
\MakeTrunc{n33370}{NLTT~33370}
\MakeObj{tvlm}{TVLM\%20513-46546}{TVLM~513--46546}

\newcommand{\um}{$\mu$m}
\newcommand{\fbol}{$F_{\mathrm{bol}}$}

\newcommand{\rchisq}{$\chi^2_{\nu}$}

\newcommand\amlt{$\alpha_{\rm MLT}$} 

\newcommand\ms{M$_\odot$}

\newcommand\teff{\ensuremath{T_\text{eff}}}

\providecommand{\adsurl}[1]{\href{#1}{ADS}}

\slugcomment{ApJ accepted}

\shorttitle{Properties of M Dwarfs}
\shortauthors{Mann et al.}

\begin{document}
\title{How to Constrain Your M Dwarf:\\ measuring effective temperature, bolometric luminosity, mass, and radius}

\author{Andrew W. Mann,\altaffilmark{1,2} Gregory A. Feiden,\altaffilmark{3} Eric Gaidos,\altaffilmark{4,5} Tabetha Boyajian \altaffilmark{6}, Kaspar von Braun \altaffilmark{7}}

\altaffiltext{1}{Harlan J. Smith Fellow, University of Texas at Austin; amann@astro.as.utexas.edu}
\altaffiltext{2}{Visiting Researcher, Institute for Astrophysical Research, Boston University, USA}
\altaffiltext{3}{Department of Physics and Astronomy, Uppsala University, Box 516, SE-751 20, Uppsala, Sweden}
\altaffiltext{4}{Department of Geology and Geophysics, University of Hawai'i at Manoa, Honolulu, HI 96822, USA}
\altaffiltext{5}{Visiting Scientist, Max Planck Institut f\"{u}r Astronomie, Heidelberg, Germany}
\altaffiltext{6}{Department of Astronomy, Yale University, New Haven, CT 06511, USA}
\altaffiltext{7}{Lowell Observatory, 1400 W. Mars Hill Rd., Flagstaff, AZ, USA}
 
\begin{abstract}
Precise and accurate parameters for late-type (late K and M) dwarf stars are important for characterization of any orbiting planets, but such determinations have been hampered by these stars' complex spectra and dissimilarity to the Sun. We exploit an empirically calibrated method to estimate spectroscopic effective temperature (\teff) and the Stefan--Boltzmann law to determine radii of 183 nearby K7-M7 single stars with a precision of 2-5\%. Our improved stellar parameters enable us to develop model-independent relations between \teff\ or absolute magnitude and radius, as well as between color and \teff. The derived \teff--radius relation depends strongly on [Fe/H], as predicted by theory. The relation between absolute $K_S$ magnitude and radius can predict radii accurate to $\simeq$3\%. We derive bolometric corrections to the $VR_CI_CgrizJHK_S$ and {\it Gaia} passbands as a function of color, accurate to 1-3\%. We confront the reliability of predictions from Dartmouth stellar evolution models using a Markov Chain Monte Carlo to find the values of unobservable model parameters (mass, age) that best reproduce the observed effective temperature and bolometric flux while satisfying constraints on distance and metallicity as Bayesian priors. With the inferred masses we derive a semi-empirical mass--absolute magnitude relation with a scatter of 2\% in mass. The best-agreement models over-predict stellar \teff s by an average of 2.2\% and under-predict stellar radii by 4.6\%, similar to differences with values from low-mass eclipsing binaries. These differences are not correlated with metallicity, mass, or indicators of activity, suggesting issues with the underlying model assumptions e.g., opacities or convective mixing length.
\end{abstract}


\maketitle

\section{Introduction}\label{sec:intro}
Very-low-mass stars ($0.1_{\odot}<M_\star<0.6M_{\odot}$), i.e., late K and M dwarf stars, have become prime targets in the search for exoplanets, particularly Earth-like planets that orbit in circumstellar "habitable zones" where the level of stellar irradiation and a planet's equilibrium temperature permit stable liquid water. M dwarfs dominate the stellar population, comprising more than 70\% of all stars in the Galaxy \citep{Henry:1994fk}, and thus their planets weigh heavily on the overall occurrence of planets in the Milky Way. Because transit and Doppler signals increase with decreasing stellar radius ($R_*$), stellar mass ($M_*$), and orbital period, Earth-size planets in the habitable zone are much easier to detect around M dwarfs. 

Results based on observations by the NASA {\it Kepler} mission suggest that M dwarfs are teeming with rocky planets, with $\simeq$1 planet per low-mass star with orbital periods of less than 50 days \citep{Cassan2012, Morton2014a, Gaidos2014}. Future space-based transit surveys such as TESS \citep{Ricker2014} and PLATO \citep{Rauer2013} are expected to detect hundreds of rocky planets around {\it nearby} stars, including M dwarfs. Bright host stars make transit spectroscopy to detect the planets' atmospheres and search for biosignatures possible with observatories such as the \textit{James Webb Space Telescope}. 

Estimates of planet properties depend directly on properties of the host stars as well as observables such as transit depth or Doppler radial velocity amplitude, thus it is critical that we have accurate stellar parameters to match the precise measurements achievable with current and planned instrumentation. The radius of the host star $R_*$ is required to establish the radius of a transiting planet. The stellar density $\rho_*$ is required to determine the probability that a planet with a given orbital period is on a transiting orbit and infer the occurrence of such planets around a set of stars targeted by a transit survey. The mass of a planet detected with a given orbital period and radial velocity reflect amplitude scales with mass of the star as $M_*^{2/3}$. The orbit-averaged irradiance on a planet is proportional to stellar luminosity $L_*$ and thus the equilibrium temperature scales as $L_*^{1/4}$. Among other applications, this information is required to determine if a planet falls within a ``habitable zone'' bounded by runaway greenhouse conditions and the maximum greenhouse effect that a CO$_2$ atmosphere can provide \citep{Kopparapu2013}. 

Stellar parameters are also useful when comparing planetary systems in search of clues to the origins of their diversity. The occurrence of giant planets around FGK dwarfs is a strong function of metallicity  \citep{Santos2004,Fischer:2005yq}, and this correlation may also hold for M dwarfs \citep{Johnson:2010lr, Mann:2013vn, 2014ApJ...791...54G}. Giant planet occurrence also likely scales with $M_*$ \citep{Johnson:2010lr,Gaidos2013}, and the occurrence of small, close-in planets appears to increase with decreasing $M_*$ \citep{Howard:2012yq}. 

Parameters of M dwarfs can be estimated by direct comparison with model predictions \citep[e.g., ][]{Casagrande2008,Gaidos2013b, Dressing2013}. Although models of M dwarf atmospheres have continued to advance \citep{Allard2013, Rajpurohit:2013}, there remain important molecular bands that are poorly described or completely omitted. Such an approach is also ultimately unsatisfactory as the interiors and atmospheres of these stars are dissimilar to the Sun and we wish to test the models, not trust them. 

Studies of low-mass eclipsing binary systems (LMEBs) find that stellar models systematically under-predict stellar radii and over-predict \teff s by up to 4\% \citep[e.g.,][]{Kraus2011,Feiden2012a, Spada2013}. The leading hypothesis is that magnetic phenomena (magneto-convection or spots) are responsible for the observed discrepancies, motivated by theoretical predictions \citep[e.g.,][]{Mullan2001} and bolstered by observational evidence \citep[e.g.,][]{LopezMorales2007}. Further investigations have lead to contradictory results, with some studies suggesting magnetic fields are adequate to reconcile models and observations \citep[e.g.,][]{MacDonald2012,Feiden2013,Torres2014}, while others question the validity of the magnetic hypothesis \citep[e.g.,][]{Chabrier2007, Feiden2014a}. It is clear that, if magnetic fields are to blame, single inactive field stars are not expected to display significant disagreements with model predictions. However, results from \citet{Boyajian2012} indicate that single inactive fields stars may in fact show similar disagreements. 

Recent observational advances have led to more precise, {\it model-independent} estimates of some M dwarf parameters. Long-baseline optical interferometry (LBOI) has been used to measure the angular diameters ($\theta$), and with trigonometric parallaxes, physical diameters, of some very nearby, bright M dwarfs with uncertainties of 1-5\% \citep[e.g., ][]{Boyajian2012, 2014MNRAS.438.2413V}. The angular diameter plus the parallax yields the physical stellar radius. The angular diameter plus an estimate of the star's bolometric flux (\fbol) yields a direct determination of \teff\ via the Stefan--Boltzmann law:
\begin{equation}\label{eqn:stefan}
\mathrm{T}_{\mathrm{eff}} = 2341\left( \frac{{F}_{\mathrm{bol}}}{\theta^2}\right) ^{1/4},
\end{equation}
where \fbol\ is in units of 10$^{-8}$\,erg\,s$^{-1}$\,cm$^{-2}$. These estimates are thus model-independent (with the exception of small limb darkening corrections to angular diameters) and are applicable to single stars (close binaries being obvious in interferometric data). 

{\it Effective temperatures} derived from fits to model grids can be compared to these bolometrically derived temperatures to calibrate or correct the former to within ~60 K \citep{2013ApJ...779..188M}. Likewise, improvements in the accuracy of photometric zero points and system passbands \citep{Mann2015a} allow {\it bolometric fluxes} to be measured to an accuracy of $\lesssim1\%$ using well-calibrated, high signal-to-noise (S/N) spectra. {\it Metallicities} of M dwarfs have previously eluded precise determination because analysis of spectra visible wavelengths is complicated by confusion of many overlapping lines and the lack of a well-defined continuum. However, the isolated lines in infrared spectra, combined with calibration by observations of common proper motion pairs of solar-type stars and M dwarfs, have been recently to estimate [Fe/H] to a precision as good as 0.07 dex \citep[e.g., ][]{2012A&A...538A..25N, Terrien:2012lr, 2013ApJ...779..188M, Newton:2014}. These estimates are model-independent to the extent that the metallicities of the solar-type calibrators are derived independently of any model assumptions but will nonetheless be affected by any systematic errors in the latter \citep{Mann2013a}. 

Precise parameter values for calibrator M dwarfs can be used to construct empirical relations between parameters that are, in principle, applicable to fainter, more distant stars for which distances or angular diameters are currently unobtainable \citep{2013ApJ...779..188M}. They can also be used to test models of M dwarf interiors and atmospheres \citep[e.g.,][]{Boyajian2015}. However, only a small ($\simeq$30) number of M dwarfs are near and bright enough for observations by LBOI. This restricts the calibration to hotter values of \teff\ and sparsely samples the relevant range of [Fe/H]. For example, it has been difficult to evaluate whether $R_*$ increases with [Fe/H] for a fixed \teff, a basic prediction of models.  

If \teff\ and \fbol\ can be determined without interferometry, Equation~\ref{eqn:stefan} can be inverted to estimate $\theta$, and with a trigonometric parallax, physical radius. Because precise spectroscopic temperatures, bolometric fluxes, and trigonometric parallaxes can be obtained for a much larger number of stars than are reachable with existing LBOI instrumentation, this greatly expands the potential pool of calibrators. A similar method called Multiple Optical-Infrared Technique (MOITE) has been used to determine the radii of M dwarfs \citep{Casagrande2008}. However, that method is sensitive to the model temperatures and photometric zero points. 

In this work we present estimates of \fbol, \teff\, and [Fe/H] for 183 bright nearby M dwarfs with precise trigonometric parallaxes, which we use to determine $R_*$. We construct improved empirical relations between these parameters and confront the predictions of models of M dwarfs with these values. In Section~\ref{sec:sample} we define the sample. We describe our spectroscopic observations in Section~\ref{sec:obs} and how we turn spectra into determinations of \fbol, \teff, [Fe/H], and $R_*$ in Section~\ref{sec:tefffbol}. We compare our estimates to values available from the literature in Section~\ref{sec:empcomp}. Using our derived M dwarf parameters, we develop empirical relations between observables (i.e., color, \teff, and $M_{K_S}$) and fundamental parameters ($R_*$, $L_{\rm{bol}}$) in Section~\ref{sec:relations}, and provide empirical bolometric corrections in Section~\ref{sec:bcorr}. We compare predictions from the Dartmouth stellar evolutionary model \citep{Dotter2008} to our values in Section~\ref{sec:models}, including exploring the impact of changes in the model physics. We conclude in Section~\ref{sec:discussion} with a summary of our work, a brief discussion of some of the issues with our analysis, and potential applications of our methods with the advent of extremely precise trigonometric parallaxes form the ESO {\it Gaia} astrometric mission.

\section{Sample Selection}\label{sec:sample}
M dwarf stars were selected from the CONCH-SHELL \citep{Gaidos2014} or LG11 \citep{Lepine:2011vn} catalogs. We first selected stars with parallax errors $<5\%$. Trigonometric parallaxes were taken from \citet{Harrington:1993}, \citet{Jao:2005}, \citet{2006AJ....132.2360H}, \citet{van-Leeuwen:2007yq}, \citet{Gatewood:2008}, \citet{Gatewood:2009}, \citet{2009AJ....137.4109L}, \citet{Riedel2010}, \citet{Jao:2011}, and \citet{2014ApJ...784..156D}. For targets with multiple parallax measurements we adopted the distance from the most precise measurement. This selection yielded 923 stars, about two thirds of which are visible to our instruments/telescopes (Section~\ref{sec:obs}). We prioritized our spectroscopic observations for targets based on: (i) error in stellar parallax, (ii) brightness, and (iii) estimated spectral type or color (since late-type stars tend to be removed by the first two criteria. Existing optical and/or near-infrared (NIR) spectra \citep{Lepine:2013, 2013ApJ...779..188M, Gaidos2014} were used where available.

Our prioritization coupled with available observing time yielded 191 stars. To this we added an additional 36 M dwarfs in wide ($\gtrsim5\arcsec$) binaries from \citet{Mann2013a} and \citet{Mann2014}. These M dwarfs did not make the original cut because they do not have trigonometric parallaxes or a sufficiently precise trigonometric parallaxes. However, the their primaries have parallaxes, which we use to calculate the distance to the M dwarf companion. We also added three M dwarfs with LBOI $\theta$ measurements that were not in the LG11 or CONCH-SHELL catalogs. We then removed three stars because of insufficient ($<5$ points) photometry to absolutely calibrate the spectrum (see Section~\ref{sec:cal}) or because there was poor agreement between photometry and spectra, i.e., reduced $\chi^2>10$ ($\chi^2_{\rm{\nu}}$).

\subsection{Binary Contamination}\label{sec:binary}

Unresolved binaries are problematic for our analysis. Unlike photometric variability, which changes the photometry and any derived stellar properties randomly, an unresolved and uncorrected companion will always increase \fbol, and hence systematically bias the radii of the overall sample. It has been estimated that $\simeq40\%$ of M dwarfs in the solar neighborhood have at least one companion within 1000~AU \citep{Fischer1992,Lada2006,Raghavan2010}. The median distance to our targets is 10~pc, so the majority of these were sufficiently separated ($>5\arcsec$ or $>50$~AU) such that the photometry and spectroscopy is unaffected. We flagged all binaries that are resolved in the spectrographs finder cameras (Section~\ref{sec:obs}), which is capable of detecting companions with $\Delta K\lesssim 2$ and as close as $1-2$\arcsec\ to the primary, depending on conditions. Systems with high contrast ratios ($\Delta K \gg 2$) which were more likely to be missed have a negligible effect on the \fbol\ (see below).

\begin{figure}[htbp] 
\begin{center}
\includegraphics[width=3.5in]{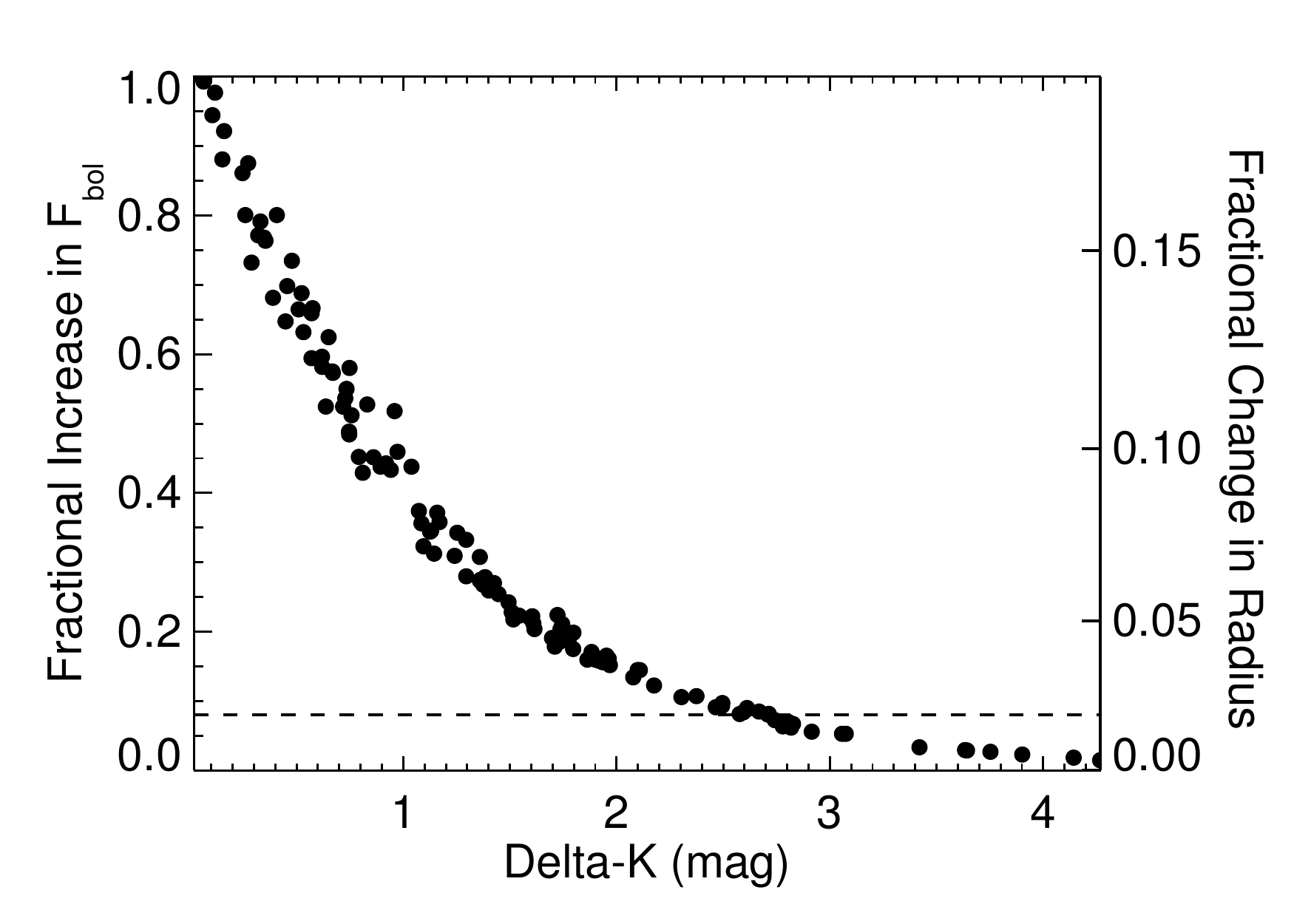}
\caption{The effect on the derived \fbol\ and $R_*$ from an unresolved binary as a function of the $K$-magnitude contrast ratio. The dashed line indicates the maximum change in \fbol\ (8\%) from a binary, above which we remove the target from our sample. The scatter in the locus of points is larger than expected from measurement errors because we are combining spectra of of a range of metallicities, and therefore have varying \fbol's for a given $M_K$. More details can be found in Section~\ref{sec:binary}. \label{fig:add_binary}}
\end{center}
\end{figure}

To identify tighter binaries we crosschecked our sample with the Fourth Catalog of Interferometric Measurements of Binary Stars \citep{Hartkopf:2001}, the Washington Double Star Catalog \citep{Mason:2001}, and SIMBAD. Each of these catalogs reports heterogeneous information, e.g., contrast ratios are not always given in the same filter and sometimes just spectral types are given rather than flux differences. To deal with this heterogeneity, we use our sample to estimate the impact of binarity on our derived parameters as a function of contrast ratio or spectral type. We scaled all spectra in our sample so that they are at the same distance (10\,pc, but the exact value is arbitrary) then added random pairs of spectra together, calculating \fbol\ before and after addition (Figure~\ref{fig:add_binary}). We removed any target with a companion that would increase \fbol\ by more than $8\%$. This cut was selected because an 8\% increase in \fbol\ results in a 2\% increase in derived $R_*$, which is still less than our typical measurement errors. This binarity criterion removed 41 targets, or 18\% of the sample.

Using the binary period distribution from \citet{Raghavan2010} and the overall binarity rate for M dwarfs from \citet{Fischer1992} we estimated that $\simeq$17\% of our sample contains binaries that are tight enough to be unresolved in our photometry/spectroscopy and not resolved in SpeX/SNIFS finder images, but could have increased the derived \fbol\ by $\gtrsim$8\%. The actual binarity rate may differ in our sample because it is not volume-limited while the binary fraction estimates are. However, the expected binary fraction is consistent with or lower than the fraction of targets that were removed because they are known binaries (18\%), suggesting binary contamination will not be a significant problem for our analysis. Most of our targets are close, and have been studied extensively for companions \citep[e.g.,][]{Janson2012, 2014ApJ...789..102J, Ansdell2015}, and hence the agreement between the estimated and observed binarity is not surprising.

The final sample contains 183 stars with spectral types ranging from K7 to M7 (median spectral type of M3), [Fe/H] from -0.60 to +0.53 (median [Fe/H] of -0.03), and distances from 1.8\,pc (Barnard's Star) to 38\,pc (median distance of 11\,pc). 

\section{Observations and Reduction}\label{sec:obs}

\subsection{Optical Spectra from SNIFS and STIS}\label{sec:SNIFS}

Optical spectra of all stars were obtained with the SuperNova Integral Field Spectrograph \citep[SNIFS,][]{Aldering2002,Lantz2004} on the University of Hawai'i 2.2m telescope on Mauna Kea. SNIFS provides simultaneous coverage from 3200--9700\,\AA\ by splitting the beam with a dichroic mirror onto blue (3200-5200\,\AA) and red (5100-9700\,\AA) spectrograph channels. Although the spectral resolution of SNIFS is only $R\simeq1000$, the instrument provides excellent spectrophotometric precision \citep{Mann:2011qy, 2013ApJ...779..188M}. This is in part a consequence of the stability of the atmosphere over Mauna Kea \citep{1994SoPh..152..351H, Buton:2013} and because the instrument is an integral field spectrograph and does not suffer from wavelength dependent slit losses that can be difficult to accurately remove. 

Spectra were obtained between 2010 and 2014 under varying conditions, although most nights were photometric or nearly photometric (few to zero cloud cover). Integration times varied from 2 to 600~s, yielding S/N $>100$ per resolution element in the red channel for all targets, while avoiding the non-linear regime of the detector. For exposure times $<5$\,s, multiple exposures were taken and stacked after reduction to suppress scintillation noise. 

SNIFS data reduction was broken into two parts: initial reduction from the SuperNova Factory (SNF) team pipeline, which was performed concurrently with all observations, and a more precise flux calibration and telluric correction, which we applied to the SNF pipeline reduced data. Detailed information on the SNF pipeline can be found in \citet{Bacon:2001} and \citet{Aldering:2006}. To summarize, the SNF pipeline performed dark, bias, and flat-field corrections, and cleaned the data of bad pixels and cosmic rays. It then fit and extracted each of spectra from the $15 \times 15$ "spaxels" in the red and blue channels and converted these into spectral-spatial image cubes. The pipeline applied a wavelength calibration to the cubes based on arc lamp exposures taken at the same telescope pointing and time as the science data, and a rough flux calibration based on an archive correction. Lastly, the pipeline extracted one-dimensional spectra from each of the cubes using an analytic model of the point spread function. 

Our part of the reduction is described in some detail in \citet{Lepine:2013} and \citet{Gaidos2014}. We restate most of the process here as there has been some improvements in our data reduction since the publication of those papers. Over several years of observing with SNIFS we collected more than 400 observations of 42 spectrophotometric standards\footnote{Full list at \href{http://snfactory.lbl.gov/snf/snf-specstars.html}{http://snfactory.lbl.gov/snf/snf-specstars.html}}. We first identified all standard star observations taken on photometric nights based on extinction estimates from the CFHT skyprobe\footnote{\href{http://www.cfht.hawaii.edu/Instruments/Skyprobe/}{http://www.cfht.hawaii.edu/Instruments/Skyprobe/}}. We divided the "true" spectrum of each standard star, as taken from the literature \citep{Oke1990,Hamuy1994,Bessell1999,Bohlin2001}, by the corresponding spectra from the SNF pipeline. We then found the best-fit correction as a function of airmass:
\begin{equation}\label{eqn:airmass}
\kappa(\lambda) = a(\lambda) + b(\lambda)z + c(\lambda)z^2,
\end{equation}
where $\kappa$ is the true atmospheric (including telluric) and instrumental correction, $z$ is the airmass, and  $a$, $b$, and $c$ are fit variables that also depend on wavelength ($\lambda$). We compared this model derived from all photometric nights to a correction derived using just standards taken in a single night. We found that the differences between these corrections is only $\simeq$1\% in the red and $\lesssim$2\% in the blue, excluding regions of significant telluric contamination. For a more accurate flux calibration we took the median derived correction from all standard stars observed in a single night (without an airmass fit), which we used to derive an additional wavelength dependent adjustment we applied to $\kappa$ (in addition to Equation~\ref{eqn:airmass}). The final correction was then applied to all stars from that night. This assumes that that while the atmospheric transparency ($a$) varies between nights, the effect of airmass ($b$, $c4$) does not.

Five of our stars are included in the Next Generation Spectral Library \citep[NGSL,][]{Gregg2006,Heap2007}, which consists of spectra obtained with the Space Telescope Imaging Spectrograph (STIS) on the {\it Hubble Space Telescope} ({\it HST}) using the G230LB, G430L, and G750L gratings. NGSL spectra cover 2000-10000\AA\ with spectrophotometric precision of $\simeq0.5$\% \citep{Bohlin2001,Bohlin:2004, 2004AJ....127.3508B}. We used the NGSL instead of the SNIFS data for these five stars. 

To test the quality of our spectrophotometric calibration we compared our SNIFS data to the NGSL spectra for the five overlapping targets. We found that we can reproduce the HST spectra with a scatter (1$\sigma$) of 1.1\% in the red and 1.5\% in the blue, with median offsets of 0.1\% and 0.3\% respectively. We added this estimate of the systematic error in quadrature to the formal errors in each spectrum for our error analysis of \fbol\ (Section~\ref{sec:tefffbol}).

\subsection{Near-infrared (NIR) Spectra from SpeX}

Between 2012 and 2014 we obtained NIR spectra of all stars using the SpeX spectrograph \citep{Rayner:2003lr} attached to the NASA Infrared Telescope Facility (IRTF) on Mauna Kea. As with the SNIFS observations, data were taken under mixed conditions, but most nights were photometric or near-photometric. SpeX observations were taken in the short cross-dispersed (SXD) mode using the 0.3$\times15\arcsec$ slit, yielding simultaneous coverage from 0.8 to 2.4\um\, with a small gap at $\simeq$1.8\um, and at a resolution of $R\simeq2000$. Each target was placed at two positions along the slit (A and B) and observed in an ABBA pattern in order to subsequently subtract the sky background. For each star we took 6-10 exposures following this pattern, which, when stacked, provided a S/N per resolution element of $>100$, and typically $>150$ in the $H$- and $K$-bands. In 2014 July the SpeX chip was upgraded (now called uSpeX), which resulted in better wavelength coverage (0.7-2.5\um), including coverage of the region between the $H$ and $K$ bands that SpeX lacked. In total, 18 of 183 stars were observed with uSpeX. Observation procedures and reduction were nearly identical for the two detectors.

SpeX and uSpeX spectra were extracted using the SpeXTool package \citep{Cushing:2004fk}, which performed flat-field correction, wavelength calibration, sky subtraction, and extraction of the one-dimensional spectrum. Multiple exposures were combined using the IDL routine \textit{xcombxpec}. To correct for telluric lines, we observed an A0V-type star within 1~hr and 0.1 airmass of the target observation (usually much closer in time and airmass). A telluric correction spectrum was constructed from each A0V star and applied to the relevant spectrum using the \textit{xtellcor} package \citep{Vacca:2003qy}. Separate orders were stacked using the \textit{xcombspec} tool, which also shifts the flux level in each orders to match. These corrections are generally 1\% or less per order.

\citet{Rayner:2009kx} noted that using the 0.3\arcsec\ slit can lead to errors in the slope of the spectrum by 1-3\% due to changes in seeing, guiding, and differential atmospheric refraction between target and standard star observation. We reduced the impact of this problem by observing at the parallactic angle and minimizing the time between target and standard star observations. We compared the slopes of unstacked observations of the same star and found that slope errors are $1-2\%$, which is backed up by a comparison $JHK$ magnitudes from the literature \citep{Mann2015a}. This source of error was included in all analyzes.

\subsection{Absolute Flux Calibration of Spectra}\label{sec:cal}

SpeX spectra have a gap at $\simeq1.8$\um\ between the $H$ and $K$ bands, although there is no gap for uSpeX data. Also, several regions in the NIR are strongly affected by telluric absorption. Although our observations of A0V stars enabled us to correct for telluric absorption, the S/N in some regions was too low for accurate reconstruction of the spectra. We replaced these regions with the best-fit atmospheric model from the BT-SETTL grid \citep{Allard2011, Allard2013}. We did the same for 2.4--10\um. Our method for finding the best model spectrum is explained in Section~\ref{sec:tefffbol}. As a test, we examined the spectra of the 12 targets with S/N $>500$ in the NIR, and found that using a model to replace the regions of telluric contamination changes the derived \fbol\ by 0.15$\pm$0.26\%, and hence adds negligible error.

Blueward of our cutoff at 0.32\um\ (0.2\um\ for targets with STIS spectra) we assumed the spectrum follows Wein's approximation and redward of 10\um\ we assumed it follows the Rayleigh-Jeans law. We fit these functions using the data at 0.3--0.4\um\ and 2.0--10\um, respectively. Flux in each of these regions represented $\ll1\%$ of the total flux from a typical star, and comparison to photometry from the {\it Wide-field Infrared Survey Explorer} \citep[{\it WISE},][]{Wright:2010fk} and UV spectra from STIS \citep{Heap2007} indicated that these approximations are negligible compared to other measurement errors. 

For each star we retrieved (where available) $JHK_S$ photometry from the Two-Micron All-Sky Survey \citep[2MASS,][]{Skrutskie:2006lr}, $BV$ photometry from the AAVSO All-Sky Photometric Survey \citep[APASS,][]{Henden:2012fk}, $H_P$ from {\it Hipparcos} \citep{van-Leeuwen:2007yq}, $V_T$ and $B_T$ from Tycho-2 \citep{Hog2000}, W1 and W2 from {\it WISE} \citep{Wright:2010fk}, and all photometry from the General Catalog of Photometric Data \citep{Mermilliod:1997qe}. We discarded APASS data for stars brighter than $V=10$ where images were potentially saturated. Much of the {\it WISE} photometry was saturated or had measurement quality flags and hence was discarded. Photometric magnitudes were converted to fluxes using the zero points from \citet{2003AJ....126.1090C} for 2MASS, \citet{Jarrett2011} for {\it WISE}, and \citet{Mann2015a} for all other photometry. The number of photometric points per star varied from 5 to 179, with a median of 22. 

We calculated fluxes from the spectrum (synthetic photometry) corresponding to the available photometry using the formula:
\begin{equation}\label{eqn:synthflux}
f_{spec,x}  =  \frac{\bigintss{f_{\lambda}(\lambda) S_x(\lambda) d\lambda}} {\bigintss{S_x(\lambda) d\lambda}}, 
\end{equation}
where $f_{\lambda}(\lambda)$ is the spectrum (radiative flux density) as a function of wavelength ($\lambda$) and $S_x(\lambda)$ is the system throughput the filter $x$ (e.g., $U$, $B$, $V$) in energy units. As with the zero points, filter profiles were taken from \citet{2003AJ....126.1090C}, \citet{Jarrett2011}, and \citet{Mann2015a}. We calculated the ratio of the synthetic flux to that derived from the photometry for each band as well as corresponding uncertainties accounting for errors in the spectrum, photometry, and photometric zero points.

Excluding the five stars with STIS spectra, each of our stars had a spectrum composed of three components from the SNIFS blue channel, the SNIFS red channel, and SpeX. The blue and red channels were relatively flux-calibrated independently from our reduction explained in Section~\ref{sec:SNIFS}. However, repeat measurements and comparison with spectra from other instruments suggested that the relative flux calibration between these two channels is only accurate to 3\%, while the calibration is accurate to 1\% within a channel. The overlapping region covers only $\simeq50$\AA\, and this spectral region of the instrument has little throughput and hence low S/N. For the SpeX to SNIFS transition there is significantly more overlap ($>600$\AA), but this region is complicated by telluric H$_2$O absorption, low throughput for SpeX, and small but significant shape errors ($\lesssim1\%$ in the optical and 2\% in the NIR) above the Poisson and measurement errors (typically $\ll1\%$). We used the overlapping region between SNIFS and SpeX (excluding 50\AA\ on each end) data to calculate an offset and error, but the result is dominated by errors in the shape of each spectral component.

\begin{figure*}[t]
\begin{center}
\includegraphics[width=3.5in]{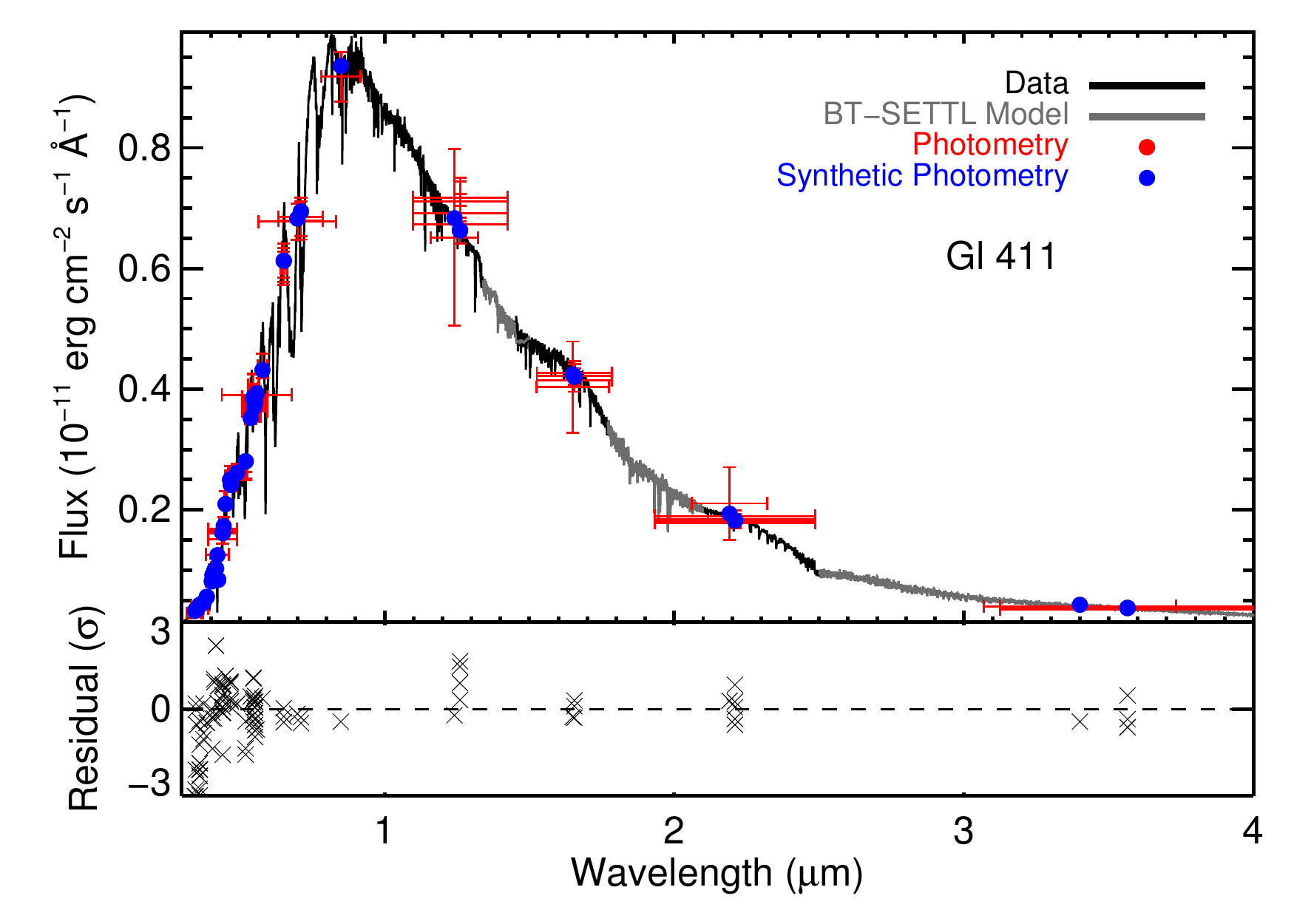}
\includegraphics[width=3.5in]{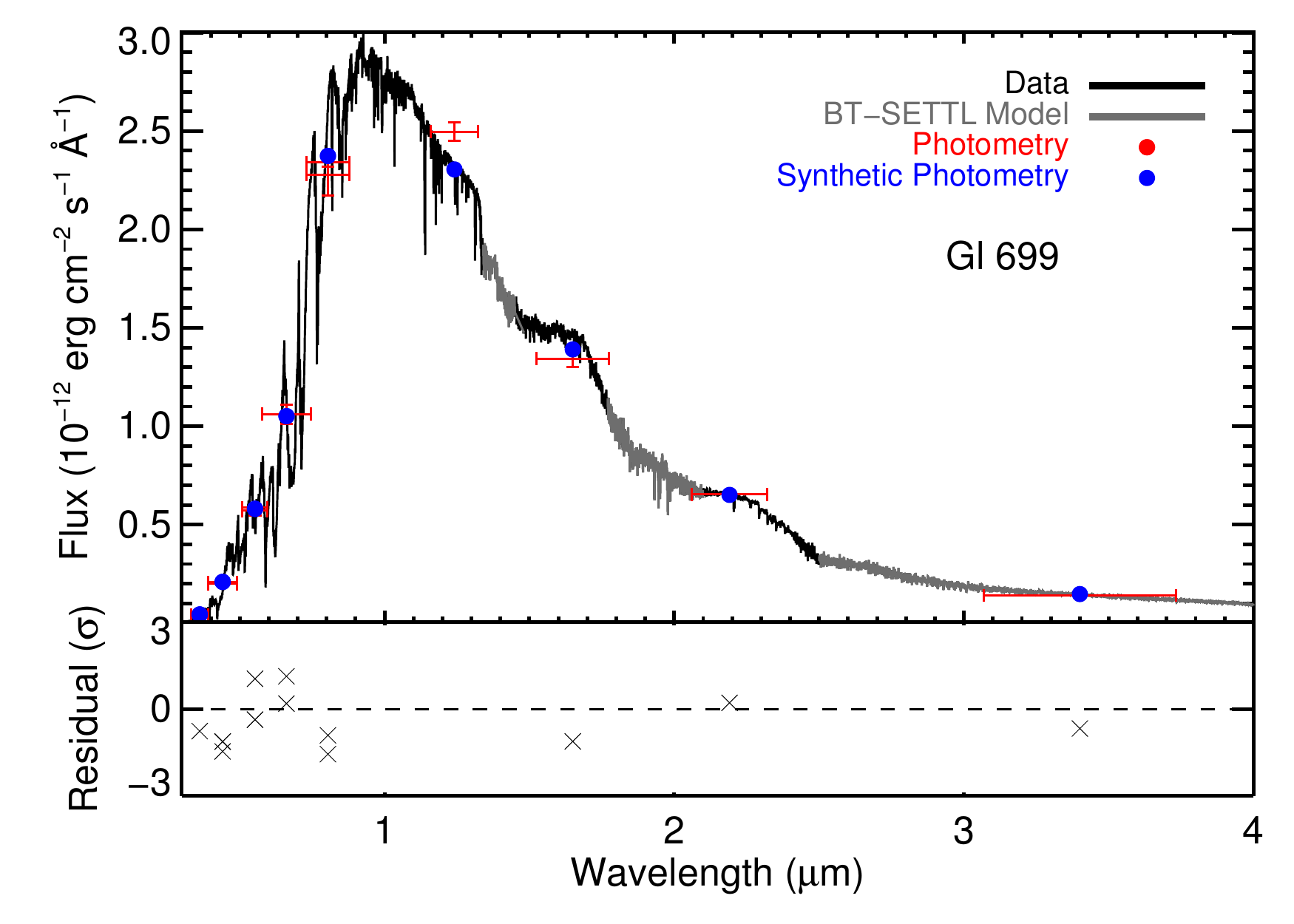}
\includegraphics[width=3.5in]{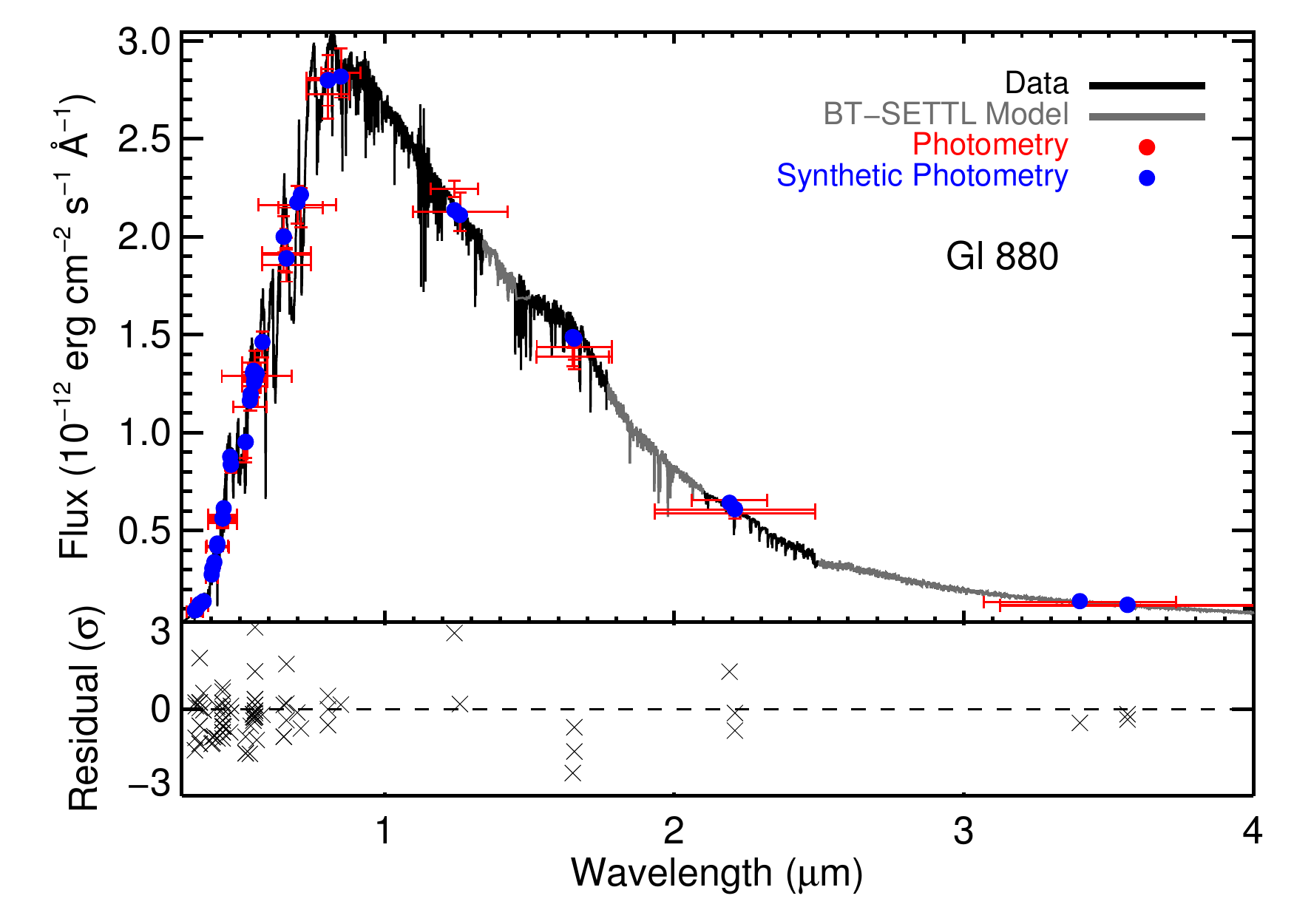}
\includegraphics[width=3.5in]{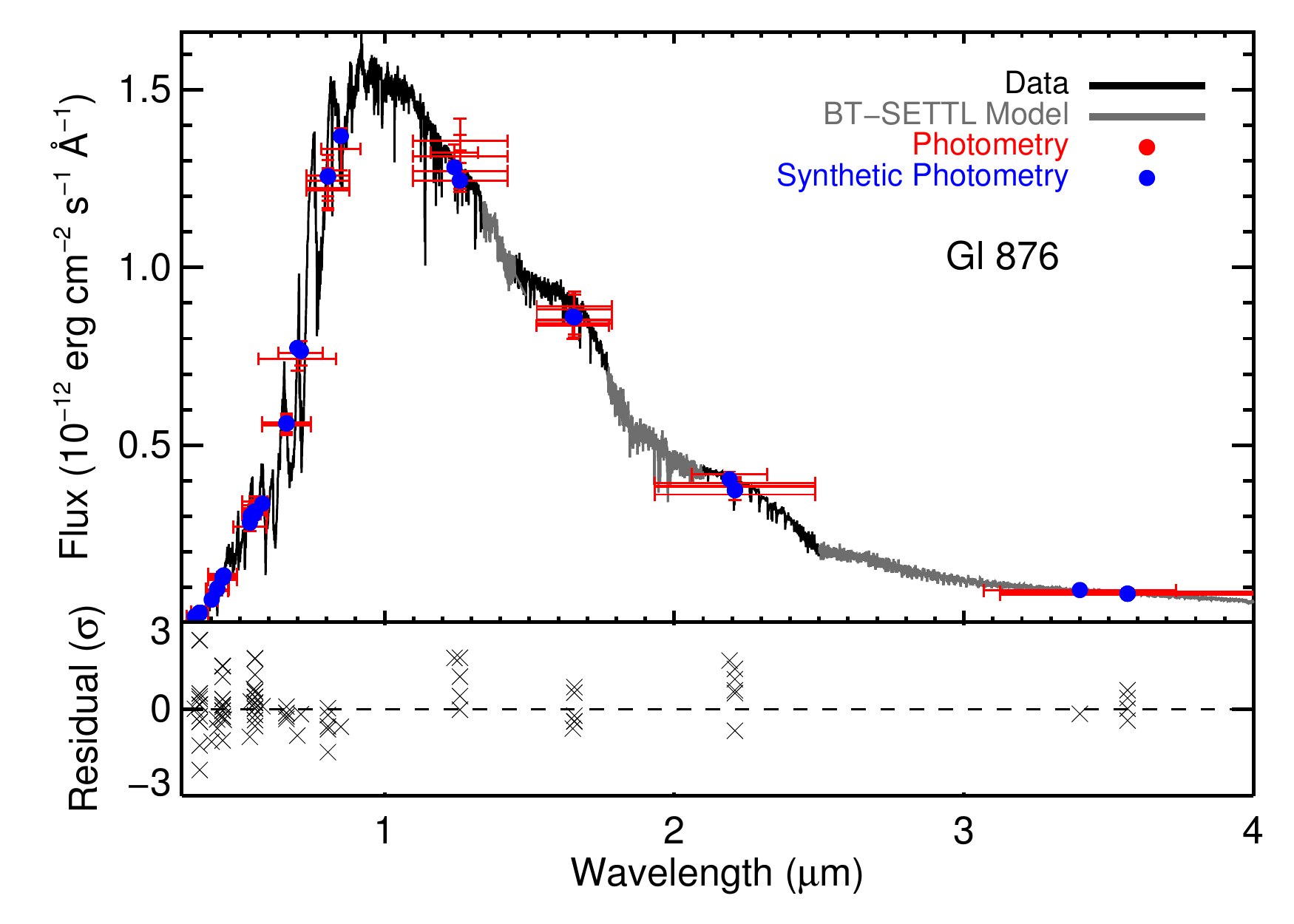}
\caption{Combined and absolutely flux-calibrated spectra of four representative stars in our sample. The spectra are shown in black, with regions replaced by models in gray. Photometry is shown in red, with the horizontal `error bars' indicating the width of the filter, and vertical errors representing combined measurement and zero point errors. Blue points indicate the corresponding synthetic fluxes calculated using Equation~\ref{eqn:synthflux}. Residuals are plotted in the bottom subpanels in units of standard deviations. More details on observations can be found in Section~\ref{sec:obs} and absolute flux calibration in Section~\ref{sec:cal}. These stars are selected because they roughly span the parameter space in metallicity and mass of our whole sample. \label{fig:spectra}}
\end{center}
\end{figure*}

The absolute fluxes of the three components of each spectrum were adjusted to minimize the difference (in standard deviations) between SNIFS and SpeX overlapping regions, the difference between photometric and spectroscopic fluxes (as described above), and the independent flux calibrations of the SNIFS red and blue channels. This renormalization process served to both combine the three components and absolutely flux calibrate each complete spectrum. For the targets with STIS data there is no separate blue channel, so the analysis had one fewer free parameter, but otherwise the method was the same. We show combined and absolutely calibrated spectra of four representative stars (Gl 699, Gl 876, Gl 880, and Gl 411) in Figure~\ref{fig:spectra}. These four are highlighted because they span a wide range of parameter space in terms of stellar mass and metallicity; two are below the fully convective boundary (Gl 699 and Gl 876), two above (Gl 411 and Gl 880), two metal-poor (Gl 699 and Gl 411), and two metal-rich (Gl 876 and Gl 880). 

The median reduced $\chi^2$ (\rchisq) value for our fits was 1.5, and the highest was 7.0 (for Gl 725B). Many of the \rchisq\ values $>1$  may be due stellar variability, which was not accounted for in our error analysis. Conversely, the low \rchisq\ of the sample suggests that either the number of highly variable stars in the sample is relatively low, or that errors are somewhat overestimated. 

\section{Derivation of Stellar Parameters}\label{sec:tefffbol}

Using the absolutely flux-calibrated spectra (Section~\ref{sec:cal}) and trigonometric parallaxes from the literature (Section~\ref{sec:sample}) we calculated \fbol, \teff, [Fe/H], $R_*$, $M_*$, synthetic photometry (Equation~\ref{eqn:synthflux}), and spectral types for all targets in our sample. For these calculations we ignored interstellar extinction, as all stars are within 40\,pc and \citet{2009MNRAS.397.1286A} showed that reddening is $\simeq0$ for stars within 70\,pc due to the Local Bubble. Allowed reddening to float had negligible effects on the \fbol\ determinations, reinforcing this assumption. We also assumed every star is single, or a resolvable binary, i.e., each spectrum and photometric measurement is from just one object (also see Section~\ref{sec:binary}). Details of our calculations follow. Values and formal errors for all stellar parameters are reported in Table~\ref{tab:sample}, with the synthetic photometry given in Table~\ref{tab:photsample}.

\subsection{Spectral Type}

Spectral types were determined from our optical spectra following the procedure from \citet{Lepine:2013}. We calculated CaH2, CaH3, TiO5, TiO6, VO1, and VO2 band indices for each of the optical spectra following the definitions given in \citet{Reid:1995lr} and \citet{Lepine:2003qy}. We then applied the empirical relations between these indices and stellar spectral type provided by \citet{Lepine:2013}. The assigned spectral type was the weighted (by the measurement error of the index) mean of spectral types from each empirical relation (six in total, but VO1 and VO2 were only used if the star is later than M3). This method gives consistent spectral types to those assigned by-eye and has an accuracy of $\pm$0.3 subtypes \citep{Lepine:2013}. The use of indices instead of template matching allowed us to assign fractional subtypes, which is justified given the uncertainties.

\subsection{Bolometric Flux}

The bolometric flux (\fbol) was calculated by integrating over the radiative flux density (the spectrum). Errors on \fbol\ were calculated from Monte Carlo randomization of our flux calibration. Specifically, we randomly varied the photometry, spectra, and photometry zero points according to their estimated errors. We then repeated the process described in Section~\ref{sec:cal} 10,000 times, recalculating \fbol\ each time. We used the standard deviation of these values as the error on \fbol. While many of our stars have S/N$\gg100$ and $\gg20$ photometric points, other sources of error (e.g., flux calibration, photometric zero points) are typically 0.5-2\% and usually dominate the error budget. As a result no star has a \fbol\ error $<0.5$\%, and typically the error is $\gtrsim1\%$. 

Due to slight modifications in our procedure for combining optical and NIR spectra, updates to our SNIFS reduction pipeline, and changes in the photometry zero points and filter profiles \citep{Mann2015a}, our \fbol\ values are slightly different from those presented in \citet{2013ApJ...779..188M} for the interferometry stars. However, these differences are almost all $<1\sigma$. Further, our \fbol\ values are in good agreement with those determined using the (similar) procedure outlined in \citet{van-Belle:2008lr} and \citet{von-Braun:2011bh} provided the same photometric zero points are used. 

\subsection{Effective Temperature}

\teff\ was calculated by comparing our optical spectra with the CFIST suite\footnote{\href{http://phoenix.ens-lyon.fr/Grids/BT-Settl/CIFIST2011/}{http://phoenix.ens-lyon.fr/Grids/BT-Settl/CIFIST2011/}} of the BT-SETTL version of the PHOENIX atmosphere models \citep{Allard2013}. More details of this procedure are given in \citet{2013ApJ...779..188M} and \citet{Gaidos2014}, although we modified their methods slightly. Unlike \citet{Gaidos2014} we did not fit and remove a polynomial trend in wavelength to correct for slit losses, as the flux calibration of SNIFS is better than many of the instruments used in the \citet{Gaidos2014} study. Like \citet{Gaidos2014} we used linear combinations of the three best models to interpolate between grid points, while \citet{2013ApJ...779..188M} used only two. \citet{2013ApJ...779..188M} used [Fe/H] to restrict which models are allowed in the fit, while this work and \citet{Gaidos2014} added [Fe/H] as a term in the equation for $\chi^2$. These differences resulted in almost negligible changes to the overall sample \teff\ (median change of 8~K) between the three papers, although several stars changed by $>$100\,K. 

Errors in \teff\ were calculated as the quadrature sum of the calibration error found by \citet{2013ApJ...779..188M} and the scatter in \teff{} from the model comparison. Generally the first term dominates, resulting in typical errors in \teff\ of $\simeq60$\,K.

\subsection{Metallicity}
Metallicities were calculated from the NIR spectra using the empirical relations from \citet{Mann2013a} for K7-M4.5 dwarfs and from \citet{Mann2014} for M4.5-M7 dwarfs. \citet{Mann2013a,Mann2014} provide relations between the equivalent widths of atomic features (e.g., Na, Ca) in NIR spectra and the metallicity of the star, calibrated using wide binaries with an FGK primary and an M dwarf companion. Errors on [Fe/H] were calculated by adding (in quadrature) measurement errors in the spectra and scatter in the empirical calibration. Comparison with other methods of measuring M dwarf metallicities \citep[e.g., ][]{Terrien:2012lr, 2013A&A...551A..36N, Newton:2014} suggests that our internal errors are only 0.04-0.06~dex \citep{2014ApJ...791...54G}, which is significantly lower than the error estimated by \citet{Mann2013a} and \citet{Mann2014}. This might be due to overlap in the calibration sample of the aforementioned studies, or underestimated errors in the FGK star metallicities \citep{2012ApJ...757..161T, 2014AJ....148...54H}. We conservatively adopt the larger value as the calibration error, which, because the S/N of our spectra is generally very high, dominate the total error in [Fe/H]. 

For 17 of the wide wide binaries among our current sample we adopted the metallicity reported for the primary. These metallicities are more precise (errors typically 0.03-0.05~dex) than those derived from our NIR spectra for the M dwarf. 

Metallicities reported here differ slightly from those in \citet{Gaidos2014}. \citet{Gaidos2014} based their [Fe/H] values on the optical metallicity calibration of \citet{Mann2013a}, while ours are based on the NIR calibration. The optical calibration gives [Fe/H] accurate to 0.10--0.13~dex, depending on spectral type, while the NIR calibration is good to 0.08~dex. For the stars with [Fe/H] measurements in both samples the standard deviation of the differences is 0.10~dex with a mean offset of 0.017$\pm$0.007, consistent with the expected errors.

\subsection{Synthetic Photometry}
We synthesized $VR_CI_CJHK_S$ photometry using the absolutely flux-calibrated spectra using Equation~\ref{eqn:synthflux} and procedure from Section~\ref{sec:cal}. We generated SDSS $griz$ magnitudes using the zero point and filter profiles from the Sloan Digital Sky Survey\footnote{We followed the procedure outlined at \href{http://classic.sdss.org/dr7/algorithms/fluxcal.html}{http://classic.sdss.org/dr7/algorithms/fluxcal.html}, so these are true SDSS magnitudes.} \citep{Fukugita:1996}. We also calculated {\it Gaia} $G$, $G_{BP}$, and $G_{RP}$ magnitudes \citep{Jordi2010}. The {\it Gaia} system profiles were provided by C. Bailer-Jones (2015, personal communication), and the zero point was estimated using a calibrated STIS spectrum of Vega \citep{Bohlin2007} and assuming $G_{\rm{Vega}}=0.03$. Readers are cautioned that the system throughput we used for {\it Gaia} passbands may prove to be inaccurate once the mission does its own on-sky calibrations. Hence our derived magnitudes could require corrections for small color terms or zero point errors when real {\it Gaia} magnitudes become available. 

Generating synthetic photometry ensured that we had a homogeneous set of {\it all} relevant magnitudes for every star. This also provided us with SDSS $griz$ magnitudes, which are not available for nearly all the stars in our sample. Further, in many cases our synthetic photometry was superior to the literature photometry in terms of precision and accuracy. While almost all stars have 2MASS $JHK_S$ magnitudes, many of the objects are brighter than the linearity/saturation limit, and hence the 2MASS $JHK_S$ magnitudes are unreliable.

\subsection{Mass}
Masses for each target were derived using the empirical mass-luminosity ($M_K$--mass) relation in \citet{Delfosse2000}, which was based on mass measurements of M dwarf eclipsing and astrometric binaries. We converted our synthetic $K_S$ magnitudes to CIT $K$ magnitudes using the corrections in \citet{2001AJ....121.2851C}, i.e., a shift of $\simeq0.02$\,mag. Based on the scatter between the relation and the binaries in \citet{Delfosse2000} we assumed 10\% errors on our masses derived this way. 

\subsection{Angular Diameter and Physical Radius}
Angular diameters were calculated from \teff\ and \fbol\ using Equation~\ref{eqn:stefan}. Radii were then determined from $\theta$ and the trigonometric parallax for each star. Errors on $R_*$ and $\theta$ were estimated by Monte Carlo simulation. We randomly perturbed the \teff\ and \fbol\ according to estimated errors (see above), and the parallax according to errors reported in the literature. We then recalculated all relevant parameters for each star. We repeated this process 10,000 times. We adopted the standard deviation of these 10,000 iterations as the error in each parameter. There derived values show significant correlations between parameters (e.g., \teff\ and $R_*$), which are accounted for by the MC error estimate. 

Errors on $R_*$ are typically $\simeq3-4\%$, although they tend to be larger with decreasing $R_*$. Our \fbol\ values are generally very precise (1--2\%) as are our constraints on \teff\ ($\simeq$2\%). However, because $\theta \propto $\,\fbol$^{1/2}$ and $\theta \propto $\,\teff$^{-2}$, errors in $\theta$ (and hence $R_*$) are primarily driven by errors in \teff. Trigonometric parallax errors are typically 2\%, and hence also noticeably contribute to errors in $R_*$.

\section{Comparison with Previously Published Parameter Values}\label{sec:empcomp}

\subsection{Interferometric \teff\ and $\theta$}\label{sec:inf}

Twenty-nine stars in our sample have interferometric radii in \citet{Lane:2001fr}, \citet{Segransan:2003ul}, \citet{Berger2006}, \citet{Kervella:2008pd}, \citet{Demory:2009qy}, \citet{von-Braun:2011bh}, \citet{von-Braun:2012lq}, \citet{Boyajian2012}, \citet{2014MNRAS.438.2413V}, or Boyajian et al. (2015, in preparation). We derived radii for these stars in the same manner as was done for every other star, i.e., we took SNIFS and SpeX spectra of these stars, absolutely flux-calibrated the spectra, measured the temperature by fitting to atmospheric models, and then derived $\theta$ and $R_*$. We also recalculated interferometric \teff\ for these stars using the LBOI $\theta$ measurements and our \fbol\ values using Equation~\ref{eqn:stefan} (although our conclusions do not change significantly if we use the \teff\ and \fbol\ values from the original reference).

\begin{figure}[t]
\begin{center}
\includegraphics[width=3.4in]{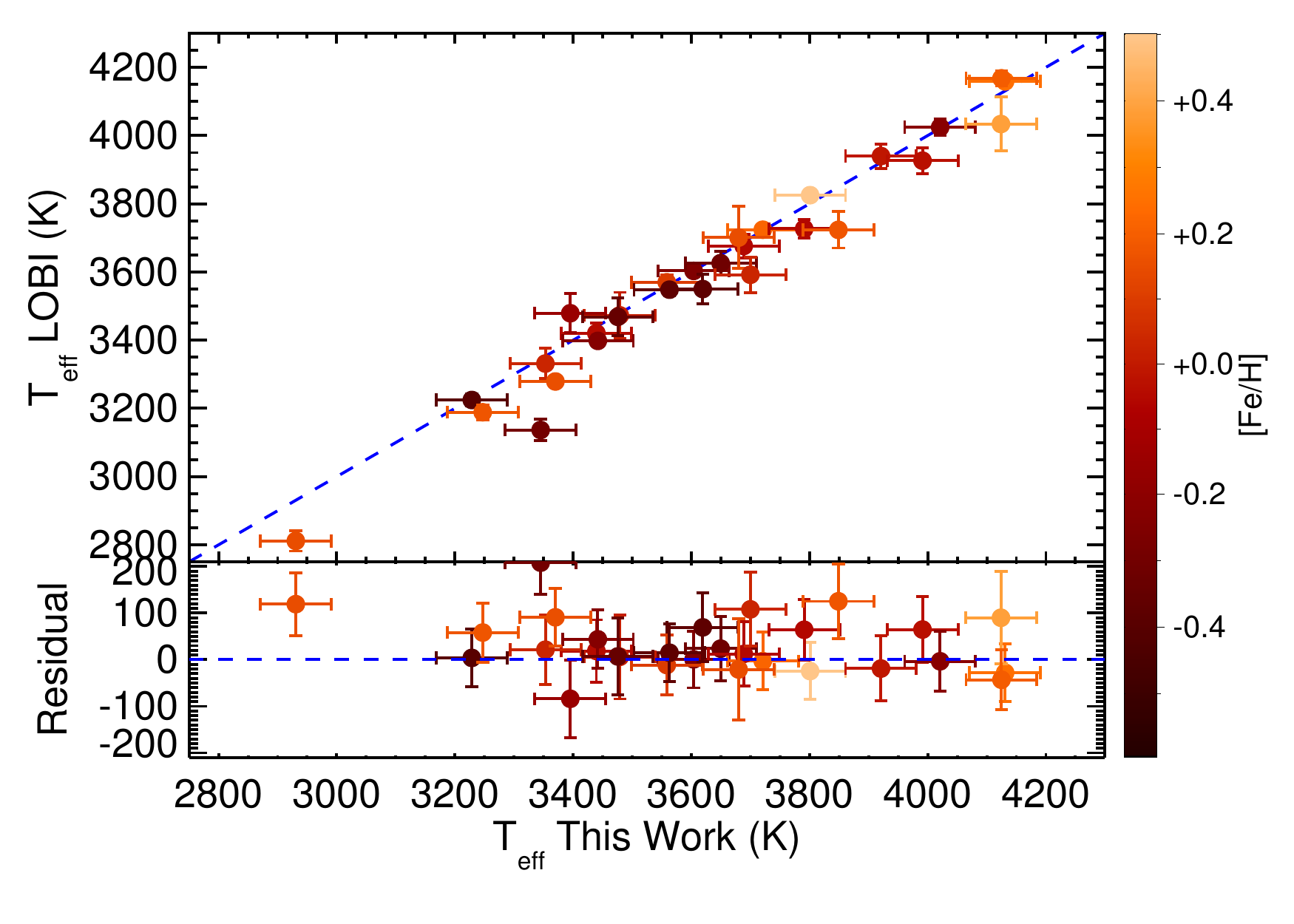}
\includegraphics[width=3.4in]{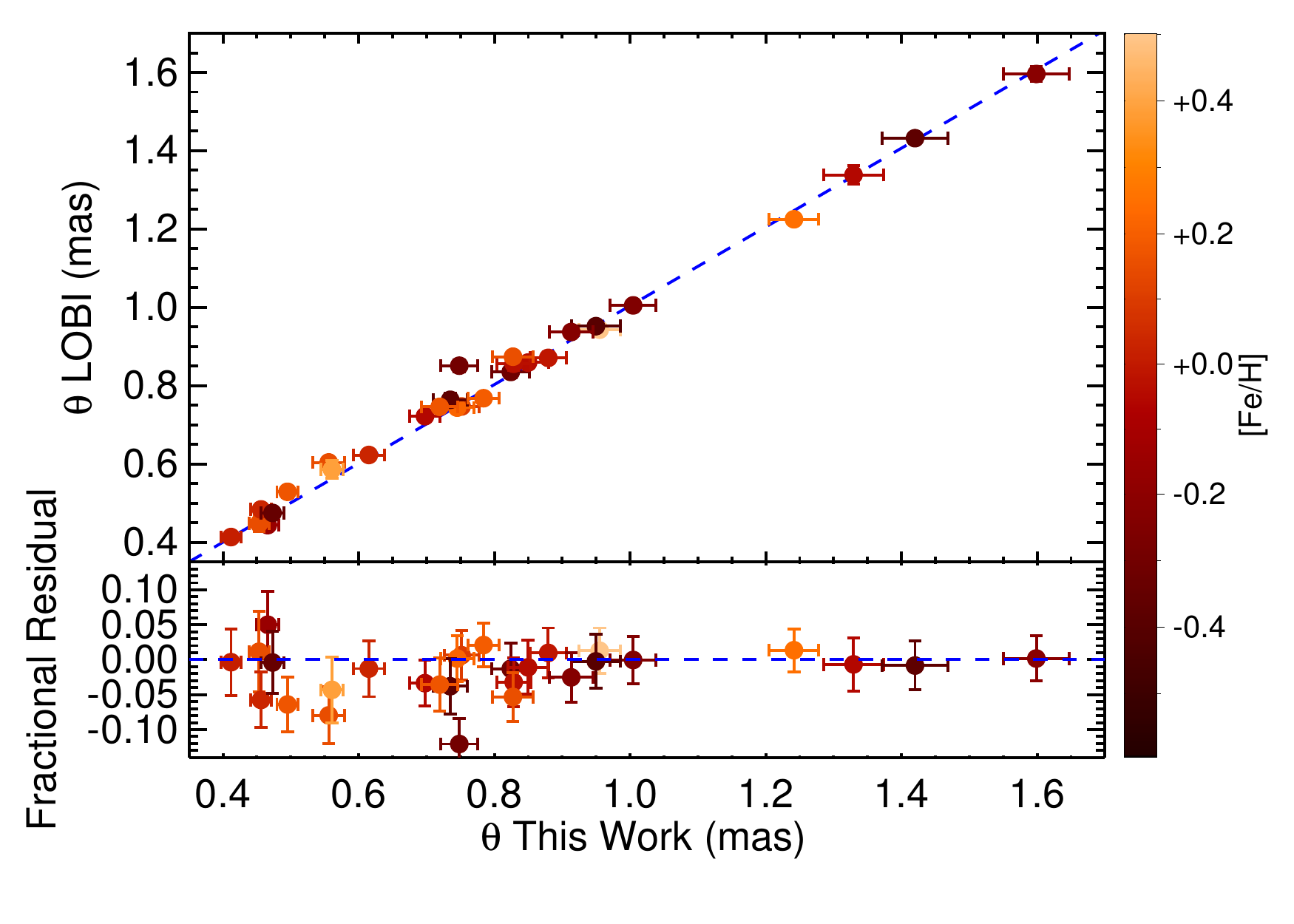}
\caption{Comparison of \teff\ (left) and angular diameter ($\theta$, right) determined through our methods (Section~\ref{sec:tefffbol}) compared to those determined from interferometry (LBOI). Bottom panels in both plots show the residuals. All points are color-coded by metallicity. Details are in Section~\ref{sec:inf}.\label{fig:comp_inf}}
\end{center}
\end{figure}

Agreement between the \teff\ and $\theta$ determinations is excellent (Figure~\ref{fig:comp_inf}). The mean difference in \teff\ is $20\pm11$\,K and the \rchisq\ is 0.90. Similarly, the mean difference in $\theta$ is 1.4$\pm$0.7\% with a \rchisq\ of 1.01. Our method for measuring \teff\ was tuned (choice of atmospheric model and spectral regions utilized) to match LBOI measurements \citep[see][for more details]{2013ApJ...779..188M}, so consistency is not surprising. However, \citet{2013ApJ...779..188M} had only 18 LBOI stars in this \teff\ range, while the expanded sample used here includes 29 LBOI stars, so this comparison is still demonstrative. The near-unity \rchisq\ values suggest that our assigned errors are reasonable. Importantly, there is no statically significant correlation between the $R_*$ residuals and [Fe/H], or \teff, suggesting that our $R_*$ determinations are clean of systematic errors.

The largest discrepancy between our values and those from LBOI is for the star Gl~725B. Our \teff\ was 3.6$\sigma$ warmer, and our radius was 3.4$\sigma$ smaller than LBOI estimates. Assuming the interferometric parameters are correct, Gl~725B has a significantly larger radius than stars of similar \teff. Photometry is also in relatively poor agreement with the spectrum compared to most stars (\rchisq$\simeq$7, see Section~\ref{sec:cal}). Photometry for this star may be inaccurate, which would have affected our derived radius and the LBOI \teff. However, \fbol\ would need to be increased by $\gtrsim20\%$ to reach $\le1\sigma$ agreement between the two determinations (although increased by just 4\% increase to achieve agreement at 3$\sigma$). Variability cannot explain the difference; the scatter in $V$-band measurements for Gl~725B is $\lesssim0.02\%$ and the LBOI visibility curve shows no sign of multiplicity. 

\subsection{$M_*$-$R_*$ from Low-mass Eclipsing Binaries}\label{sec:lmebs}

\begin{figure}[t]
\begin{center}
\includegraphics[width=0.45\textwidth]{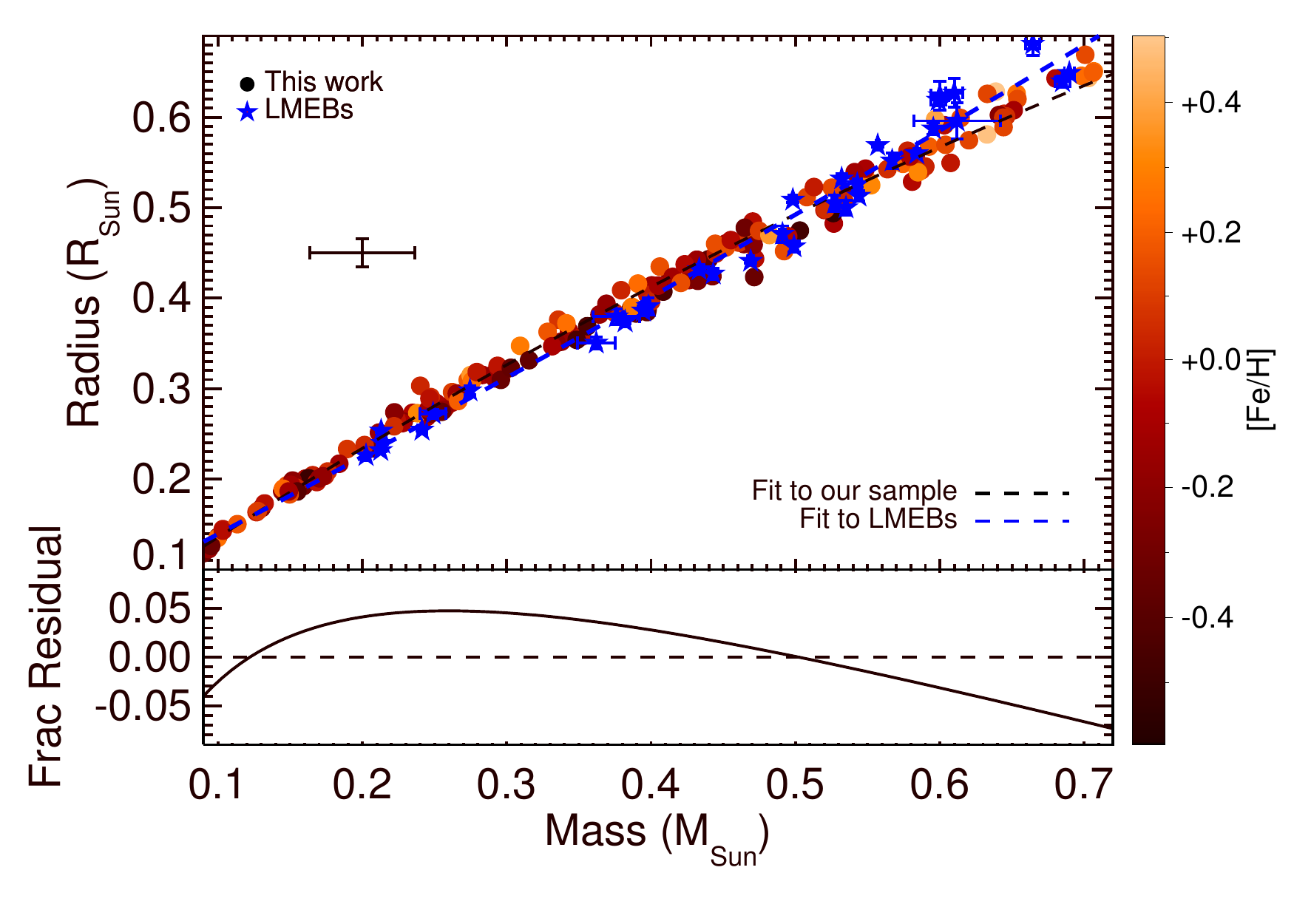}
\caption{Mass--radius diagram for stars in our sample (red circles) and those from low-mass eclipsing binaries (LMEBs, blue stars). A typical error bar on our measurements is shown to the left. Stars in our sample are color-coded by their metallicity. The fit to both samples is shown as a dashed line. The bottom panel shows the fractional residual between these two fits. More details can be found in Section~\ref{sec:lmebs}. \label{fig:mass_radius}}
\end{center}
\end{figure}

We compared our derived masses and radii to those from low-mass eclipsing binaries (LMEBs). Since unresolved binaries hamper our \fbol\ and \teff\ estimates, and most LMEBs do not have trigonometric parallaxes (or sufficiently precise parallaxes) we did not make a direct comparison as we did with the interferometry targets. Instead we could only see if the trends in the data are in agreement. We compare with a collection of detached, double-line eclipsing binaries with mass and radius errors less than 5\% in Figure~\ref{fig:mass_radius}. LMEB parameters are taken from \citet{Torres2002}, \citet{Ribas2003}. \citet{Lopez2005}, \citet{Lopez-Morales2006}, \citet{Lopez-Morales2007}, \citet{Morales2009a}, \citet{Irwin2009}, \citet{Morales2009}, \citet{Carter2011}, \citet{Irwin2011}, \citet{Kraus2011}, \citet{Doyle2011}, \citet{Orosz2012}, \citet{Orosz2012a}, \citet{Heminiak2012}, \citet{Bass2012}, and \citet{Torres2014}.

We fit the $M_*$-$R_*$ relations for both our stars and the LMEBs with second order polynomials. We show the fractional difference between these fits in Figure~\ref{fig:mass_radius}. There is a notable discrepancy at the masses $M_*\gtrsim0.65M_{\odot}$. Below 0.65$0.65M_{\odot}$ agreement is better than 5\%. As we show in Section~\ref{sec:semiemp}, model-inferred masses for our sample better reproduce the LMEB mass-luminosity relation across the sample. Thus the disagreement is most likely due to (expected) errors in the mass-luminosity relation from \citet{Delfosse2000}, which was based on only 2-3 objects with masses in this range.

\subsection{Temperatures and Bolometric Magnitudes Based on the Infrared Flux Method}\label{sec:casa}

\begin{figure}[t]
\begin{center}
\includegraphics[width=3.5in]{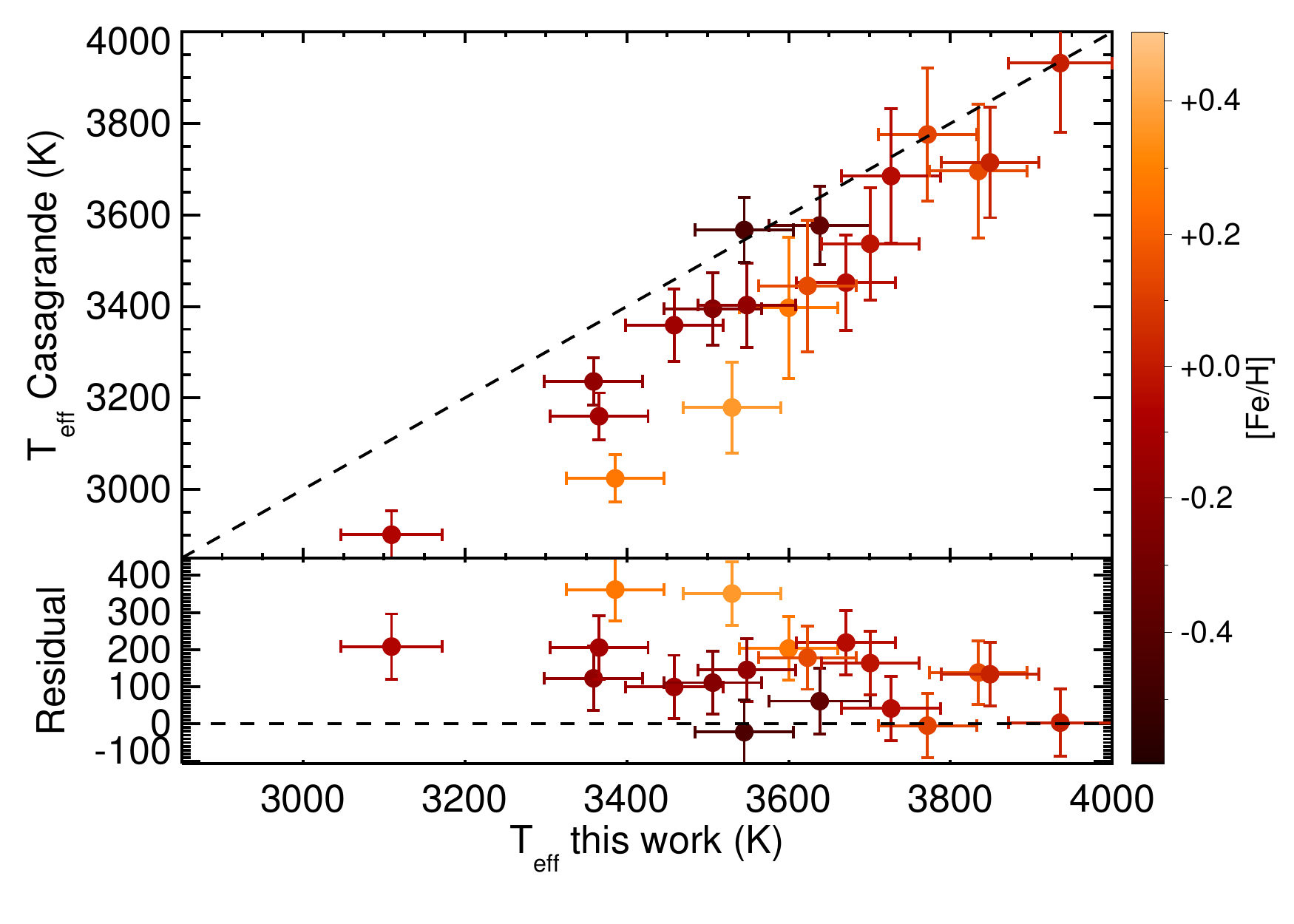}
\includegraphics[width=3.5in]{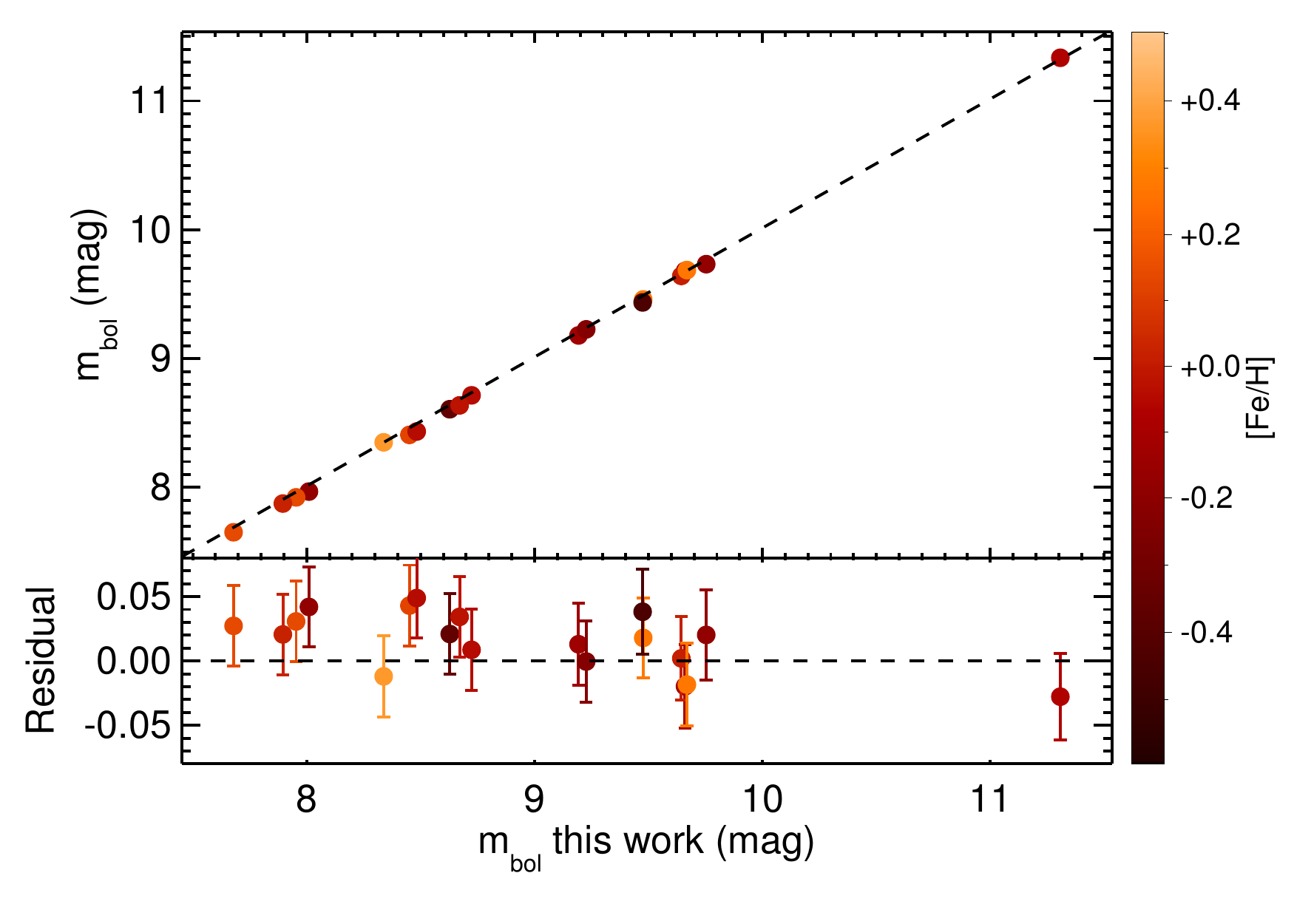}
\caption{Comparison between \teff\ (left) and $m_{\rm{bol}}$ (right) from this work to those derived in \cite{Casagrande2008}. Dashed line indicates equality. The difference between the measurements is shown in the bottom panel of each figure. Points are color-coded by our metallicities. See Section~\ref{sec:casa} for more information. \label{fig:casa}}
\end{center}
\end{figure}

\citet{Casagrande2008} extended the infrared flux method for FGK dwarfs from \citet{2006MNRAS.373...13C} to M dwarfs with a method they called the multiple optical-infrared technique (MOITE). There are 19 stars in our sample with parameters from \citet{Casagrande2008}. We found significant systematic differences between MOITE derived parameters and our values for these 16 stars, which we show in Figure~\ref{fig:casa}. Our \teff\ are on average 140$\pm$24\,K hotter, with increasing disparity at lower \teff\ and higher metallicity. Our bolometric magnitudes are $0.020\pm0.05$~mag higher. For any individual star the difference in the bolometric magnitude is small, but taken together the systematic offset is significant. The MOITE values for \teff\ and \fbol\ convert to $\theta$ values that were on average $8.1\pm1.5\%$ larger than ours. 

IRFM and MOITE assume that a star can be approximated as a blackbody beyond $\simeq2$\um. While this is reasonable for warmer stars, M dwarfs have significantly more flux in the NIR than predicted by a blackbody \citep{Rajpurohit:2013}. Hence MOITE \teff\ values tend to be systematically too low, with growing disparity at cooler temperatures where stars deviate more strongly from the Rayleigh-Jeans law. Our \fbol\ determinations also assumed a blackbody, but we only invoked this at $\lambda > 10$\um, where models suggest this is a safe assumption. Further, while this assumption could change our \fbol\ values, it has no effect on our \teff\ determinations. MOITE also assumes that the PHOENIX model \teff\ scale is accurate. While the more recent CIFIST version reproduces the \teff\ scale from LBOI determinations, the older versions (i.e., the best available to \citet{Casagrande2008}) differ systematically from LBOI derived \teff s \citep{2013ApJ...779..188M}.

\citet{Casagrande2008} argued that MOITE is in agreement with LBOI determinations. However, this was based on just nine stars, many of which have since had their parameters revised by more precise LBOI measurements. Further, at least four of these stars have problematic 2MASS magnitudes (saturated or near saturated). Since MOITE depends on reliable infrared magnitudes, it could not be directly applied to these stars. We conclude that our \teff\ and $\theta$ determinations are more reliable, and that the MOITE values are more likely systematically in error.

\subsection{Metallicities and Temperatures from \citet{Rojas-Ayala:2012uq} and \citet{2014A&A...568A.121N}}\label{sec:comp_metal}

\citet{2014ApJ...791...54G} compared the metallicities derived from our calibration \citep[][based on NIR spectra]{Mann2013a,Mann2014} to those from \citet[][also based on NIR spectra]{Rojas-Ayala:2012uq}, \citet[][based on absolute V, K magnitude]{2013A&A...551A..36N}, and \citet[][based on high-resolution optical spectra]{2014A&A...568A.121N}. To summarize their findings, the mean metallicity differences are 0.03$\pm$0.03, 0.08$\pm$0.02, and 0.06$\pm$0.02~dex, with \rchisq\ values of 0.23, 0.28, and 0.58, respectively. All of these methods used wide binaries to calibrate their metallicity measurements. The small systematic offset between our [Fe/H] values and those from \citet{2013A&A...551A..36N} and \citet{2014A&A...568A.121N} may be a reflection of small systematic differences in the metallicities of the primaries of these calibrators \citep{2014AJ....148...54H}. The low \rchisq\ values may be due to significant overlap in binary calibration samples, and/or overestimated errors for one or more method. 

\begin{figure}[t]
\begin{center}
\includegraphics[width=3.6in]{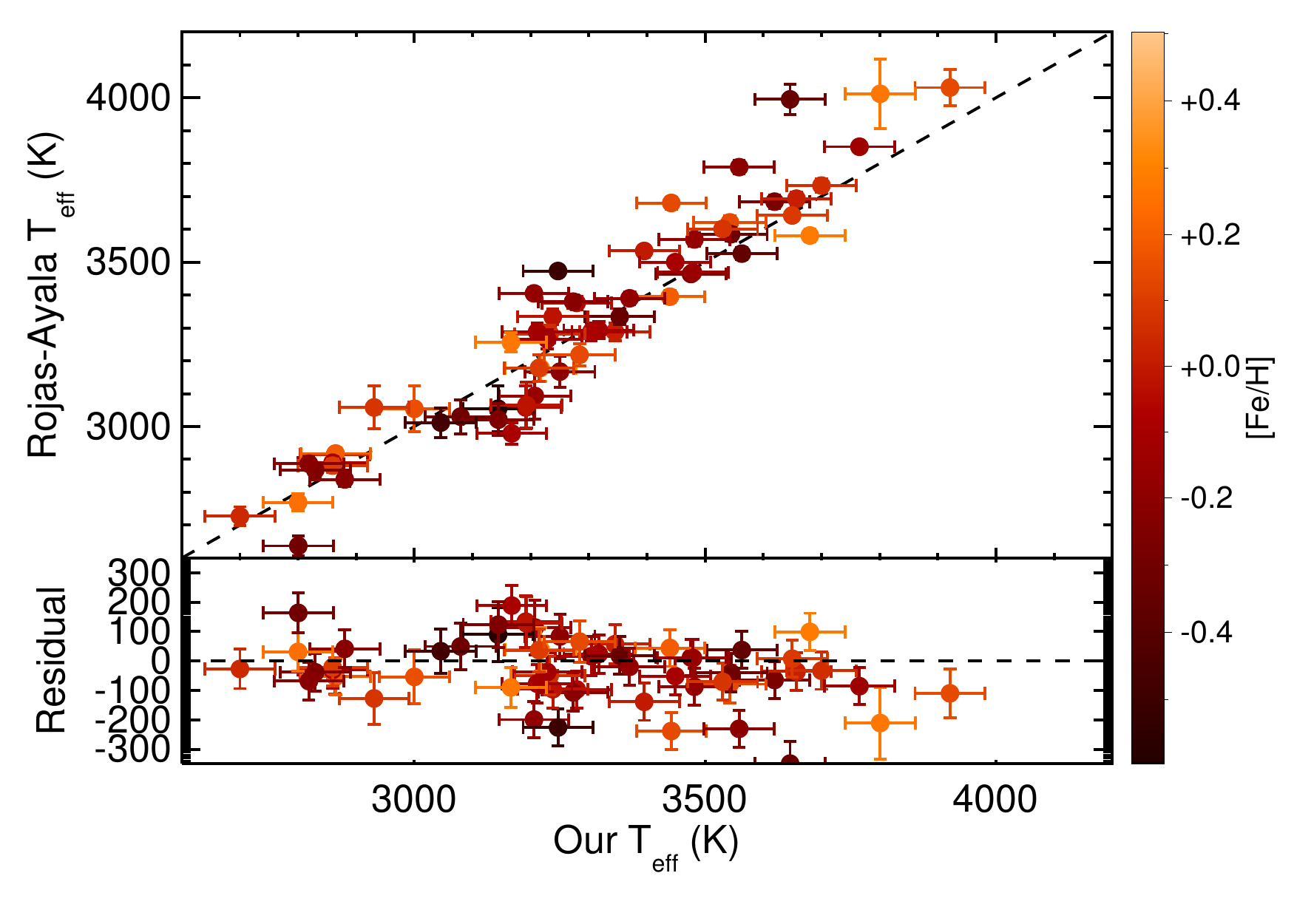}
\caption{Comparison of \teff\ from our analysis and \teff\ from \citet{Rojas-Ayala:2012uq} for the 59 overlapping stars. Points are color-coded by our metallicity determinations, although our [Fe/H] values are consistent with those from \citet{Rojas-Ayala:2012uq}. Further details can be found in Section~\ref{sec:comp_metal}.  \label{fig:rojas}}
\end{center}
\end{figure}

\teff\ values from \citet{2014A&A...568A.121N} were calibrated using the \citet{Casagrande2008} \teff\ scale, which we already discussed in Section~\ref{sec:casa} and found to be problematic. We compared our \teff\ values to those from \citet{Rojas-Ayala:2012uq} in Figure~\ref{fig:rojas}. \citet{Rojas-Ayala:2012uq} estimated \teff\ based on their measurement of H$_2$O-K2 index, which they interpolated onto BT-SETTL models. Because we also used BT-SETTL models the methods are not completely independent. However, \citet{Rojas-Ayala:2012uq} used a different model grid (AGSS; with \citet{Asplund2009} relative abundances), and based their results on NIR instead of optical spectra, so this comparison is still illuminating. The mean difference is 28$\pm$14\,K, which is not significant, but there are some notable trends. \citet{Rojas-Ayala:2012uq} appear to overestimate \teff\ for the warmest stars. The \rchisq\ of the \teff\ difference was 2.4, which is particularly high considering the methods are based on a similar set of models. \citet{Rojas-Ayala:2012uq} estimated the errors in their \teff\ values just from the measurement error in their derived H$_2$O-K2 values, and did not factor in errors in the models or shape errors in the spectra \citep{Newton:2014}. These are exaggerated by the fact that changes in M dwarf \teff s have a weaker effect on NIR spectra than optical. Hence errors from \citet{Rojas-Ayala:2012uq} are probably underestimates.

\section{Empirical Relations}\label{sec:relations}

We calculated empirical relations between observable (e.g., $M_{K_S}$) and intrinsic (e.g., $R_*$) stellar parameters. These relations are valid for $0.1 R_{\odot} \lesssim R_* \lesssim 0.7R_{\odot}$ (spectral types K7--M7, $4.6<M_{K_S}<9.8$, $2700<$\teff$<4100$\,K), and $-0.6<$[Fe/H]$<0.5$. However, because of sample selection biases, the range of metallicities is significantly smaller for stars of spectral types later than M4 (mostly slightly metal-poor) and for the late-K dwarfs (mostly metal-rich). There are also only 15 stars in total with spectral types M5-M7 and only three stars with [Fe/H]$<-0.5$. These sample biases should be considered when applying these formulae. 

For all relations we found the best-fit parameters of polynomial functions with the least-squares minimization algorithm \textit{MPFIT} \citep{2009ASPC..411..251M}. The number of parameters (polynomial order) was determined by an $F$-test; a probability of $>95\%$ that the fit is an improvement is required to increase the fit order. Fits for $R_*$ and $M_*$ can be found in Table~\ref{tab:fits}, for \teff\ (from color) in Table~\ref{tab:teffcolor}, and bolometric corrections (BC) in Table~\ref{tab:bcorr}. 

\subsection{Radius-Absolute Magnitude}\label{sec:LR}

\begin{figure}[t]
\begin{center}
\includegraphics[width=3.5in]{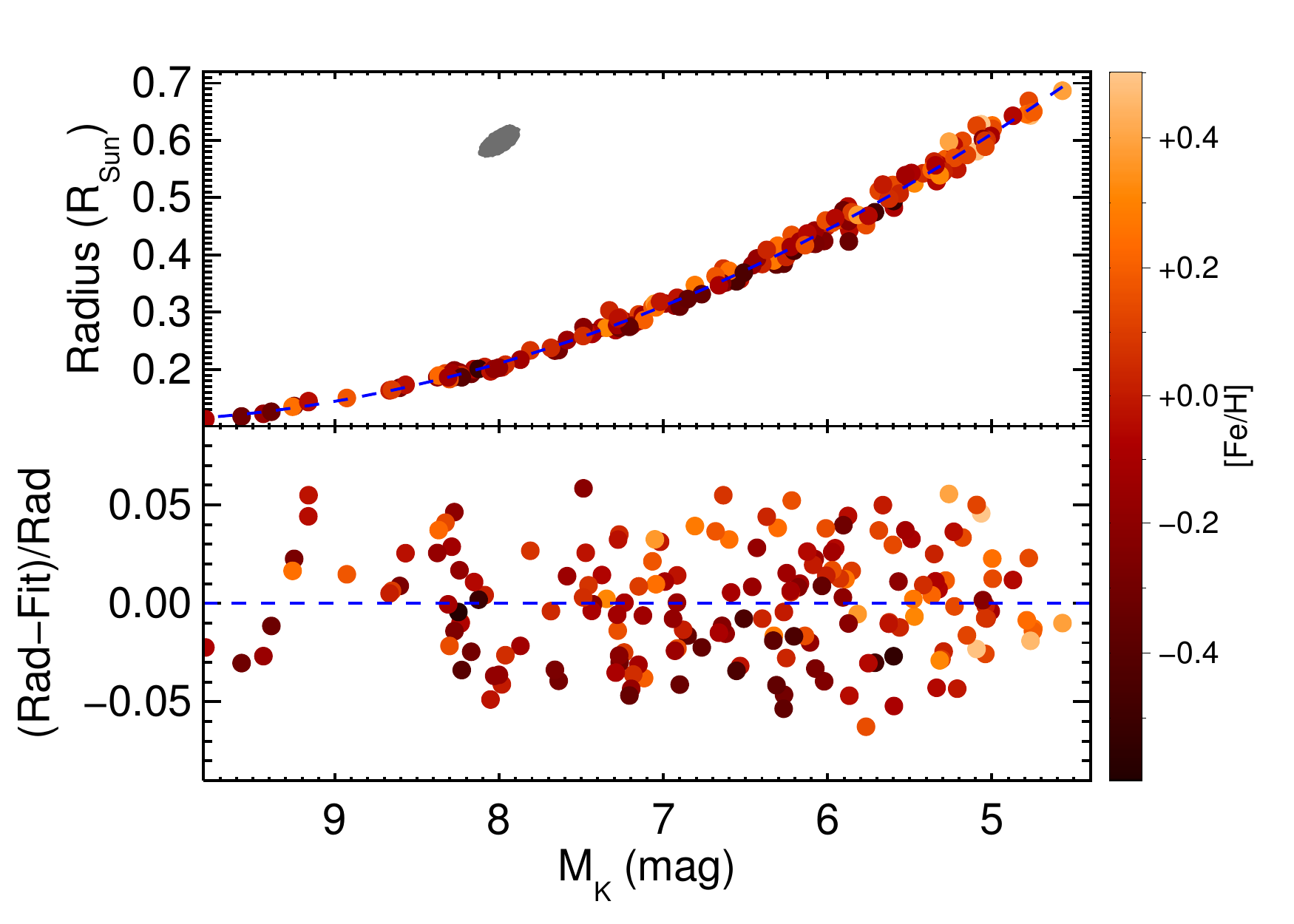}
\caption{Top: $R_*$ as a function of absolute $K_S$-band magnitude. The best-fit to the data is shown as a blue dashed line (see Equation~\ref{eqn:MKR} and Table~\ref{tab:fits}). $M_{K}$ and radius both depend on the distance, so the errors are correlated. Hence we show a one-sigma error ellipse in the top-left which indicates the typical 1$\sigma$ errors for a typical point ($M_{K_S}\simeq6.6$, $R_*\simeq0.35$). Bottom: fractional residual to the fit. All points are color-coded by metallicity. \label{fig:mk_r}}
\end{center} 
\end{figure}

In Figure ~\ref{fig:mk_r} we show the relation between absolute $K_S$ band magnitude ($M_{Ks}$) and $R_*$. We derived the best-fit of the form:
\begin{equation}\label{eqn:MKR}
Y = a+bX+cX^2+...,
\end{equation}
where X is the absolute $K$ band magnitude ($M_{K_S}$), $Y$ is the stellar radius ($R_*$), and $a$, $b$, and $c$ were parameters allowed to float to provide a better fit (reported in Table~\ref{tab:fits}). 

The fit yielded an RMS scatter of only 2.9\%, which is remarkably tight given that typical errors in our radii were 3-4\%. The scatter is low because errors in $M_K$ and $R_*$ are correlated in a direction similar to the trend (see the ellipse in Figure~\ref{fig:mk_r}), a result in turn of both being driven largely by error in the distance. For example, imagine a star with a measured distance 4\% closer than the true value. The star's derived radius will be 4\% greater than the true value, but the $M_{K_S}$ will also be $\simeq$0.1\,mag smaller (brighter). The result is that a 4\% radius error corresponds to only a 1-2\% deviation from the fit (the exact numbers depending on where the point is along the relation). 

\begin{figure*}[t]
\begin{center}
\includegraphics[width=0.9\textwidth]{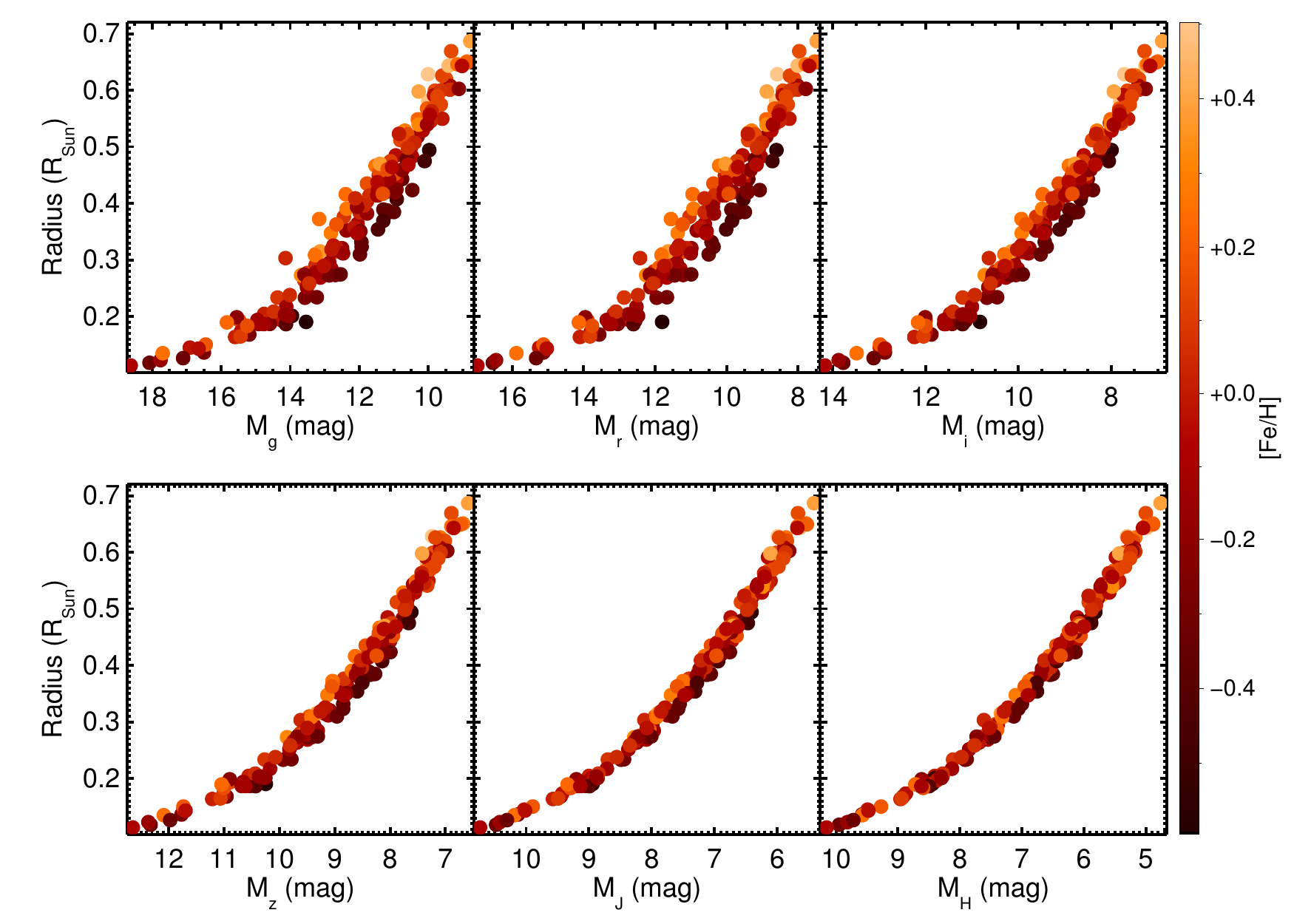}
\caption{Relation between radius and absolute magnitude (for SDSS $griz$ and 2MASS $JH$ bands). The same relation for $K_S$ is shown in Figure~\ref{fig:mk_r}. All magnitudes are generated synthetically from our flux-calibrated spectra to match 2MASS $JH$ and SDSS $griz$ (Section~\ref{sec:cal}). All Y-axis ranges are identical, but the X-axis ranges are different. \label{fig:lum_rad}}
\end{center}
\end{figure*}

We perform a fit including [Fe/H] as a second independent variable:
\begin{equation} \label{eqn:T_R2}
Y = (a + bX+cX^2+...)\times(1+f\rm{[Fe/H]}),
\end{equation}
where (as before) $Y$ = $R_*$ and $X=M_{K_S}$. This particular functional form (multiplying by 1+$f$[Fe/H] term) was used because linear changes in [Fe/H] result in fractional changes in $R_*$. A simpler polynomial fit resulted in noticeable residuals at extreme values of $R_*$ and [Fe/H]. The fit including [Fe/H] produced a small but statistically significant improvement (based on an $F$ test), with a resulting scatter of 2.7\%. This relation is especially useful when trying to reduce systematic errors with metallicity, such as studying correlations between planet size and stellar metallicity. 

We tested the absolute magnitude-radius relations for all SDSS and 2MASS magnitudes ($grizJHK_S$). We used the same formula (Equation~\ref{eqn:MKR}) but the number of parameters was allowed set according to an $F$-test. $K_S$ gave the smallest scatter (2.9\%) and lowest \rchisq\ (0.93), although $M_J$ and $M_H$ performed similarly (3.4\% and 3.2\% scatter, and \rchisq\ of 1.1 and 1.3, respectively). The scatter and \rchisq\ grew as we used filters at bluer wavelengths because of the increasing effect of metallicity on the shape of the of the spectrum \citep{Segransan:2003ul}. This effect has been seen previously when fitting mass-luminosity relations \citep{Delfosse2000}, and even exploited to estimate metallicities of M dwarfs \citep[e.g.,][]{2005A&A...442..635B, Schlaufman:2010qy, 2012A&A...538A..25N}. It manifests in our data as a strong correlation between the luminosity-radius relation for optical colors and [Fe/H], which we show in Figure~\ref{fig:lum_rad}. 

\subsection{Radius--Temperature}\label{sec:TR}
\begin{figure}[htbp]
\begin{center}
\includegraphics[width=0.45\textwidth]{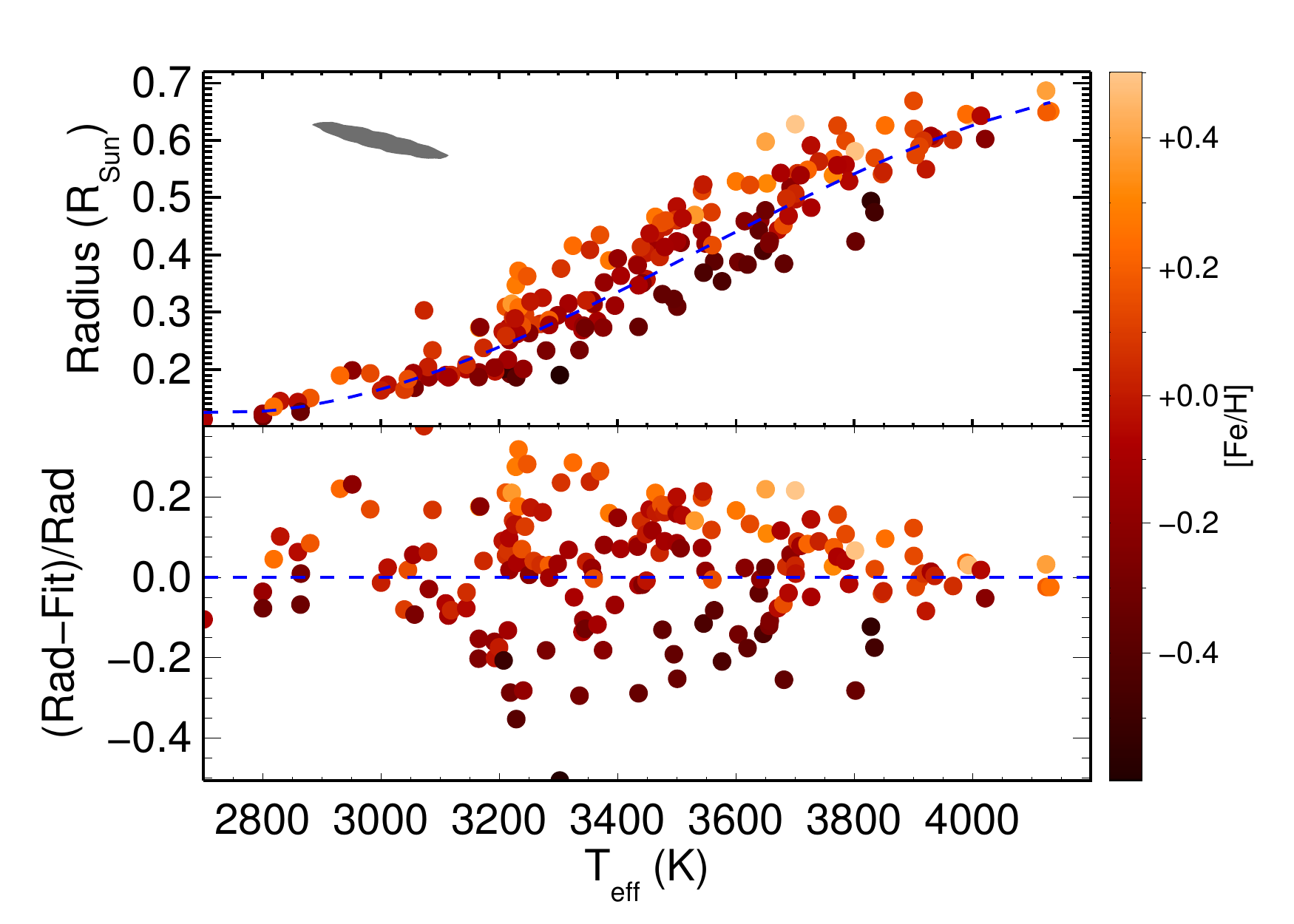}
\caption{Top: $R_*$ as a function of stellar \teff. The derived $R_*$ depends on \teff\ (Equation~\ref{eqn:stefan}) so the errors are strongly correlated. A typical error is shown a gray ellipse in the top left of the plot. The best-fit ignoring [Fe/H] (Equation~\ref{eqn:MKR}) is shown as a dashed blue line. Bottom: residual from the best-fit. Points are colored according to their metallicity. }
\label{fig:t_r}
\end{center}
\end{figure}

In Figure~\ref{fig:t_r} we show stellar radius as a function of \teff. Ignoring metallicity, we derived a fit of the same form as Equation~\ref{eqn:MKR}, but with $X = $T$_{\rm{eff}}/3500$\,K. Values for the fitted variables are in Table~\ref{tab:fits}. There is considerable scatter (13\% in radius) in the relation and the fit has a \rchisq\ of 2.4, in part owing to significant correlation in the residuals with [Fe/H]. Accordingly, we derived a fit accounting for metallicity following Equation~\ref{eqn:T_R2}, again with $X = T_{\rm{eff}}/3500$\,K. Values for the fitted variables are in Table~\ref{tab:fits}. The fit produced a scatter in $R_*$ of 9\%. Accounting for strong correlations between \teff\ and $R_*$ the \rchisq\ for Equation~\ref{eqn:T_R2} was 1.10, suggesting that all of the scatter can be explained by measurement errors, and that the precision of this formula is limited primarily by errors in \teff\ and [Fe/H].

\subsection{Color-Temperature}

\begin{figure*}[h]
    \centering
    \includegraphics[width=0.9\textwidth]{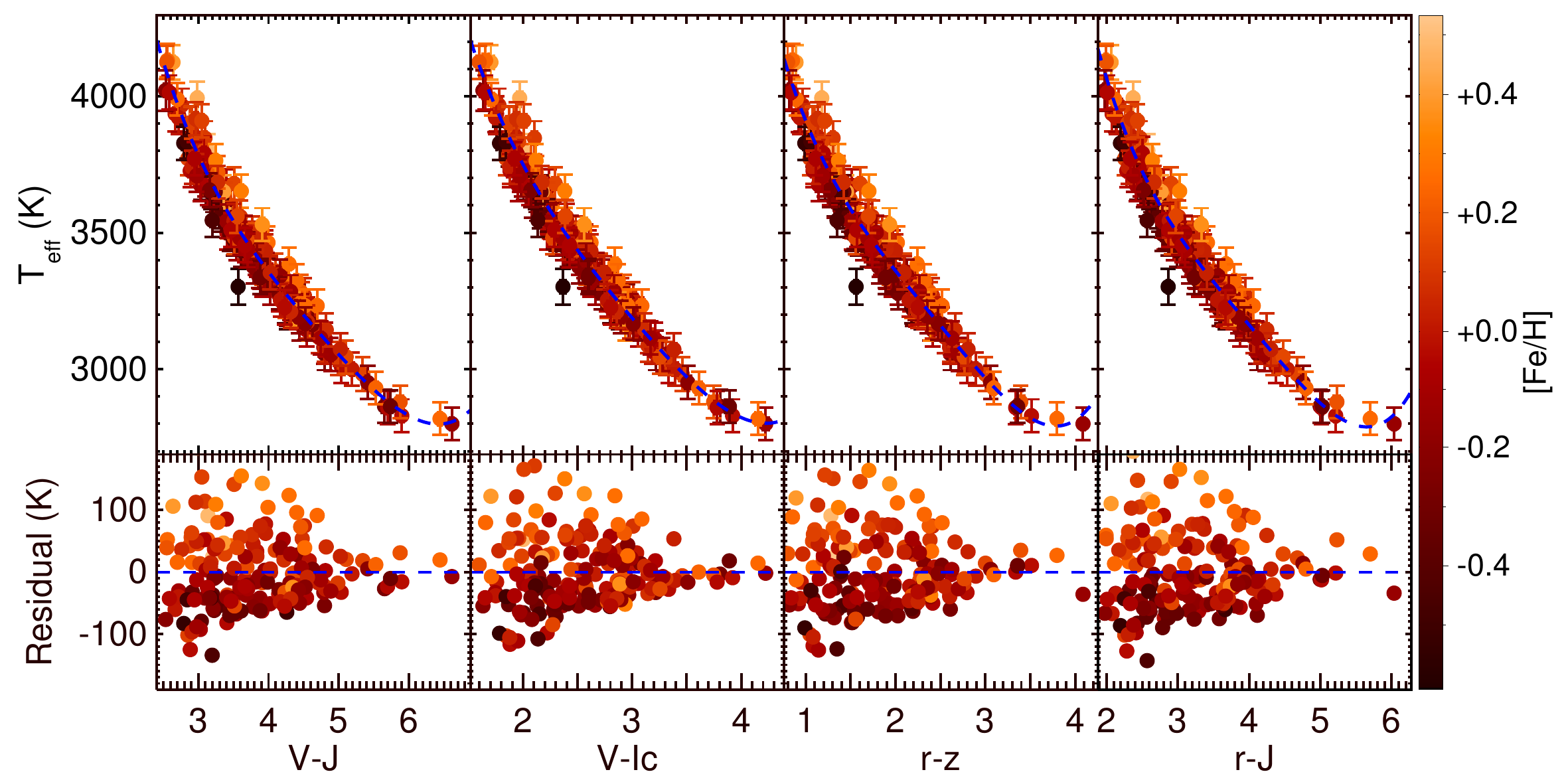} 
    \caption{Spectroscopically derived \teff\ as a function of different color combinations. The best-fit is overplotted as a blue dashed line. The bottom panels show the fit residuals. Fit coefficients are given in Table~\ref{tab:teffcolor}.}
    \label{fig:teffcolor}
\end{figure*}

We used our synthetic magnitudes to construct colors and fit color-\teff\ relations following the functional form of Equation~\ref{eqn:MKR}, with $Y=$\teff/3500 and $X$ representing the relevant color. Figure~\ref{fig:teffcolor} plots the relations for some of the colors that are most predictive of \teff\ (based on the standard deviation) and Table~\ref{tab:teffcolor} reports the fit parameters. Readers are cautioned against using our empirical relations outside the color range of our sample where empirical fits show slope changes that are probably not real (See Figure~\ref{fig:teffcolor}). Additional color relations can be derived using the \teff\ values in Table~\ref{tab:sample} and synthetic photometry provided in Table~\ref{tab:photsample}. 

All relations show significant dependence on metallicity, and hence should be not be used on stars with metallicities far from solar. For cases where the target's metallicity is known, we derive relations following the formula:
\begin{equation}\label{eqn:teff2}
\teff = a+bX+cX^2+dX^3+eX^4+f(\rm{[Fe/H]}).
\end{equation}
In the case that metallically is not known, $JHK_S$ colors can be used to approximate metallicity \citep{Leggett:1992lr, Johnson:2012fk, Mann2013a, Newton:2014}. Motivated by this, we tested different color combinations and found that the metallicity term could be best mitigated by including $J-H$ in the fit using the formula:
\begin{align}\label{eqn:teff3}
\teff =&\ a+bX+cX^2+dX^2+eX^3\\
&+f(J-H)+g(J-H)^2. \nonumber
\end{align}
Including $J-H$ color significantly reduced, but did not totally eliminate, the residuals with [Fe/H]. The scatter is notably smaller in \teff\ when using [Fe/H] then when using $J-H$ color, and therefore actual metallicities should be used over the single-color formulae if [Fe/H] can be independently determined.

\begin{figure*}[htbp]
    \centering
    \includegraphics[width=0.9\textwidth]{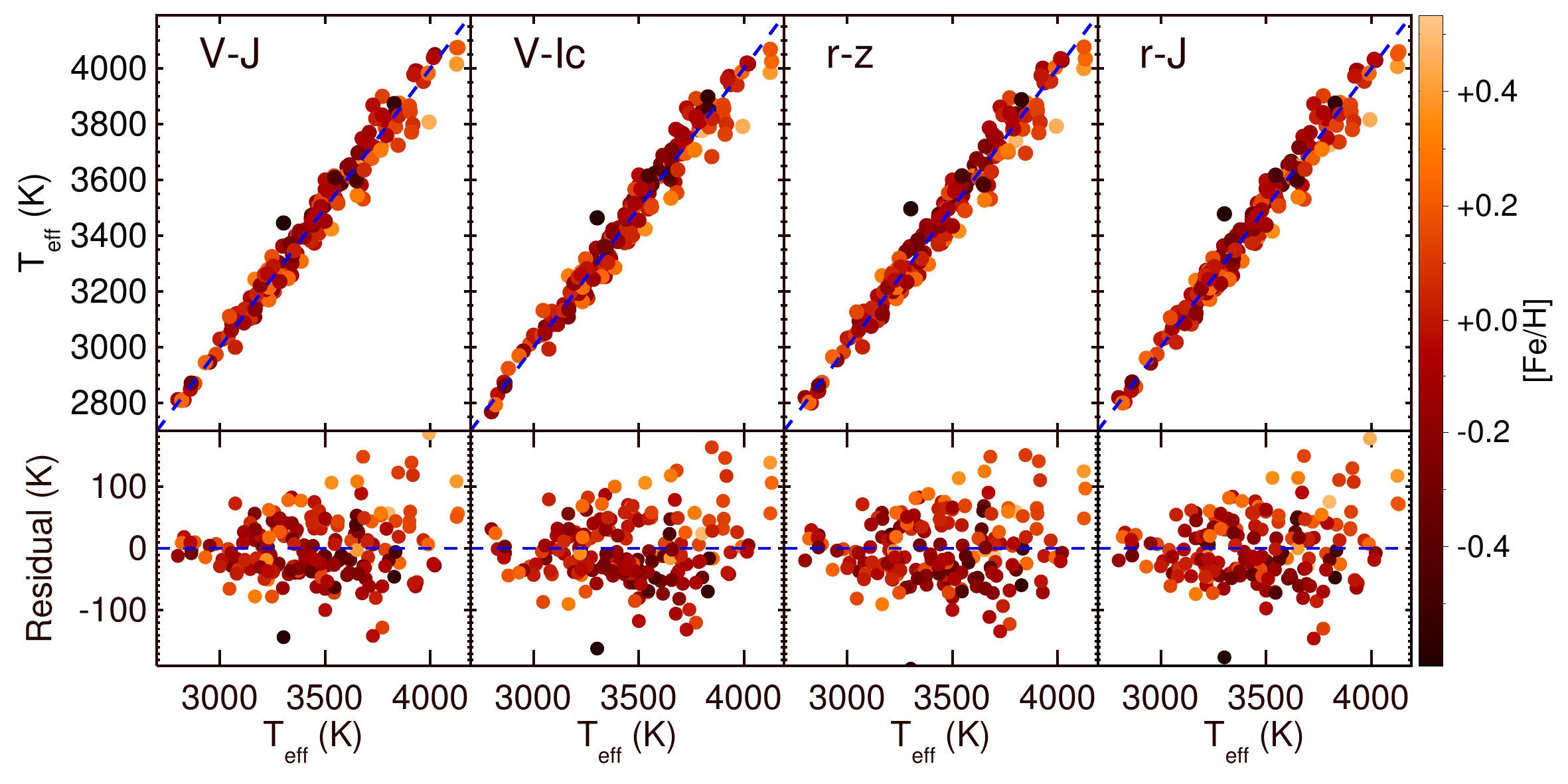} 
    \caption{Spectroscopically derived \teff\ vs. \teff\ derived from the color marked in the top left of each pane and $J-H$ to correct for metallicity. The bottom panels show the best-fit residuals. Best-fit coefficients are given in Table~\ref{tab:teffcolor}.}
    \label{fig:teff_color2}
\end{figure*}

A comparison of spectroscopically derived \teff\ vs. \teff\ from Equation~\ref{eqn:teff3} is shown in Figure~\ref{fig:teff_color2} and the coefficients are given in Table~\ref{tab:teffcolor}. 

\section{Bolometric Corrections}\label{sec:bcorr}

\begin{figure*}[t]
    \begin{center}
    \includegraphics[width=0.25\textwidth]{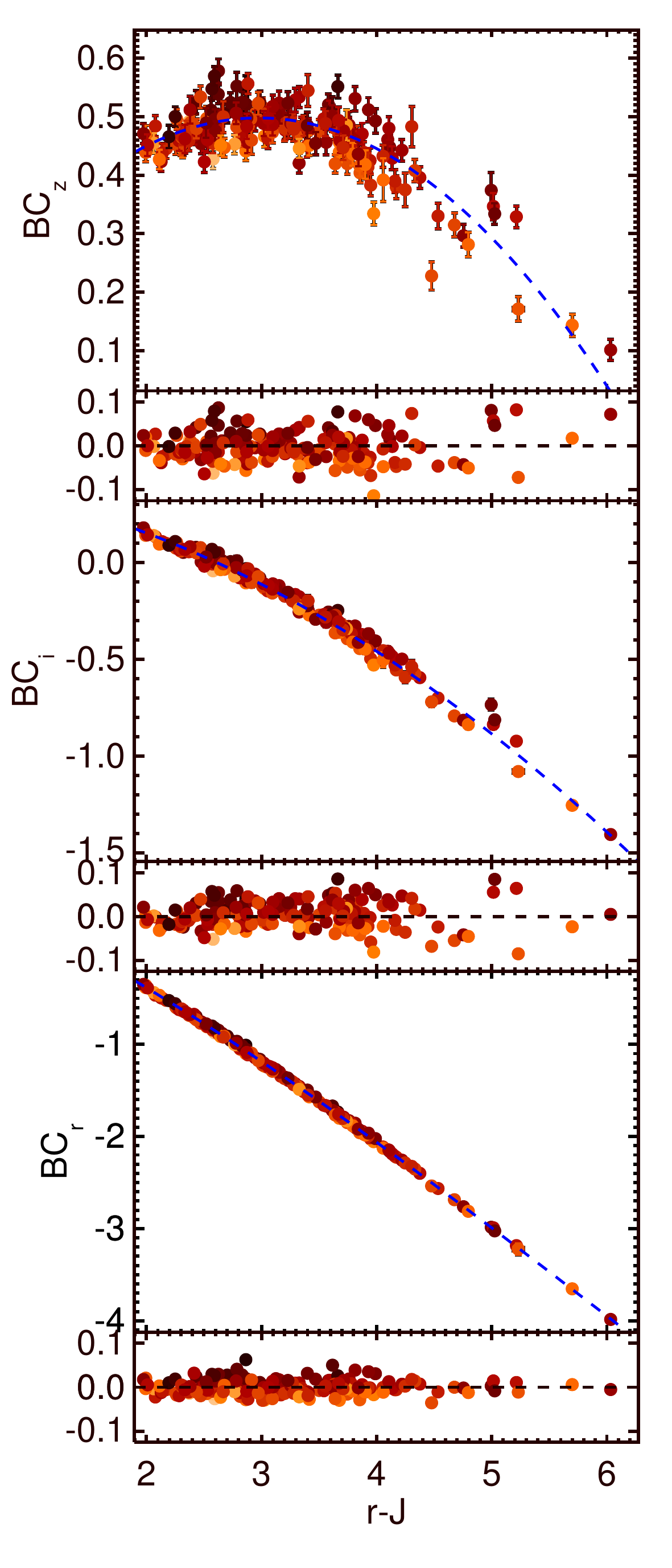}
    \includegraphics[width=0.25\textwidth]{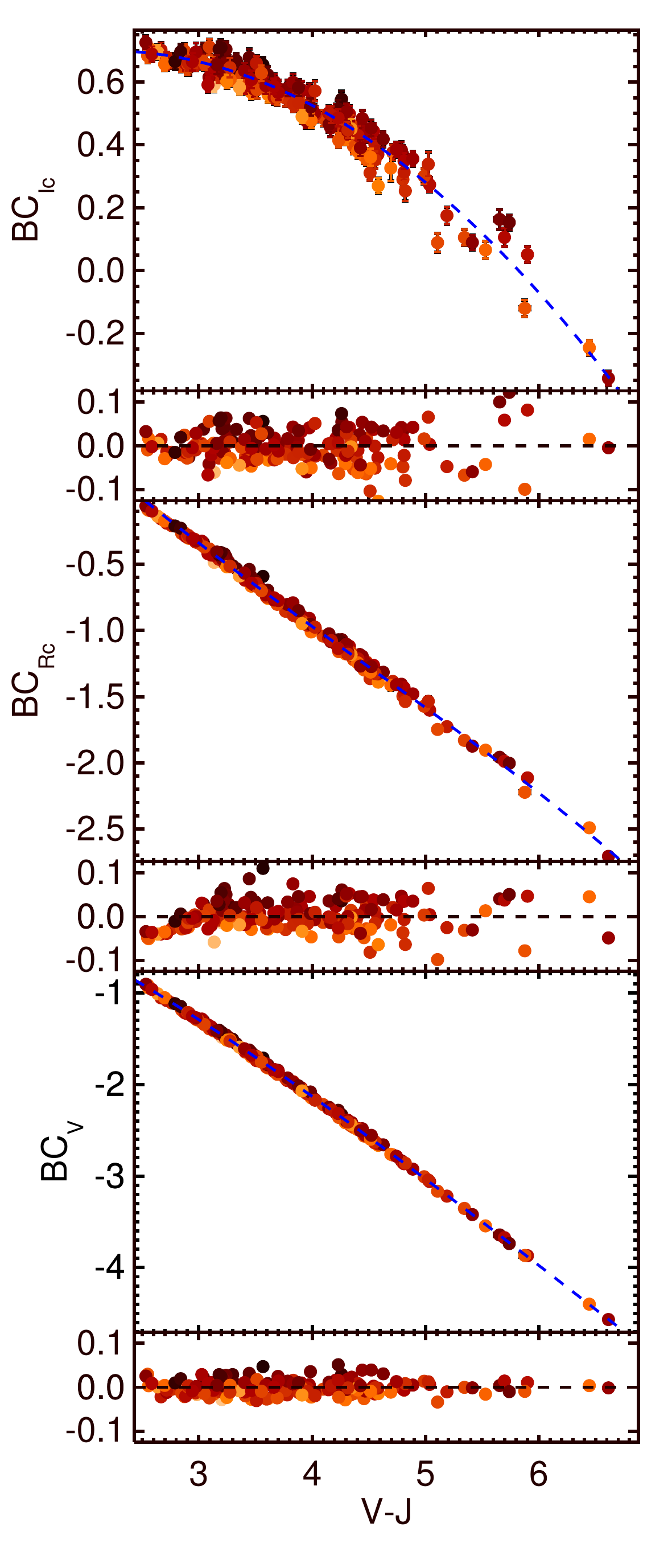}
    \includegraphics[width=0.49\textwidth]{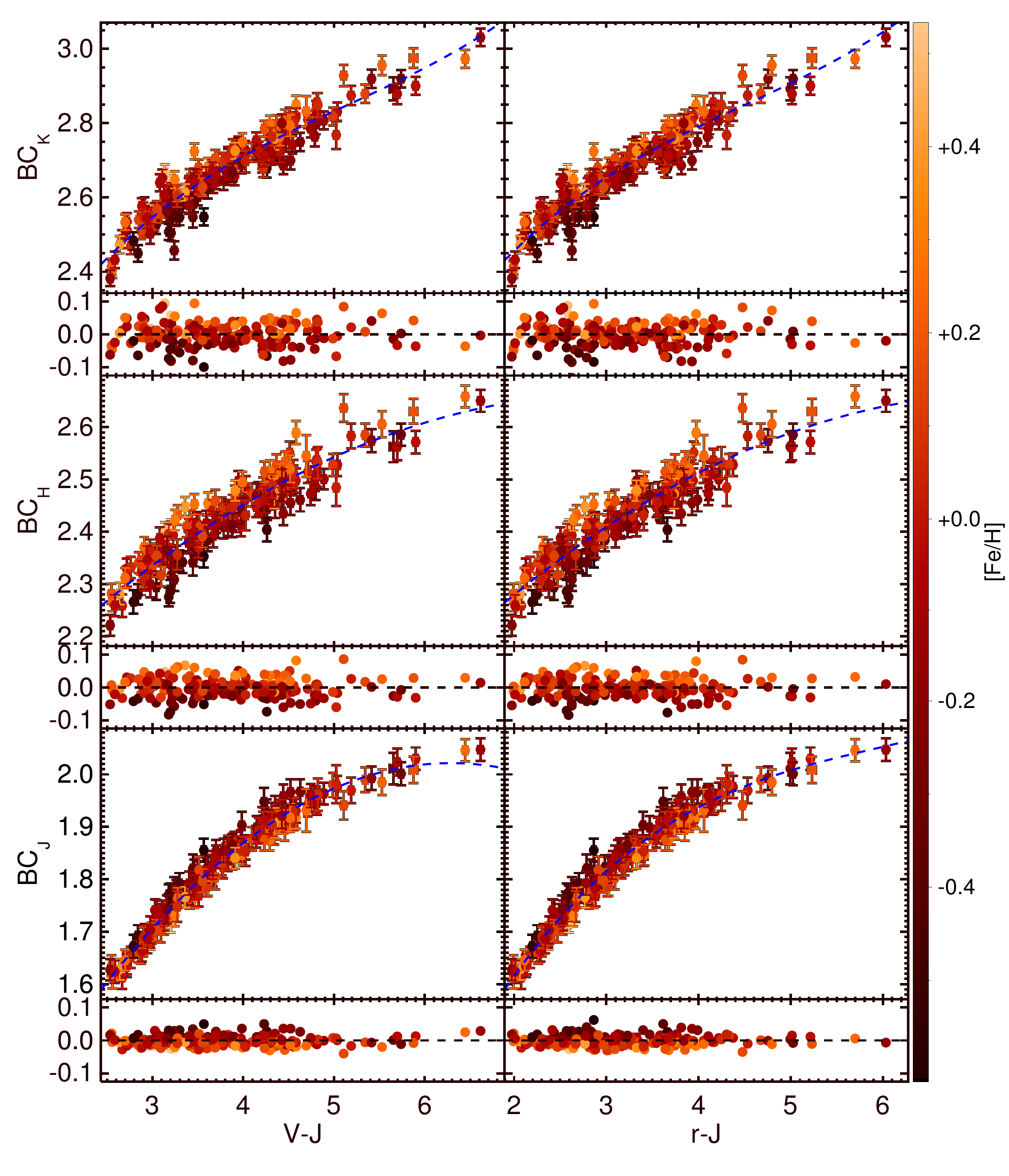}
    \caption{Bolometric corrections for SDSS ($riz$) passbands vs. $r-J$ (left), Johnson-Cousins $VR_cI_C$ vs. $V-J$ (middle left), and  2MASS $JHK_S$ vs. $V-J$ (middle right) and $r-J$ (right). Best-fits are plotted in blue (dashed line). Small panels below each of the relations show the residuals of the fit. Coefficients for each fit can be found in Table~\ref{tab:bcorr}. Note that the scales for Y-axes in the residual plots are all the same. Points are color-coded according to metallicity. See Section~\ref{sec:bcorr} for more details. \label{fig:bc1}}
    \end{center}
\end{figure*}

\begin{figure}[htbp]
    \begin{center}
    \includegraphics[width=0.445\textwidth]{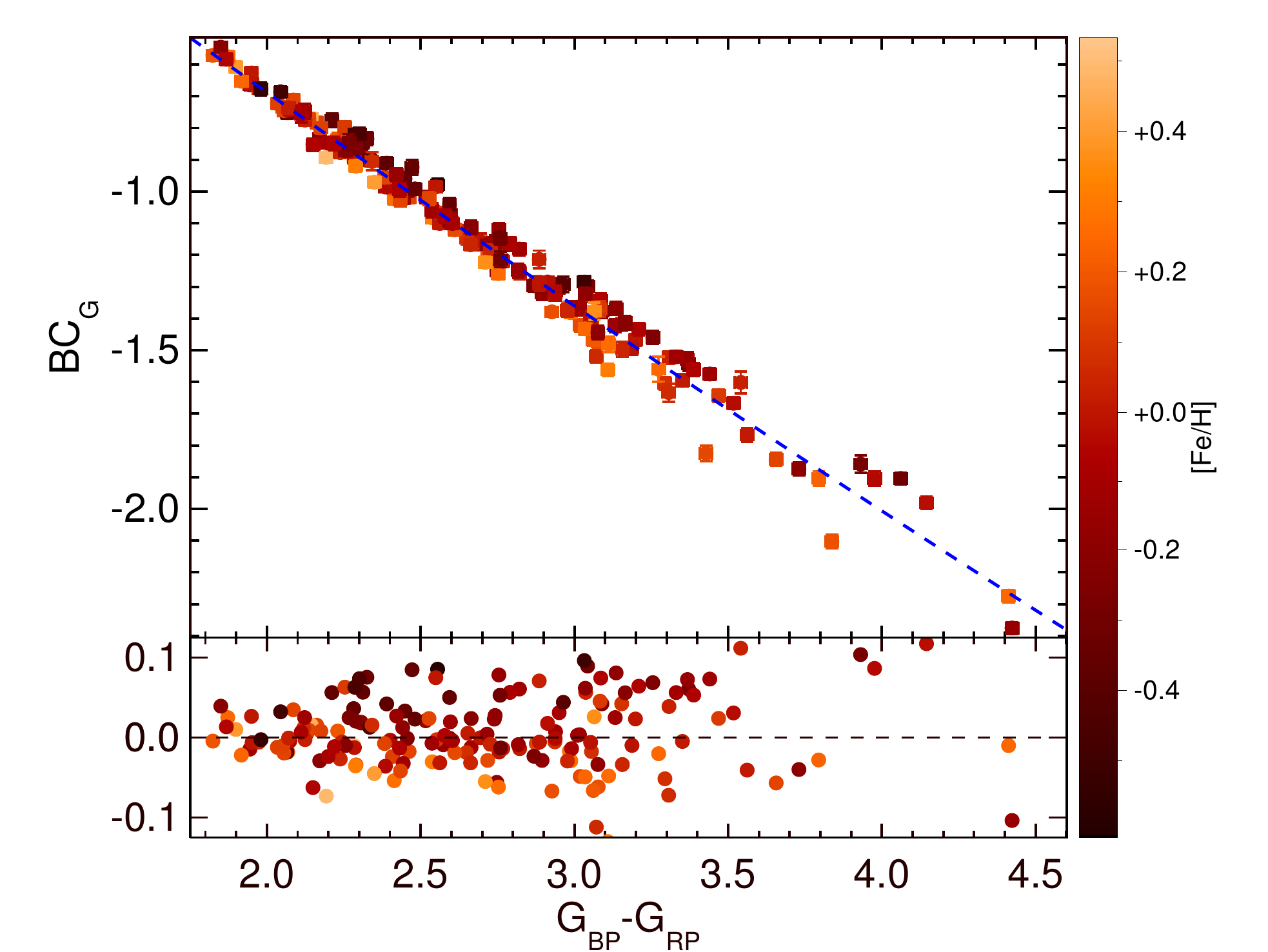}
    \caption{ame as Figure~\ref{fig:bc1} but for the {\it Gaia} $G$ passband vs. $G_{BP}-G_{RP}$ color. \label{fig:bc2}}
    \end{center}
\end{figure}

We calculated apparent bolometric magnitudes from our \fbol\ values using $m_{\odot}=-26.8167$ \citep{2012ApJ...754L..20M, 2013ApJS..208....9P}\footnote{see \href{http://sites.google.com/site/mamajeksstarnotes/bc-scale}{http://sites.google.com/site/mamajeksstarnotes/bc-scale}.}. We then derived bolometric corrections for each filter for which we have synthetic magnitudes ($VR_CI_CgrizJHK_S$ and {\it Gaia} $G$). We fit BCs with a second order polynomial in a color and performed these fits for all major passbands for which we have synthetic magnitudes ($VR_CI_CgrizJHK_S$), using all possible color combinations. Relations using $V-J$ and $r-J$ colors exhibited the smallest standard deviations between fit and observed BC, and we show these for the most relevant passbands in Figures~\ref{fig:bc1} and \ref{fig:bc2}. Coefficients for the polynomial fits are given in Table~\ref{tab:bcorr}. The scatter indicates that we can infer BC to $2-3\%$ using just a $V-J$ or $r-J$ color. 

All of the fit residuals show significant correlations with [Fe/H]. We derived additional relations useful for when the metallicity is known by adding an $e$([Fe/H]) term to the fits (given in Table~\ref{tab:bcorr}). When metallicity is included it is possible to estimate BC within 1-3\%. For cases where the metallicity of the star is not known, it is possible to mitigate these systematics by averaging over several corrections. For example, for BC$_V$, metal-rich stars land preferentially above the fit (larger BC), while for BC$_J$ they land preferentially below (for both $V-J$ and $r-J$).

\section{Comparison with Predictions of a Stellar Evolution Model}
\label{sec:models}

Predictions from the Dartmouth stellar evolution models \citep{Dotter2008,Feiden2012a,Feiden2014a} were compared with the parameters measured or estimated in Section~\ref{sec:tefffbol}. Previous investigations have shown that disagreements between model predictions and observations of low-mass stars are of similar magnitude among standard stellar evolution models that adopt appropriate physics for the physical conditions in these objects \citep[e.g., Lyon, Dartmouth, Yale, Pisa, and Padova models;][]{Feiden2012a,Boyajian2012,Spada2013,Chen2014,Torres2014}. Therefore, our results should be applicable to other model sets.

\subsection{Model Physics \& Grid Description}
Models were computed using an updated version of the Dartmouth stellar evolution code \citep{Dotter2008}. The updates have been previously described \citep{Feiden2013,Feiden2014a,Muirhead2014}, but here we summarize changes relevant to low-mass stars. First, nuclear reaction cross sections were updated to the recommended values from the Solar Fusion {\sc ii} review \citep{Adelberger2011}. This largely affects the cross section for the primary channel of the  proton-proton chain, although nuclear reaction rates in low-mass stars are not significantly altered. Second, the specification of surface boundary conditions was moved deeper into the interior model to the location where the optical depth $\tau = 10$ to more accurately treat regions in low-mass stars where convection occurs in optically thin layers. Compare this with the formulation in \citet{Dotter2008} where surface boundary conditions are defined where $T = \teff$, which leads to boundary conditions being specified in layers where $\tau \ll 1$. As described in \citet{Muirhead2014}, moving the fit point to an optical depth $\tau = 10$ makes computations more reliable below 0.20 \ms, but also causes all low-mass models to have hotter \teff s by up to 60~K. Model radii are not affected significantly by this update, meaning there is an increase in luminosity owing to the increase in \teff. Finally, \citet{Feiden2014a} revised the grid of model atmospheres used to define surface boundary conditions to provide a finer sampling with metallicity. This was achieved by interpolating within the existing grid of model atmosphere structures, and allows for greater accuracy when calculating models with metallicities between the values in the previous grid. Note that the models still adopt PHOENIX Next-Gen model atmospheres \citep{Hauschildt1999} with the \citet{GS98} relative abundance pattern for the Sun.

A high resolution grid was constructed with the aforementioned models to permit more reliable interpolation of observed stellar properties. Specifically, models were computed with masses in the range of 0.10 -- 0.80 \ms\ in steps of 0.02 \ms\ and metallicities in the range of $-0.5$ to $+0.5$ dex with spacing of $0.1$ dex. The helium mass fraction was assumed to be linearly proportional to the heavy element mass fraction with slope $\Delta Y / \Delta Z \approx 1.6$ and $y$-intercept equal to the primordial helium abundance $Y_p = 0.2488$ \citep{Peimbert2007}, and assuming a solar-calibrated mixing length parameter (\amlt\ = 1.884). The influence of relaxing these assumptions is explored for a few characteristic systems in Section~\ref{sec:knob_fiddle}.

\subsection{Stellar Parameter Estimation}
\label{sec:parest}
Stellar model parameters were inferred using a Markov Chain Monte Carlo (MCMC) method implemented with \texttt{emcee} \citep{Foreman-Mackey2013}. The \texttt{emcee} package adopts an affine--invariant ensemble sampling algorithm \citep{Goodman2010} to evolve a set of random walkers that sample the posterior probability distribution (PPD) within the model grid parameter space. Our sampler explores the mass, age, metallicity, and distance PPDs to find an optimal fit to the observations (\teff, \fbol, distance, metallicity) using model outputs (\teff, luminosity) and applying constraints from observations on observable parameters (metallicity, distance). The probability that a random model realization corresponds to the properties of a star in our sample is then evaluated using a likelihood function:
\begin{multline} 
\label{eqn:likelihood}
    \mathcal{L}({\bf \Theta}; {\bf X}) = \prod_{n = 1}^{N}\left( \frac{1}{2\pi\sigma_n^2}\right)^{1/2} \times \\
    \exp\left(-\beta\sum_{n = 1}^{N} \left[ \frac{\vartheta_n({\bf \Theta}) - X_n}{\sigma_n} \right]^2\right),
\end{multline}
where ${\bf X}$ is a vector of length $N$ containing the observed data (stellar \teff\ and \fbol) and ${\bf \vartheta (\Theta)}$ is a vector of model predictions for a given set of unknown parameters ${\bf \Theta}$, which contains both observable (e.g., metallicity, distance) and unobservable (e.g., mass, age) parameters. Prior information about the stellar metallicity and distance is known from our observations, allowing us to further constrain the PPD. We describe the prior probability distribution for those quantities assuming the true value is normally distributed about the observed value with a standard deviation equal to the formal error. Priors on mass and age are taken to be uniform over the range allowed by the definition of our model grid given in Section~\ref{sec:models}. 

For each sample from the joint PPD, the corresponding set of model observables is obtained by linearly interpolating onto a 4D convex hull\footnote{A convex hull for a set of points is defined as the smallest region in parameter space that contains the set of points within which any pair can be connected by a straight line segment without crossing the boundary of that region.} formed by the full grid of model parameters through Delaunay triangulation.\footnote{Given a set of points that have been triangulated, a Delaunay triangulation optimizes the minimum angle within the set of triangulations by ensuring that the circumcircle of each triangle contains no points from the original set within it.} We use the Quickhull algorithm for computing the convex hull \citep{Barber1996}, included in the SciPy interpolation package as the \texttt{LinearNDInterpolator} routine. We tested the accuracy of the interpolation routine by computing several 5~Gyr model isochrones using the 4D interpolation and then comparing it with 5 Gyr isochrones generated directly from the model mass tracks using cubic-spline interpolation. Results from the 4D interpolation were in agreement to better than 0.1\%, except in the region between 0.14 and 0.18\ms\ at higher metallicities, where there are gaps in the model grid due to convergence issues. In this region, the 4D interpolation was accurate to within 2\%.

Parameters from the MCMC analysis were tested to ensure they were not biased by the specific MCMC algorithm. We compared results from a standard single-walker Metropolis--Hastings algorithm to results produced by \texttt{emcee} for a set of four well-characterized stars in different regions of the model grid parameter space. In all cases, the two approaches yielded identical results, with the exception that \texttt{emcee} produced larger estimates for the model uncertainties. This is most likely a consequence of any single-walker becoming stuck in local minima in the vicinity of the global minimum. This was supported by separate experiments on the single-walker MCMC, where we varied the input parameters such as widths of Gaussian priors, the initial conditions, and the chain length. Most telling was that the use of a rudimentary simulated annealing produces errors more consistent with those of \texttt{emcee}. Different values of the factor $\beta$ in Equation \ref{eqn:likelihood} were tested and found to not influence the results. Therefore, we adopted a single value $\beta = 1/2$. Finally, for \texttt{emcee}, we settled on using 400 walkers with 100 iterations. The ensemble sampler relaxes to a stable distribution after approximately 40 iterations. We discard the first 10 ``burn in'' steps of each walker, which are affected by the initial conditions. Otherwise, we do not reject any additional steps taken by the walkers.

\begin{figure*}[htbp]
    \centering
    \includegraphics[width=0.45\linewidth]{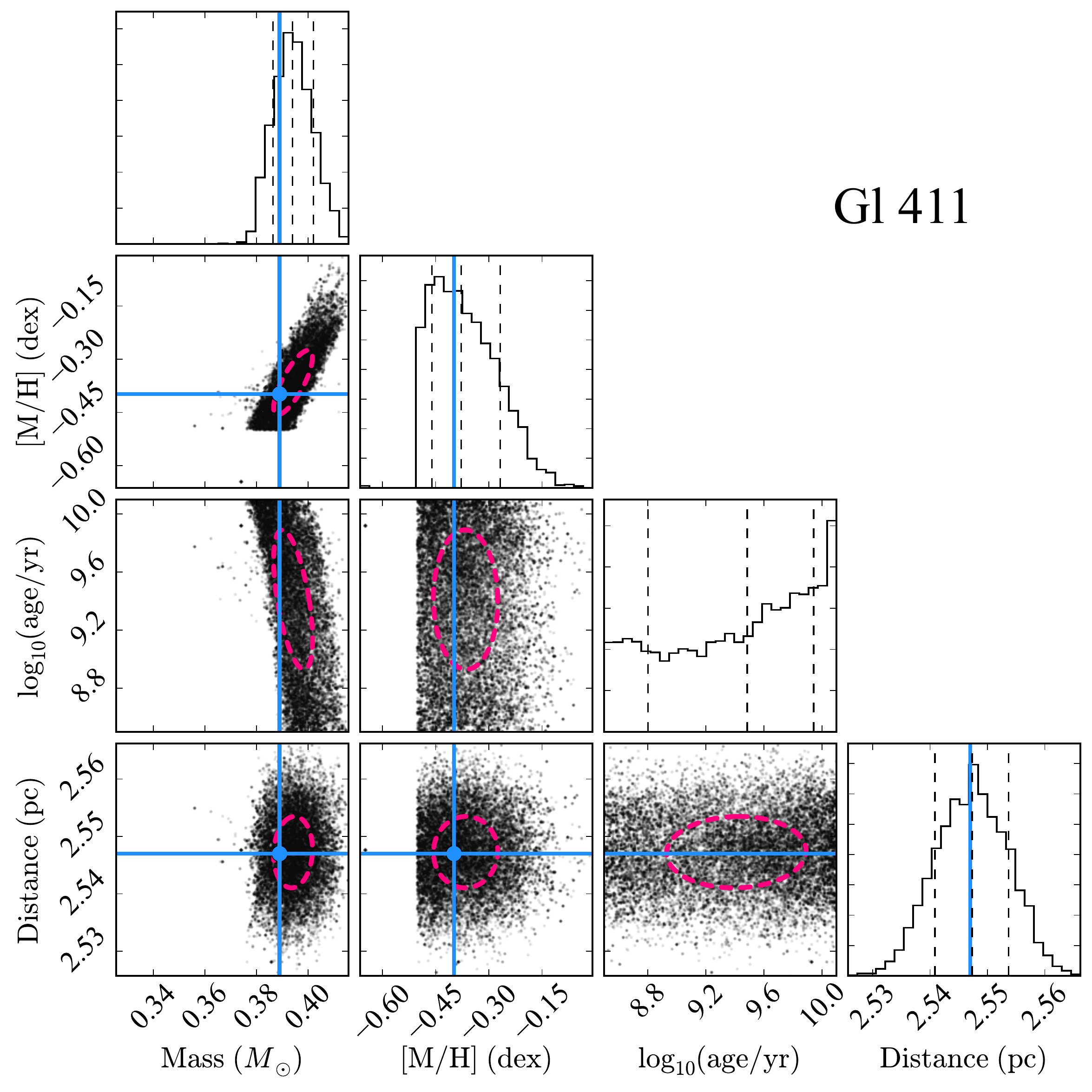} \qquad
    \includegraphics[width=0.45\linewidth]{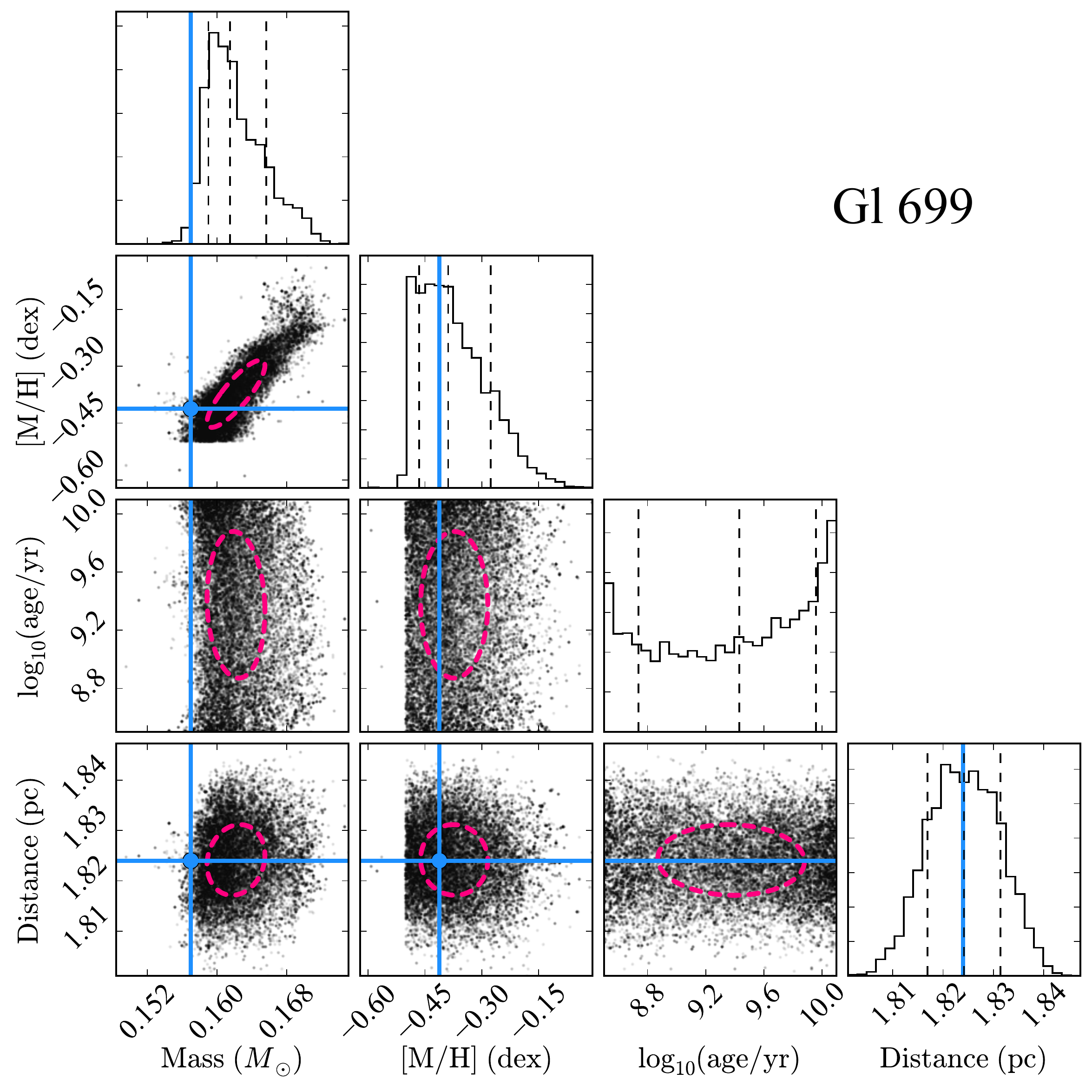} \\
    \includegraphics[width=0.45\linewidth]{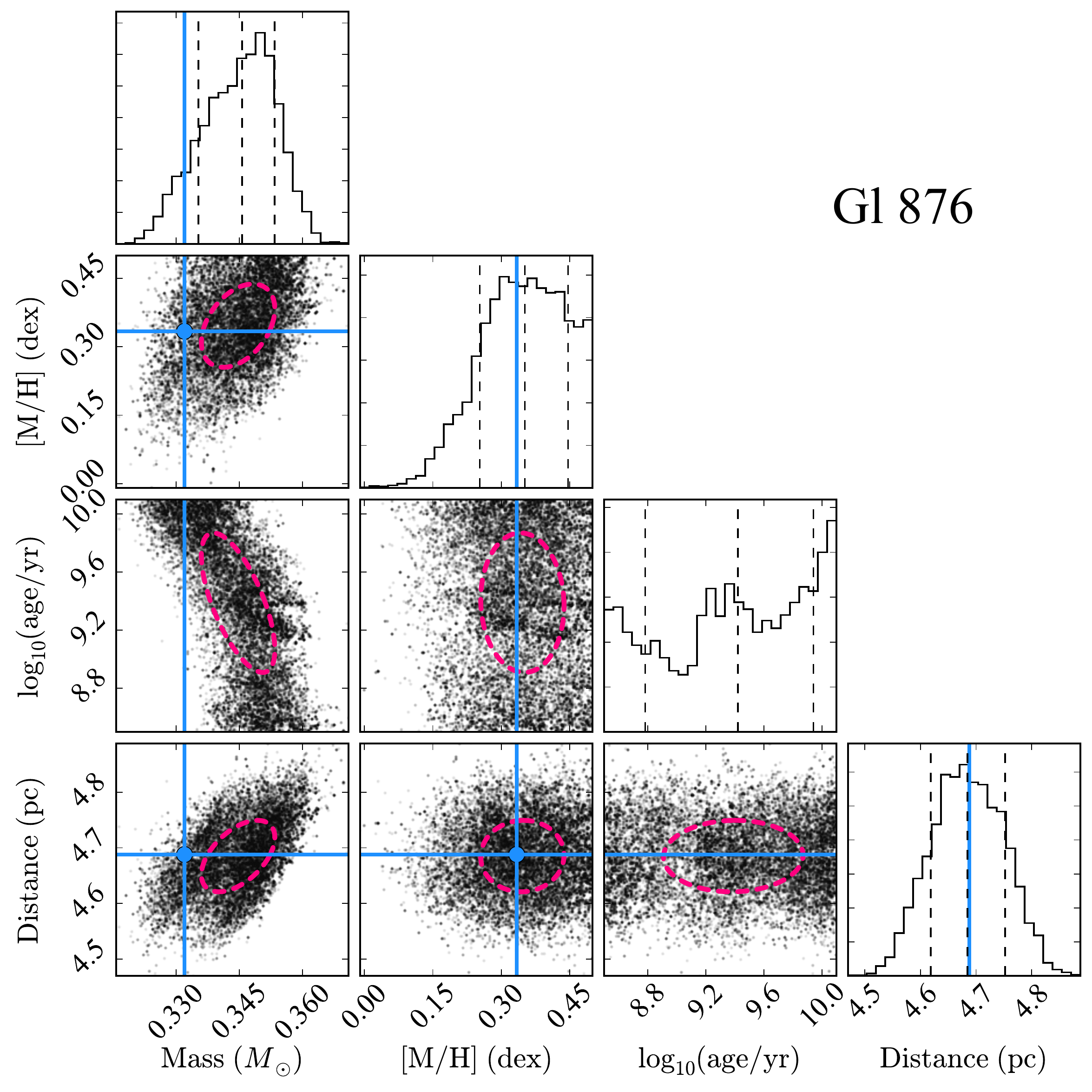} \qquad
    \includegraphics[width=0.45\linewidth]{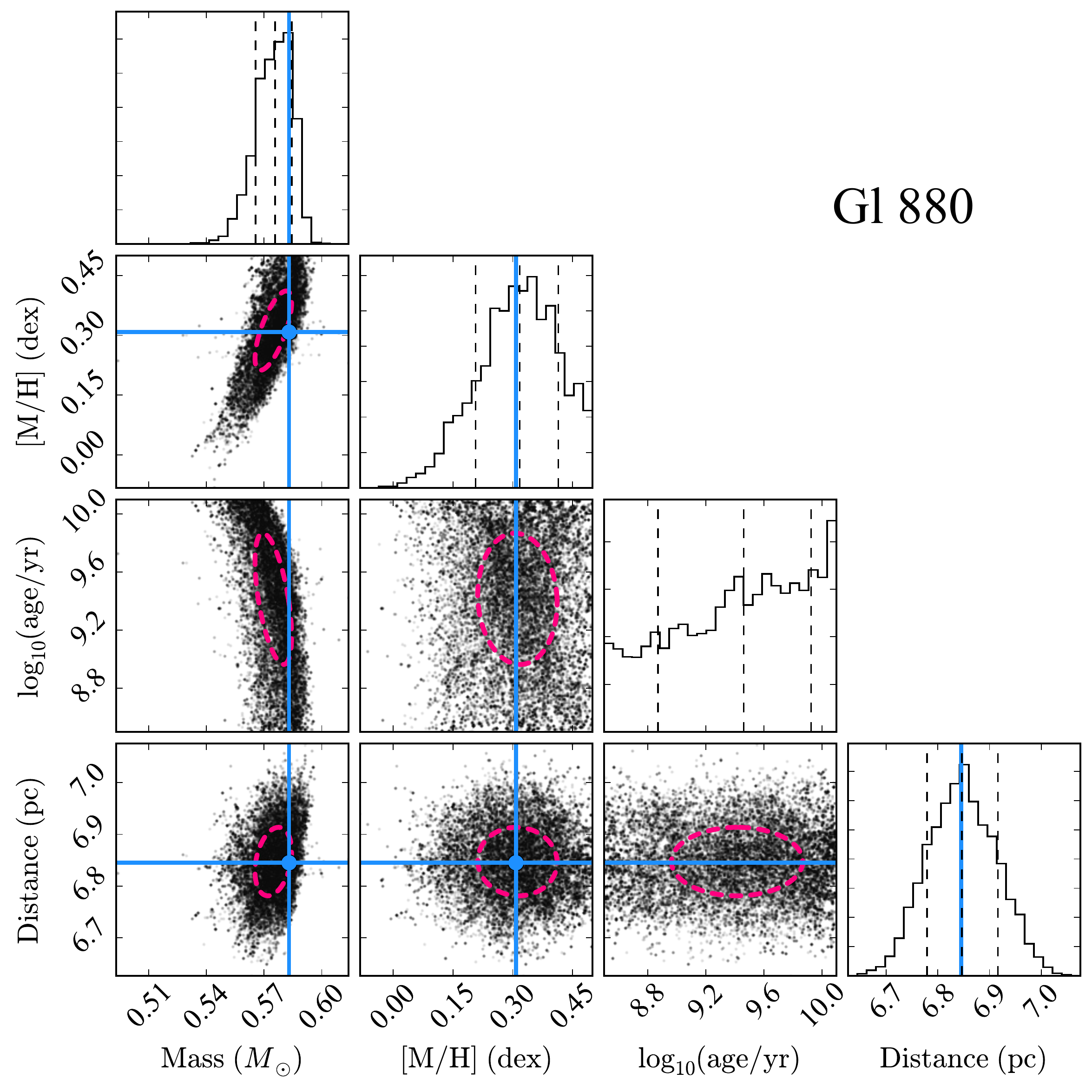}
    \caption{Mass, metallicity, age, and distance posterior probability distributions (PPDs) for Gl~411 (top, left), 
    Gl~699 (top, right), Gl~876 (bottom, left), and Gl~880 (bottom, right). Vertical dashed lines define the 16th,
    50th, and 84th quantiles of the distributions. Error ellipses are shown with pink dashed lines. Blue lines mark
    the observed value for the different quantities, except for the stellar age. Masses are derived from the 
    \citet{Delfosse2000} empirical relation. }
    \label{fig:ppds}
\end{figure*}

Examples of the PPDs for four stars are shown in Figure~\ref{fig:ppds}, i.e., the same four stars whose spectra are shown in Figure~\ref{fig:spectra}. Recall that Gl~699 and Gl~876 are anticipated to be fully convective, while Gl~411 and Gl~880 are expected to possess a radiative core. Resulting metallicity PPDs are also roughly Gaussian, however, they exhibit sharp cutoffs at either the high or low metallicity end of the distribution, depending on the star's metallicity. This is a consequence of the limits of the model grid, but does not appear to strongly affect the mass or distance  PPDs. One can also see there is little dependence on stellar age for the best-fit model parameters describing these stars, although the age distribution for any given star appears to be mildly skewed toward older ages greater than about 8~Gyr.

\subsection{Modeling Results}
\label{sec:model_results}

Stellar parameter values inferred from our comparison to the Dartmouth models are reported in Table~\ref{tab:model_results}. We were able to derive model parameters for 178 of 183 of the stars in our sample. Five of the stars have observed properties outside of the convex hull used for interpolation. PM~I09437-1747 is too metal-poor for our model grid, while PM~I02530+1652, GJ~1111, Gl~444~C, and Gl~412~B have \teff s that are too cool for the model grid.

\begin{figure}[t]
    \centering
    \includegraphics[width=0.4\textwidth]{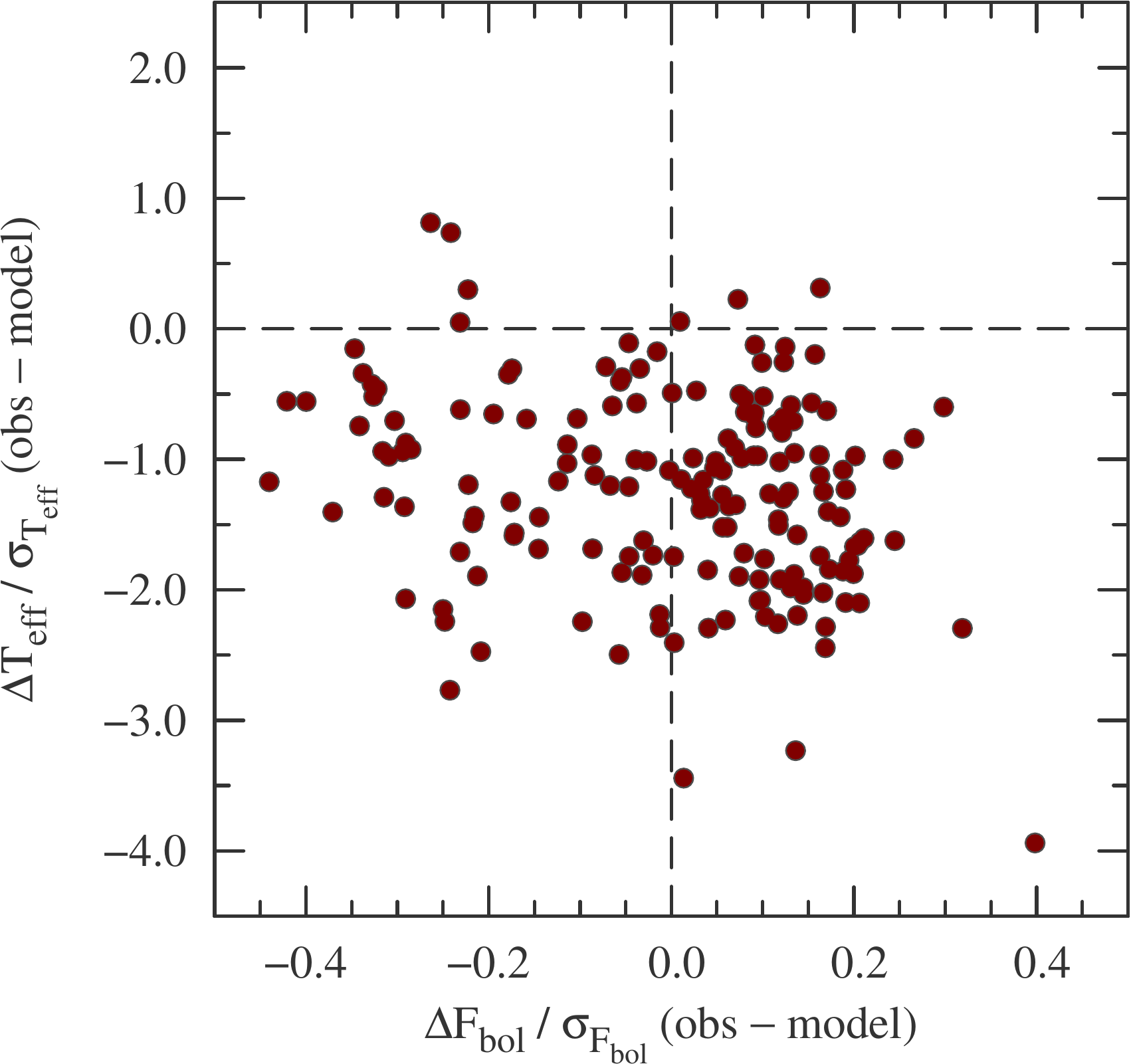}
    \caption{Luminosity and temperature residuals between observations and best-fit model predictions normalized
     to the observed $1\sigma$ observational uncertainties. Zero points are marked with dashed lines.}
    \label{fig:lum_teff_resids}
\end{figure}

The overall quality of model fits are shown in Figure~\ref{fig:lum_teff_resids}. Relative errors\footnote{by errors we mean difference between model and empirical determinations} of the model-predicted \teff s are plotted against the relative errors of the model-predicted \fbol. Errors are quoted in number of standard deviations from the observed mean value. There is broad agreement between the model and observed bolometric fluxes, with all stars formally fit within $1\sigma$ of their quoted observational errors. The average offset of model predictions from the observed \fbol\ values is $0.01\pm0.25$\%, with the distribution around the zero point being consistent with a normal distribution.

We find a significant correlation between the model$-$observed \fbol\ and the observed \teff. Below about 3500 K, roughly where stars are expected to be fully convective, we find a small offset in the \fbol\ residuals at the 0.2\% level with models systematically underestimating \fbol s compared to observations. However, above 3500 K, models increasingly overestimate \fbol\ by up to 0.4\%. The offsets are insignificant for any given star, however a Spearman rank test indicates that a significant correlation with \teff\ is present ($\rho = -0.65$ with $p \gg 10\sigma$). This could be the result of model inaccuracies or systematic errors in the observations. Our approximations of the spectrum (e.g., use of models past 2.5\um, detailed in Section~\ref{sec:cal}) could create these small systematic errors, since the validity of these estimates changes with \teff. However, the change in behavior is coincident with the onset of full convection, suggesting that small model inaccuracies are a more likely explanation.

\begin{figure}[t]
    \centering
    \includegraphics[width=0.45\textwidth]{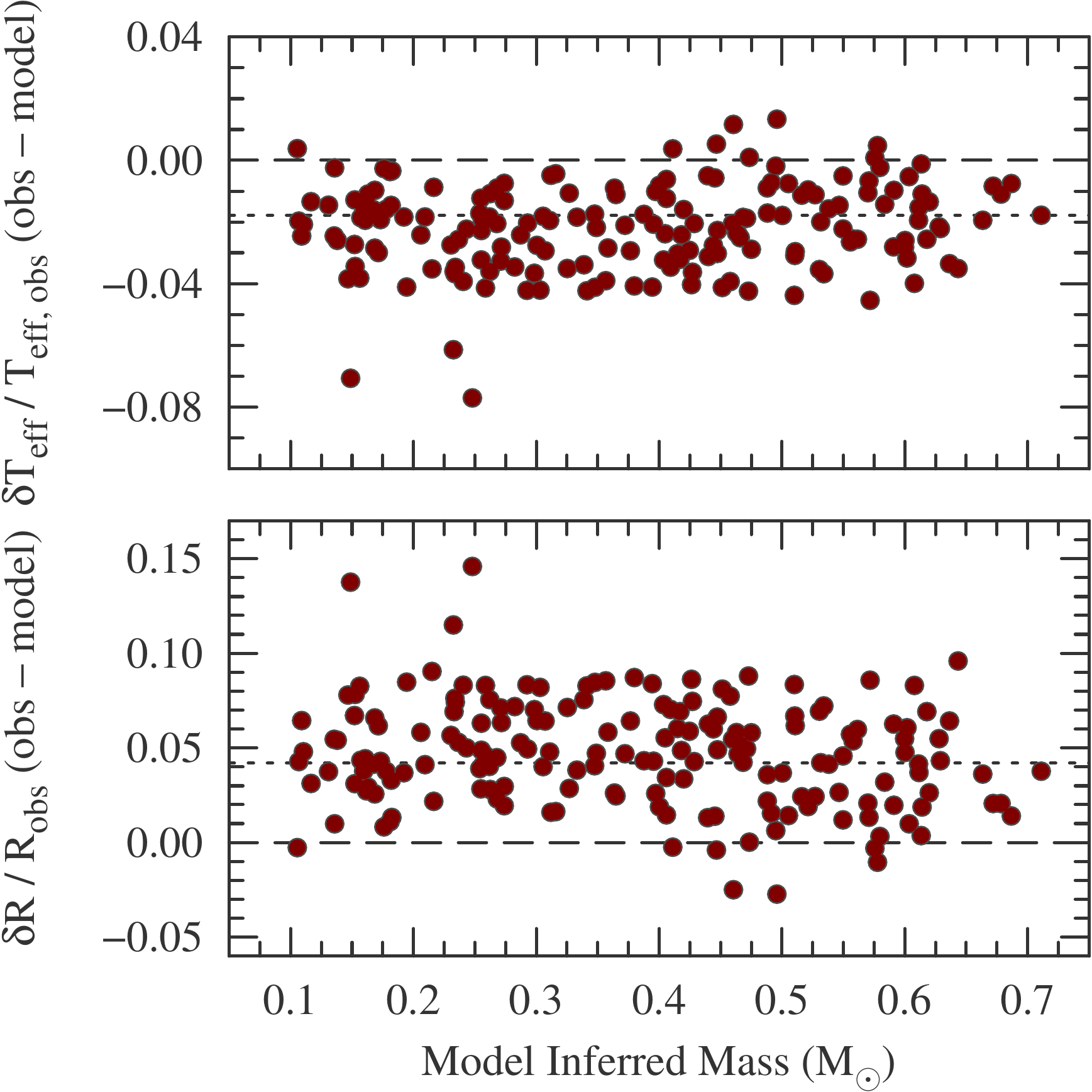}
    \caption{Relative model offsets for \teff\ (top) and radius (bottom) predictions as a function of the inferred 
     stellar mass. Typical $1\sigma$ observational uncertainties are given by dotted lines.}
    \label{fig:errors_mass}
\end{figure}

There is a systematic bias in the models toward hotter temperatures (Figures~\ref{fig:lum_teff_resids}, \ref{fig:errors_mass}). Models systematically over-predict \teff\ and under-predict $R_*$ by an average of $-2.2\pm0.1$\% and $+4.7\pm0.2$\%, respectively, echoing previous studies of EBs and single main sequence (MS) field stars (See Section~\ref{sec:intro}). Note that small systematic uncertainties are likely present in the data (e.g., in the \teff\ determination), so the standard errors quoted are likely underestimates. Comparison with LBOI determinations suggest systematic uncertainties of $\lesssim0.3\%$ in \teff\ and $\lesssim0.7\%$ in $R_*$ in the empirical determinations, still well below the observed offset with models (7.2$\sigma$ and 6.7$\sigma$, respectively). The distribution of \teff\ differences is also non-Gaussian, with an excess of cases where models are discrepant. Out of the 177 stars successfully modeled, 72 (40\%) match the observed \teff\ within $\pm1\sigma$, 79 (44\%) within $\pm1-2\sigma$, 24 (13\%) within $\pm2-3\sigma$, and 3 (2\%) with discrepancies $> \pm3\sigma$.

Unlike \fbol\ disagreements, \teff\ disagreements do not show any definite correlation with stellar properties, including \fbol, \teff, and the inferred stellar mass (Figure~\ref{fig:errors_mass}). This is illustrated for the case of the inferred stellar mass in Figure~\ref{fig:errors_mass}. We also show in Figure~\ref{fig:errors_mass} that radius disagreements do not exhibit any correlation with inferred stellar mass. With the exception of a few outliers below the fully convective boundary, there appears to be a floor in the observed \teff\ errors of $-4$\% and a ceiling in the radius errors of $+10$\%. Three exceptions lying beyond the error floor/ceiling are PM~I10430-0912, Gl~896~B (EQ~Peg~B), and Gl~166~C (40 Eri C).

\begin{figure}[t]
    \centering
    \includegraphics[width=0.45\textwidth]{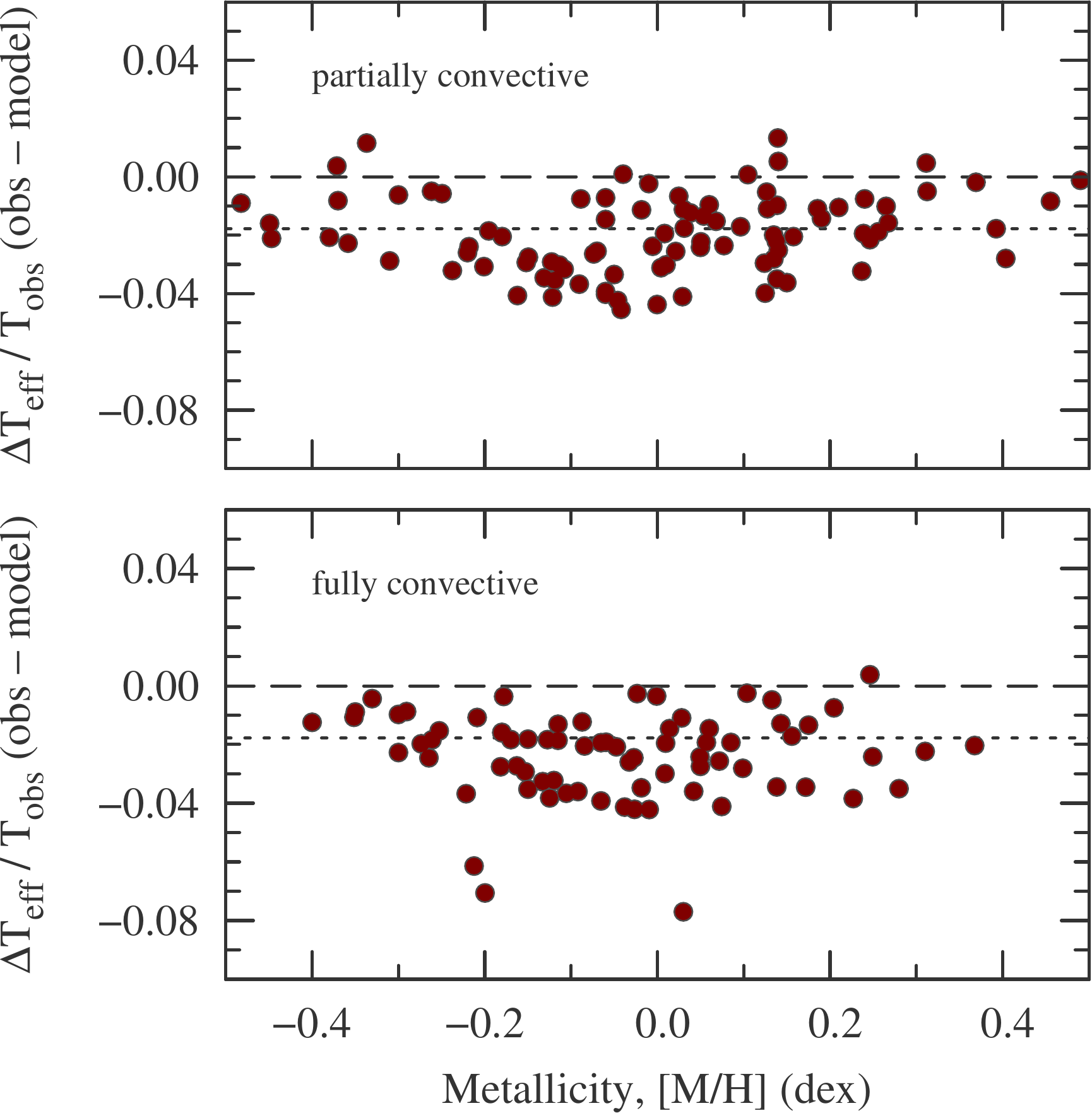} \\
    \vspace{\baselineskip}
    \includegraphics[width=0.45\textwidth]{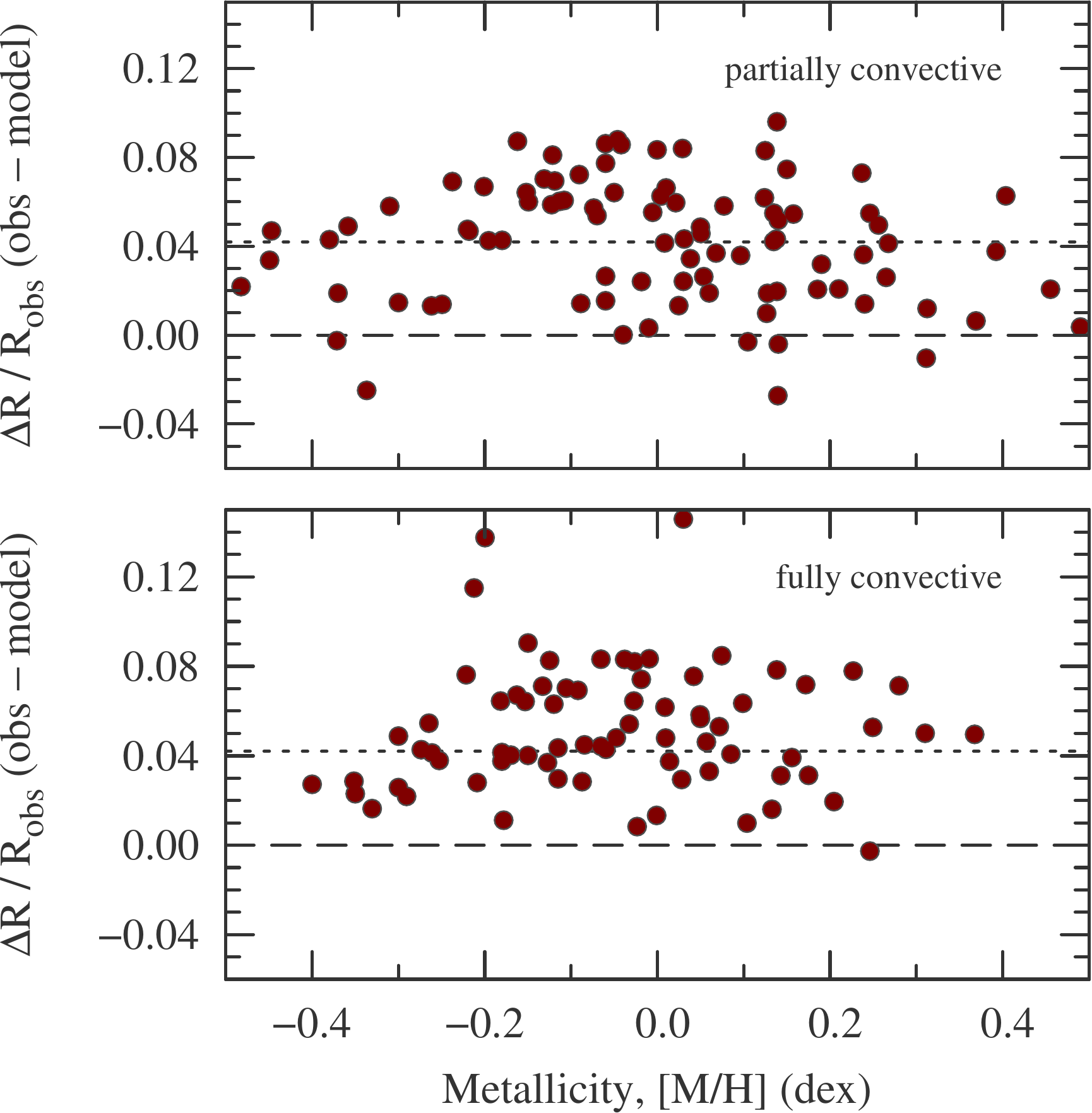}
    \caption{Relative model errors for \teff\ (left) and radius (right) predictions as a function of the observed metallicity. The sample is split into partially convective stars with a radiative core (top panels) and fully convective stars (bottom panels). Distance between the dashed and dotted line is the median $1\sigma$ uncertainty, between which models can be assumed to adequately reproduce the observation of a given star.}
    \label{fig:errors_MH}
\end{figure}

Figure~\ref{fig:errors_MH} show that neither relative errors in \teff\ nor $R_*$ show a dependence on the observed metallicity, even when the sample is split into the partially convective and fully convective regimes. The division between partially and fully convective interiors was taken to be $M_* \geq 0.37$ \ms\ and $M_* \leq 0.33$ \ms, respectively, thereby excluding stars whose convective state is uncertain. A linear regression analysis suggests that relative errors in \teff\ are not correlated with metallicity, with slopes of $0.7$\% error/dex  and $-0.2$\% error/dex for the partially convective and fully convective populations, respectively. The lack of a correlation is confirmed with Spearman $\rho$, Pearson $r$, and Kendall's $\tau$ correlation tests returning results entirely consistent with the null hypothesis.

\begin{figure}[t]
    \centering
    \includegraphics[width=0.45\textwidth]{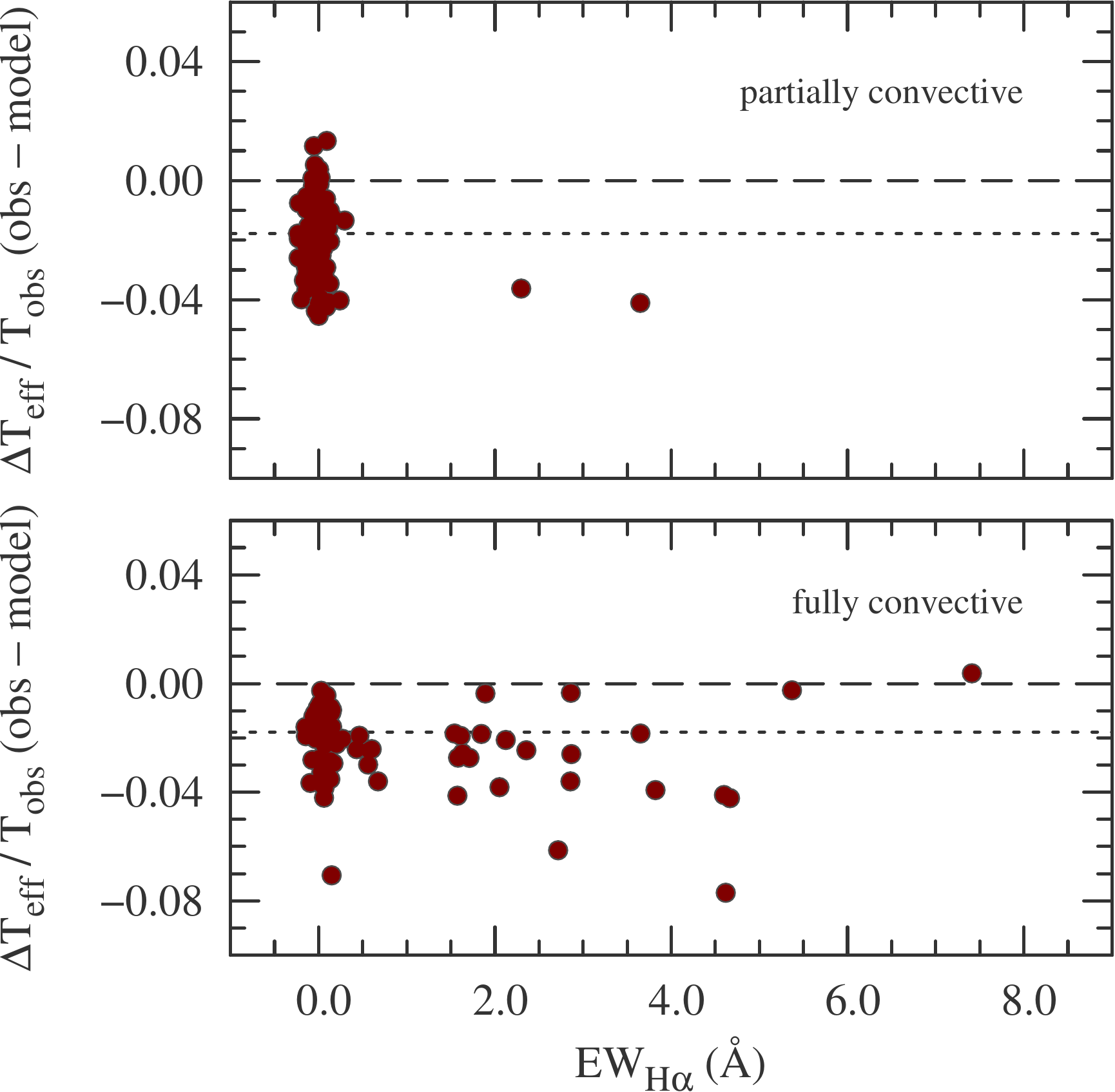} \\
    \vspace{\baselineskip}
    \includegraphics[width=0.45\textwidth]{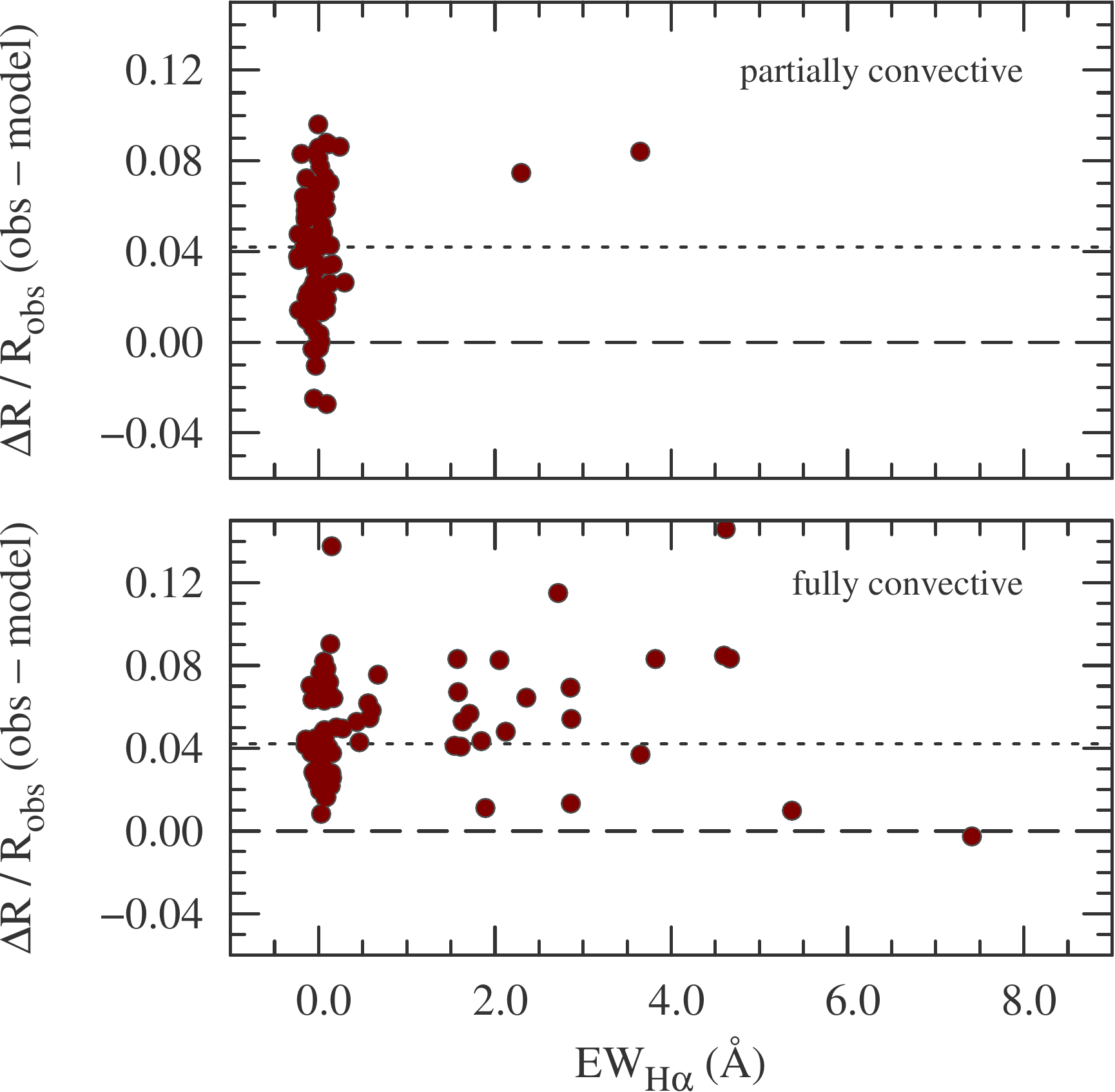}
    \caption{Relative model errors for \teff\ (left) and radius (right) predictions vs. the equivalent width of the H$\alpha$ line. The sample is split into partially convective stars with a radiative core (top panels) and fully convective stars (bottom panels). Distance between the dashed and dotted line is the median $1\sigma$ uncertainty, between which models can be assumed to adequately reproduce the observations of a given star. }
    \label{fig:errors_Ha}
\end{figure}

Plotting model errors against the measured EW of H$\alpha$ for each star also reveals no significant correlations (Figure~\ref{fig:errors_Ha}), where we again split the population of stars into partially and fully convective subsets. There are only two partially convective stars with measured H$\alpha$ EWs larger than 0.5 \AA. While they show significant disagreement with models, comparable differences are seen in stars that do not show elevated levels of H$\alpha$ emission. Model errors in the fully convective regime also appear to be independent of H$\alpha$ EW. Fully convective stars appear systematically larger and cooler, regardless of the EW of H$\alpha$. As with metallicity, different correlation tests all return results that are consistent with the null hypothesis when considering radius and \teff\ errors as a function of H$\alpha$ EW, with $p \gtrsim 0.20$. Combining the two populations also reveals no significant trend between \teff\ and H$\alpha$ EW, with Pearson $r = -0.119$ ($p = 0.112$), Spearman $\rho = -0.165$ ($p = 0.027$), and Kendall's $\tau = -0.134$ ($p = 0.008$). Notably, we were able to generate a rank coefficient of an equivalent value for each correlation test by randomly perturbing temperature errors by their quoted uncertainties, suggesting that the observed weak correlations are due to random noise in the measurements. 

H$\alpha$ equivalent width alone is not the most accurate measure of chromospheric emission. A more robust indicator is the ratio of the H$\alpha$ luminosity to the total bolometric luminosity of the star ($L_{\rm{H}\alpha}/L_{\rm{bol}}$). Following \citet{Tinney1998}, $L_{\rm{H}\alpha}/L_{\rm{bol}}$ can be computed using:
\begin{equation}
\frac{L_{\rm{H}\alpha}}{L_{\rm{bol}}} = \frac{F_{\rm{H}\alpha}}{F_{\rm{bol}}} = \frac{EW\times S_c}{F_{\rm{bol}}},
\end{equation}
where $S_c$ is the local continuum flux density. The quantity $S_c/F_{\rm{bol}}$ is simply a bolometric correction. We derived $S_c/F_{\rm{bol}}$ as a function of \teff\ and [Fe/H] as was done in Section~\ref{sec:bcorr}, but using only stars with no detectable H$\alpha$ (EW$\le0.27$, $<1\sigma$ above 0). For this we defined the band of interest as a region spanning 6500\AA--6600\AA. Our results did not change significantly by making small ($\sim$20\AA) adjustments to the width of the region. We then applied the empirical relation to calculate $S_c/F_{\rm{bol}}$ and $L_{\rm{H}\alpha}/L_{\rm{bol}}$ (as well as it's logarithm, [$L_{\rm{H}\alpha}/L_{\rm{bol}}$]) for all stars in the sample. The resulting [$L_{\rm{H}\alpha}/L_{\rm{bol}}$] values are relatively consistent with those derived from \citet[K. Stassun, 2015 private communication;][]{Stassun2012} for stars with \teff>3000\,K.

Our current observations are insensitive to small H$\alpha$ values, as poor resolution combined with strong (unresolved) molecular bands and observational uncertainties are capable of mimicking or masking weak (H$\alpha\lesssim0.8$) values \citep{Gaidos2014}, diluting the result. For these stars we set [$L_{\rm{H}\alpha}/L_{\rm{bol}}$] to a quiescent value of $10^{-6}$, although the result does not change for any reasonable assignment for the quiet stars, or even simply leaving them unadjusted and removing negative $L_{\rm{H}\alpha}/L_{\rm{bol}}$ values.

We find no significant correlation among the data when comparing \teff\ and radius errors to [$L_{\rm{H}\alpha}/L_{\rm{bol}}$]. Without accounting for uncertainties in the \teff\ values, we find Pearson $r = -0.216$ ($p = 0.004$), Spearman $\rho = -0.172$ ($p = 0.023$), and Kendall's $\tau = -0.136$ ($p = 0.007$), suggesting a weak correlation is present. Correlation coefficients of similar magnitude, but opposite in sign, and of similar statistical significance were found when considering radius. To account for measurement uncertainties, we followed the same procedure as with H$\alpha$ EWs; we randomly perturbed each \teff\ by their quoted uncertainties and performed correlation tests on each random realization of the data set. This was performed 1000 times and compared to a similar data set where \teff\ errors were perturbed, but then assigned a value of [$L_{\rm{H}\alpha}/L_{\rm{bol}}$] from the data set at random. We find that including uncertainties on the \teff\ errors produces mean correlation coefficients for the real data set that are entirely consistent with those of the randomly generated data set to within 1.7$\sigma$ (i.e., $p \geq 0.09 $).

X-ray emission is also a proxy for stellar magnetic activity. We cross-referenced our M dwarf sample with the ROSAT All Sky Survey Bright and Faint Source Catalogues \citep{Voges1999,Voges2000}. X-ray count rates and hardness ratios were converted to X-ray fluxes, $F_{X}$, following \citet{Schmitt1995}. As with H$\alpha$ EW, model errors show no dependence on $F_X$/\fbol. Finally, we also cross-referenced our sample with the source catalog from the Galaxy Evolution Explorer (GALEX) All-sky Survey \citep{Martin2005} and extracted near-ultraviolet (NUV) and far-UV (FUV) fluxes. Modeling errors show no correlation with NUV or FUV fluxes. We are therefore confident that variation in magnetic activity plays no significant role in driving the systematic offset between model predictions and observations of stellar properties.

Even if the proposed correlation is real (despite insignificant $p$-values for a variety of metrics) it must be a relatively weak effect. The linear slope of the relation between radius and temperature errors (difference between model and observed values) and [$L_{\rm{H}\alpha}/L_{\rm{bol}}$] is $\sim -0.46$\% error/dex in \teff\ and 0.84\% error/dex in $R_*$, in contrast to slopes of $-4.7$\% error/dex and 15.4\% error/dex found by \citet{Stassun2012}.

\subsection{Semi-empirical $M_{\rm K_S}$--Mass Relation}\label{sec:semiemp}
\begin{figure}[t]
    \centering
    \includegraphics[width=0.45\textwidth]{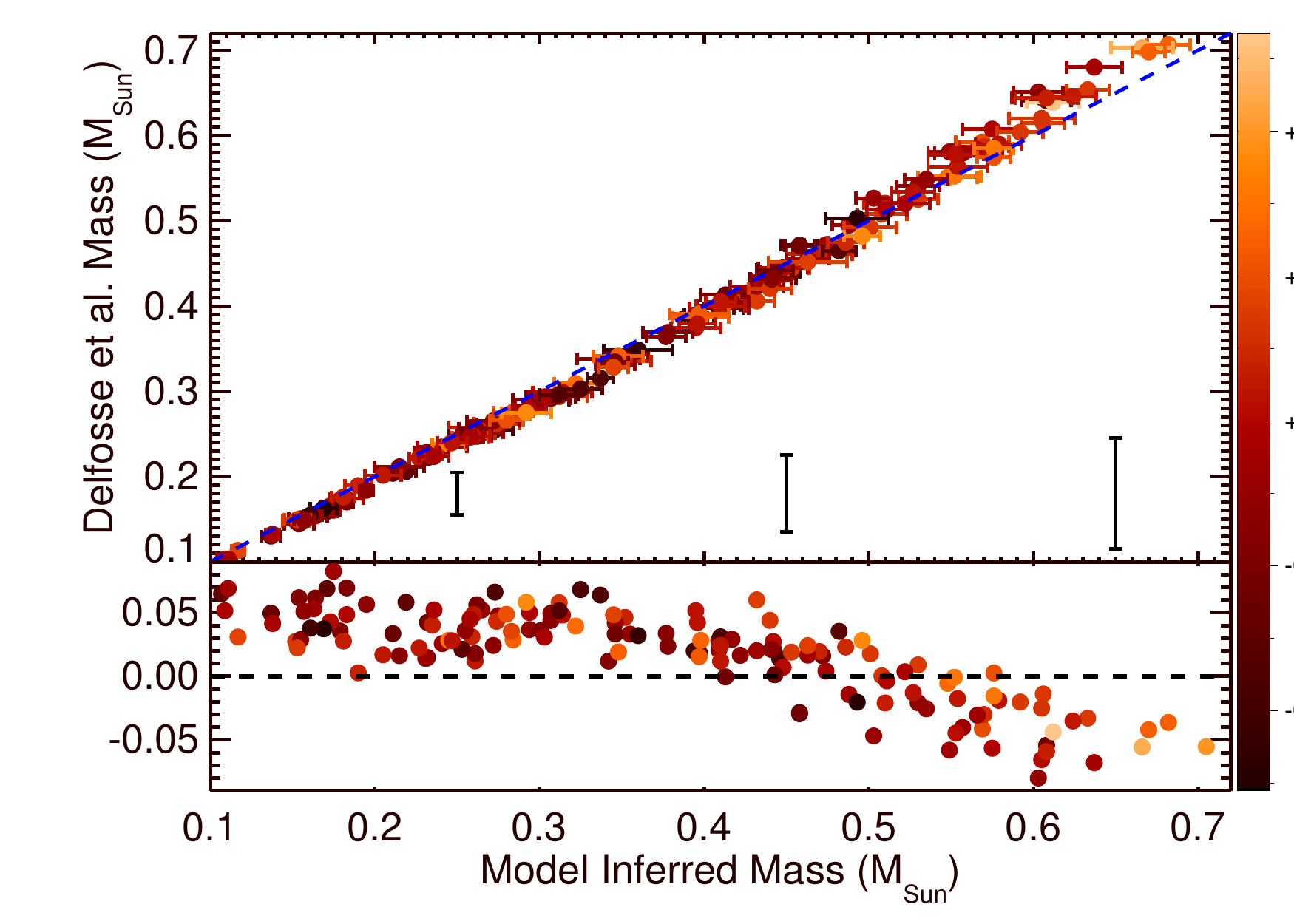}
    \caption{Mass derived from the empirical mass-$M_{K}$ relation from \citet{Delfosse2000} vs. of the mass 
    inferred from the models (Section~\ref{sec:model_results}). We show some characteristic errors for \citet{Delfosse2000} masses at the bottom of the panel. The bottom pane shows the fractional mass difference. Points are color-coded by metallicity.}
    \label{fig:mass_emp_diff}
\end{figure}

In Section~\ref{sec:lmebs} we found that our empirical radii combined with masses from the \citet{Delfosse2000} relation could reproduce the mass-radius relation from LMEBs within quoted uncertainties, but there were noticeable systematics (Figure~\ref{fig:mass_radius}). As we show in Figure~\ref{fig:mass_emp_diff}, similar systematics are seen when we compare model-based masses to those from \citet{Delfosse2000}. The model and empirical mass determinations agree given the estimated 10\% errors on the \citet{Delfosse2000} relation, however the models predict systematically lower masses above 0.50 \ms and systematically higher masses below that threshold. This disagreement cannot be explained by the offsets in \fbol\ and \teff\ noted in Section~\ref{sec:model_results}. Slightly lower model masses (\teff) would be needed to produce lower \fbol\ values, which would accentuate the deviations between the model masses and empirically derived masses at the high mass end in Figure~\ref{fig:mass_emp_diff}, albeit slightly.

The consistency between the offsets seen in Figure~\ref{fig:mass_emp_diff} and those in Figure~\ref{fig:mass_radius} suggests that the model-derived masses are equally or more reliable, and significantly more precise than those from the \citet{Delfosse2000} relation. Motivated by this, we constructed a luminosity ($M_{\rm K_S}$)--mass relation using masses derived from stellar models and observed absolute $K_S$-band luminosities. A fourth-order polynomial, i.e.,
\begin{equation}
    \label{eqn:mk_mass}
    M_{\star}/M_{\odot} = a + b\cdot M_{K_S} + c\cdot M_{K_S}^2 
       + d\cdot M_{K_S}^3 + e\cdot M_{K_S}^4
\end{equation}
was required to obtain a reasonable fit to the data. Best-fit coefficients are reported in Table~\ref{tab:fits}. Uncertainties were calculated using maximum likelihood through a MCMC method implemented with \texttt{emcee}. We used 500 walkers and 100 steps with random seed parameters for each coefficient normally distributed around the solution derived from a least-squares regression. For each walker and each step in the MCMC analysis, the adopted set of $M_{K_S}$--mass pairs were a random realization of the set of mean values, allowing individual points to be shifted within their quoted uncertainties. Parameters listed in Table~\ref{tab:fits} represent the mean value of the PPD from the final step of each walker, although the final values were independent of the number of steps adopted while sampling the PPD. 

\begin{figure}[t]
    \centering
    \includegraphics[width=0.45\textwidth]{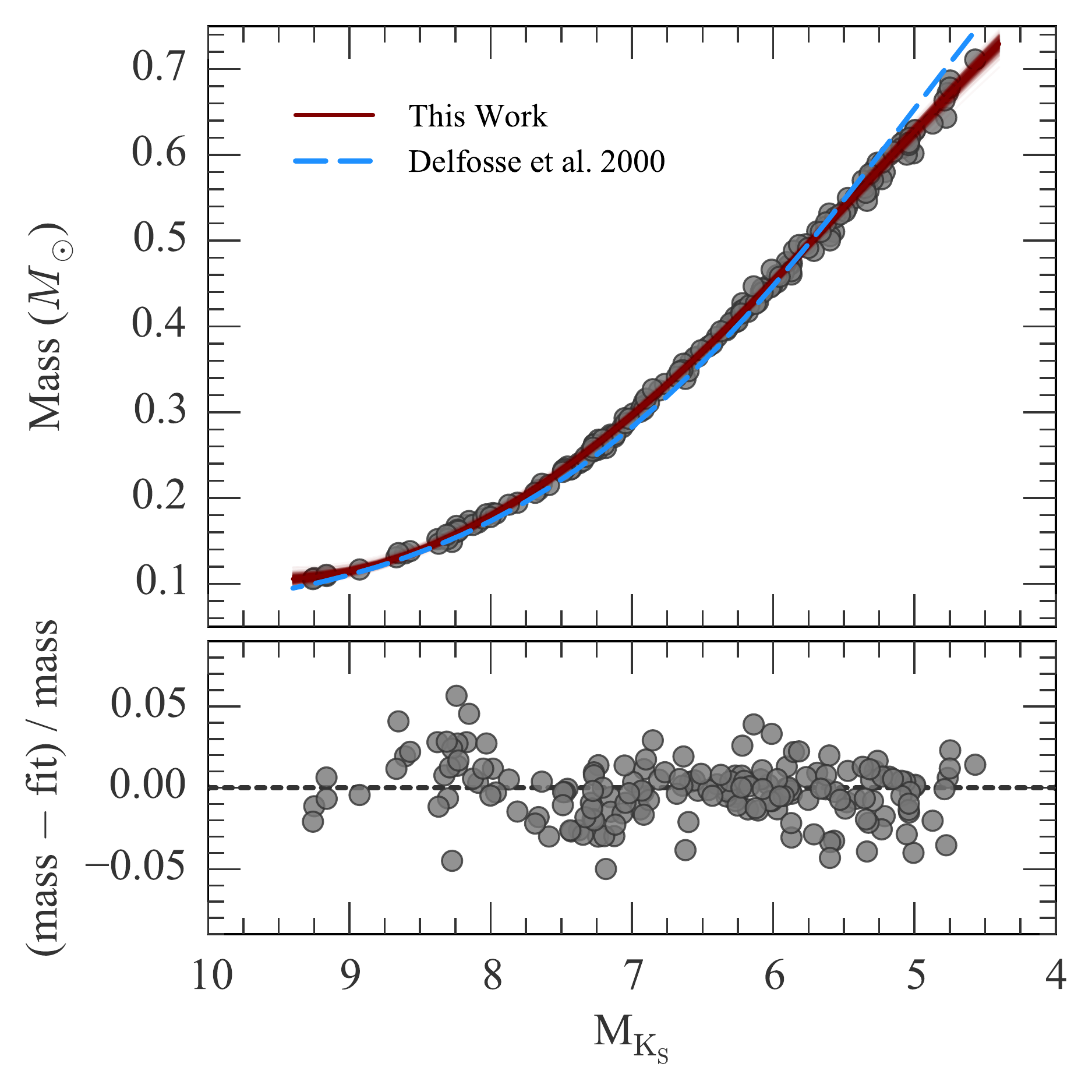} 
    \caption{(Top) Mass--luminosity ($M_{K_S}$) relationship. $M_{K_S}$ was determined from observations and paired with masses inferred from the Dartmouth stellar evolution model (Section~\ref{sec:model_results}). Red, solid lines are 500 random realizations of quartic polynomial fits to the data drawn from the joint PPD for the polynomial coefficients. The \citet{Delfosse2000} relationship is shown for reference (blue, dashed line). (Bottom) Residuals of the data with respect to a best-fit quartic polynomial whose coefficient values listed in Table~\ref{tab:fits}.}
    \label{fig:mk_mass}
\end{figure}

The best-fit polynomial is plotted in Figure~\ref{fig:mk_mass}. We plot the computed relation for each sample of the joint PPD of the fit coefficients in the domain $M_{K_S} \in [4.5,\ 9.5)$. Our relation compares well to that from \citet{Delfosse2000}, with the most significant deviations between the two occurring in the mid-$M_{K_S}$ and bright-$M_{K_S}$ regions. We note that the scatter around our relation is significantly smaller than the canonical 10\% scatter in the \citet{Delfosse2000} relation. The mean absolute error of the data about the relation, shown in the bottom panel of Figure~\ref{fig:mk_mass}, is 1.4\%, while the standard deviation of the errors about the zero point is about 1.8\%. About two-thirds (69\%) of the data have residuals within $\pm$1.8\%, whereas 95\% are located within $\pm$3.6\% of the zero point. The residuals show no correlation with metallicity. There appears to be a systematic offset of the data from the fit around $M_{K_S} = 8.25$, which results from interpolation errors around $0.16$~\ms\ that were mentioned previously. This also causes the fit to systematically overestimate masses in the range $8.0 > M_{K_S} > 7.5$. The fit has a nominal $\chi^2_{\nu} = 0.37$, suggesting uncertainties may be overestimated. This can be partially attributed to the fact that mass and $M_{K_S}$ uncertainties are treated independently, though in reality they are correlated along the observed relationship.

\begin{figure}[t]
    \centering
    \includegraphics[width=0.45\textwidth]{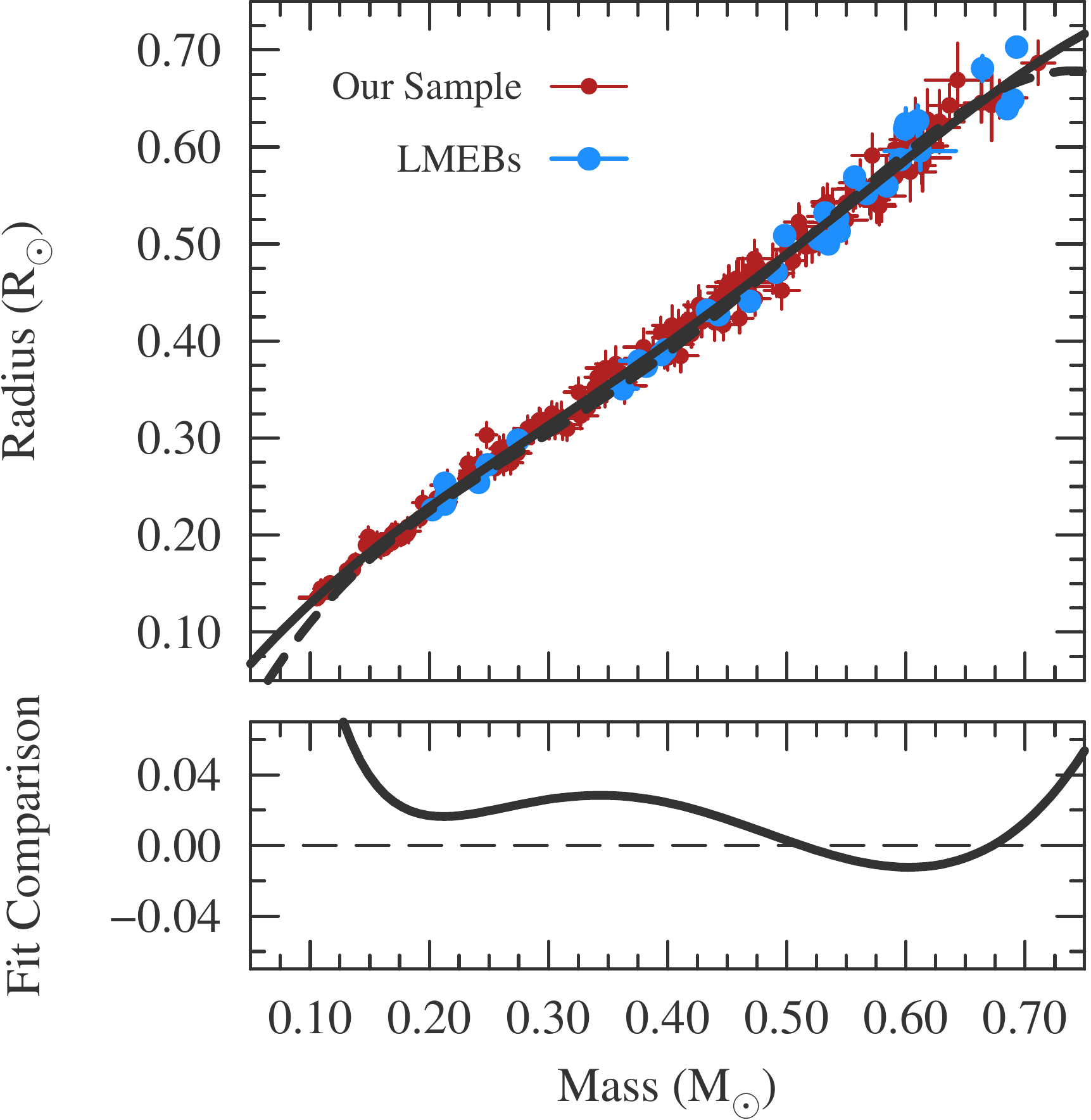}
    \caption{(Top) Mass-radius distribution from our bolometric radii and model-inferred masses compared to the distribution from LMEBs. Similar to Figure~\ref{fig:mass_radius}, but using model-inferred masses instead of those from the empirical mass-luminosity relation of \citet{Delfosse2000}. Polynomial fits to the two distributions are shown as solid and dashed lines for our sample and the LMEB sample, respectively. (Bottom) Relative error between the polynomial fits.}
    \label{fig:m_r_relation}
\end{figure}

We also compare the semi-empirical mass-radius relation for our sample to the mass-radius relation defined by LMEBs in Figure~\ref{fig:m_r_relation}. We see that the two estimates compare well in the bottom panel of Figure~\ref{fig:m_r_relation}, where we show the relative uncertainty between our semi-empirical fit and a polynomial fit to the LMEB sample. Between 0.2 and 0.7~\ms, the maximum deviation between the two fits is $\simeq$4\% and the largest deviation occurs in a region where there are no LMEB systems to constrain the LMEB polynomial fit. In addition, the polynomial fit from the LMEB sample provides a \rchisq = 0.9 with respect to the data from our single star sample. This provides support to the validity of our derived masses and our semi-empirical mass--$M_{K_S}$ relation. 

\subsection{Sensitivity to Standard Model Parameter Values}
\label{sec:knob_fiddle}
\begin{figure}[t]
    \centering
    \includegraphics[width=0.45\textwidth]{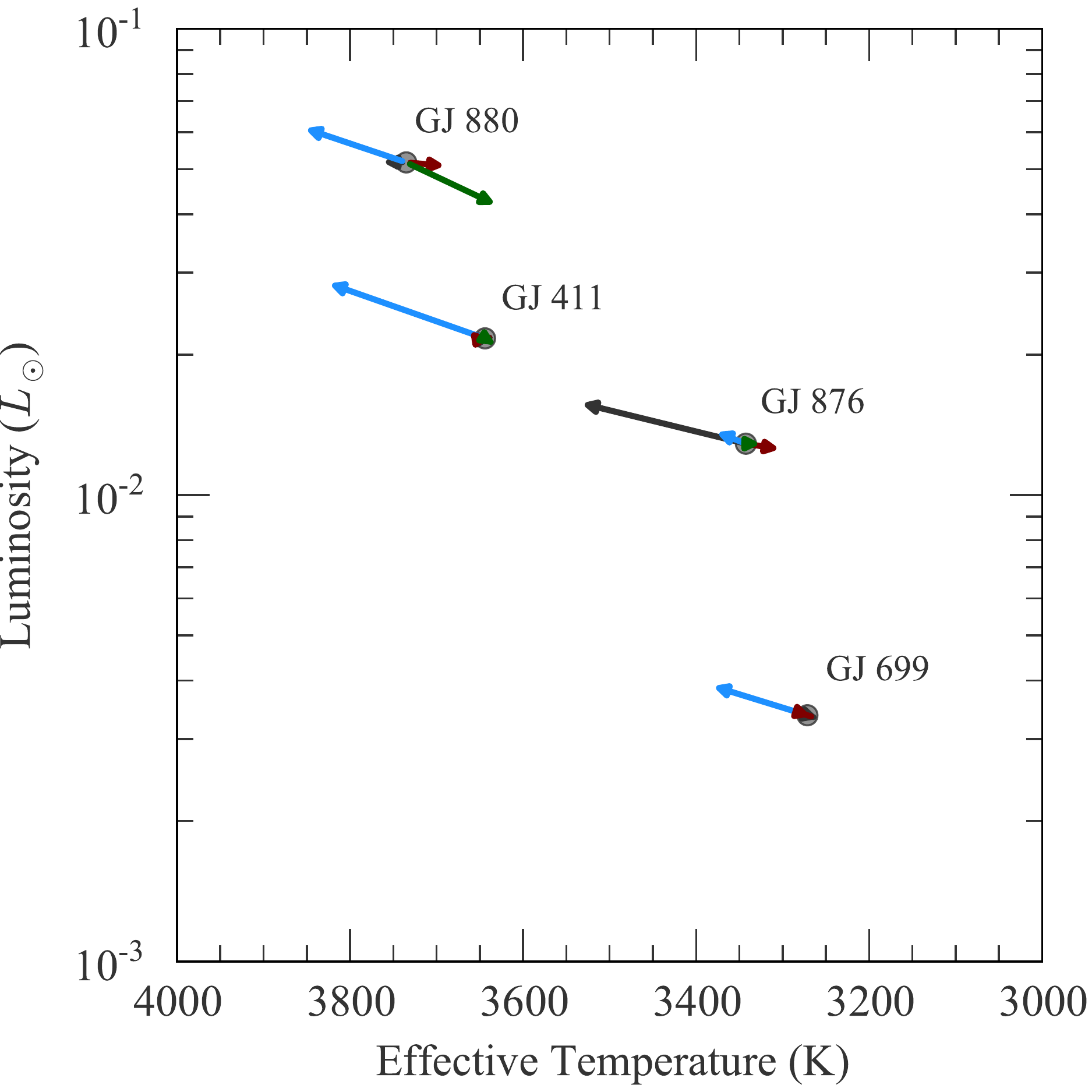} 
    \caption{HR diagram showing location of best-fit stellar models (gray points) of four representative stars: Gl~411, Gl~699, Gl~876, Gl~880. Vectors point to the location of test model results for the same stars computed with the \citet{Asplund2009} solar composition (gray), helium enhanced composition with $Y = 0.33$ (blue), reduced convective mixing length parameter \amlt = 1.0 (red), and artificially increased radiative opacities (green).}
    \label{fig:knob_fiddle}
\end{figure}

Given that model-observation disagreements appear to not be strongly correlated with either metallicity or magnetic activity indicators, we briefly explore how adopted model physics affect the location of the data in Figure~\ref{fig:lum_teff_resids} and the results that may have for the inferred stellar masses. Specifically, we address the impact of the adopted solar composition, helium mass fraction, convective  mixing length parameter, and radiative opacities on the results for the four representative systems we identified in Section~\ref{sec:parest}. Results of these investigations are given in Table~\ref{tab:knob_fiddle} and are shown in Figure~\ref{fig:knob_fiddle} as the set of displacement vectors in an HR diagram. 

Models of four stars are chosen for individual comparison: Gl~411, Gl~699, Gl~876, and Gl~880, the same four representative stars from Figure~\ref{fig:spectra}. They occupy different regions of parameter space that may lead to changes in the adopted physics having different effects on the model predictions. Two stars are fully convective (Gl~699, Gl~876) and two partially convective (Gl~411, Gl~880). At the same time, two are at the metal-poor end of the spectrum in this study (Gl~411, Gl~699) and two are more metal-rich (Gl~876, Gl~880). All test models below are computed at the precise model-inferred mass from Section~\ref{sec:model_results} and with the exact observationally determined metallicity. Differences between test models (see below) and the standard, unaltered model are taken at 5~Gyr.

\subsubsection{Solar Composition} \label{sec.composition}
The set of absolute abundances of heavy elements in the Sun is an important parameter of stellar evolution models, and there is considerable debate as to the correct values. Stellar abundance analyzes performed with sophisticated 3D radiation-hydrodynamic simulations and using the latest atomic and molecular line lists find that the abundance of heavy elements is $Z \approx 0.013$ -- $0.016$, with a surface $(Z/X)_{\odot} \approx 0.018$ -- $0.020$ \citep[e.g.,][]{Asplund2009, Caffau2011}. This is well below the solar heavy element abundance distribution adopted in our stellar evolution models, which are taken from the solar abundance study by \citet{GS98}, who found $Z \approx 0.017$ and $(Z/X)_{\odot} = 0.023$. However, it is well documented that revised solar heavy element abundances, when included in standard solar models, yield disagreements with the sound speed profile in the solar convective envelope, the lower boundary of solar convection zone inferred from helioseismic inversions, and with the solar surface helium abundance \citep[e.g., ][]{Bahcall2005,Delahaye2006,Basu2008}. Nevertheless, we test the impact on low-mass star predictions if the solar heavy element abundances are lower.

We adopt the \citet{Asplund2009} solar composition and consistently modify all of the model input physics. The \citet{Asplund2009} composition was adopted for two reasons: (1) it represents the largest shift away from \citet{GS98} among the recent values, and (2) more complete grids of PHOENIX model atmospheres and low temperature opacities exist for the \citet{Asplund2009} mixture. High and low temperature opacities  were calculated at the specified solar abundance \citep{Iglesias1996,Ferguson2005}, and new surface boundary condition tables were compiled from PHOENIX BT-Settl model atmosphere structures at a depth of $\tau = 10$ \citep{Allard2011}. Surface boundary conditions were computed with the same grid resolution as our standard  models (0.1~dex) using the same interpolation procedures. The distribution of heavy elements in FreeEOS was also updated, for consistency in the computation of the EOS. A new solar calibration was run to identify the appropriate starting conditions for our test models.

Recalibration of the solar model mostly suppresses the influence of changing absolute solar abundances by adjusting other unconstrained model parameters. We see in Figure~\ref{fig:knob_fiddle} that changing absolute solar abundances has little effect on the properties of the test models, with the exception of Gl~876. Decreasing the solar metal abundance leads to a decrease in the overall radiative opacity. Onset of nuclear fusion in the core is not affected by the change in opacity. Pre-MS stars transport energy released during gravitational contraction through near-adiabatic convection, and thus the core temperature is largely governed by an adiabatic temperature stratification. However, once fusion begins, the development of a radiative core is directly affected by the radiative opacity, with a lower opacity causing the radiative core to develop more quickly. Lower opacities lead to hotter temperatures and higher densities in the stellar core, which hastens the equilibration of the {\it p--p} chain due to a faster build up of $^3$He. This halts the stars contraction earlier, leaving it with a larger radius, hotter \teff, and higher luminosity at the zero-age MS. Evolutionary effects then come into play as the star evolves along the MS. Since we require a 1.0~\ms\ model to reproduce the luminosity and radius of the Sun at the solar age, the solar-calibrated helium abundance and convective mixing length are adjusted to offset changes imparted by lower radiative opacities. In this case, the helium abundance is decreased and the convective mixing length parameter (\amlt) is increased. Helium reduction causes the star to have a lower luminosity, while increasing \amlt shrinks the envelope convection zone, making the star smaller and hotter. By design, effects due to all three of these changes cancel in models of the Sun. At lower masses, stars react in different ways to variations in these three quantities, but overall, negligible changes are imparted to the models.

The sensitivity of Gl~876 to the solar abundance probably erroneous and not representative of the expected changes. Instead, we suspect the increased \teff\ and luminosity are due to an inaccuracy in the PHOENIX model atmosphere structures at the highest metallicities for stellar parameters of \teff~$\sim$~3600~K at $\log$~g = 3.5, precisely the parameter regime occupied by Gl~876 while it is undergoing pre-MS contraction. Specifically, there is a sharp discontinuity in gas temperature at $\tau = 10$ as a function of \teff. The abruptness is indicative of a numerical error, as a physical change of the model atmospheres at high metallicity would likely create a smoother transition. The resulting atmosphere structures appear to be approximately 250~K hotter at this optical depth than expected from an extrapolation of model atmosphere properties at lower metallicities. An offset of 250~K is very nearly the temperature offset endured by our model of Gl~876. Such an offset ultimately leads to the model arriving on the MS earlier than expected and at a hotter \teff. Therefore, we conclude that the test model of Gl~876 is in error.

\subsubsection{Helium Abundance}
We test the sensitivity to the model assumption that helium abundance scales linearly with metal abundance by artificially increasing the initial helium abundance to $Y_i = 0.33$. Models were generated with constant $Z_i$ and constant $(Z/X)$. In both cases, increasing the helium abundance moves all models toward higher luminosities, hotter \teff, and larger radii, as shown in Figure~\ref{fig:knob_fiddle}. We also see in Table~\ref{tab:knob_fiddle} that each quantity is positively correlated with $Y$. This can be understood in terms of helium's effect on the mean molecular weight. Increasing the amount of helium increases the mean molecular weight, which in turn causes the central temperature to rise, leading to a higher nuclear energy generation rate. While this primarily affects \teff\ and luminosity, it also causes the model to arrive on the MS at an earlier time with a larger radius compared to the standard case.

As a result, mass estimates from model fits would be lower by about 7\% and 3\% per 0.05 dex increase in $Y$ for stars above and below the fully convective boundary, respectively. However, model \teff s at these lower masses would increase by approximately 3\% and 1\% over the \teff s derived assuming a standard linear relation between $Y$ and $Z$. Fitted model radii, on the other hand, would decrease by about 4\% and 1\%, for partially and fully convective stars, respectively. Therefore, to provide better agreement between models and observations at constant luminosity and metallicity, the assumed helium abundance needs to be decreased. To match the systematic offsets seen in Figure \ref{fig:errors_mass}, we estimate that helium abundance needs to be systematically decreased by $\Delta Y \approx 0.03$ -- $0.04$ dex at constant metallicity. Some stars would therefore have helium abundances at or below the primordial helium abundance \citep{Peimbert2007}, which is typically understood to be unrealistic, although not necessarily precluded by observations. 

\subsubsection{Convective Mixing Length Parameter}\label{sec:mixing}
Reducing the convective mixing length parameter is akin to suggesting that convective energy transport is less efficient. It has long been noted that models of low-mass M-dwarfs require a convective mixing length below the solar-calibrated value \citep[e.g.,][]{Cox1981,Chabrier:1997}. Recent evidence from asteroseismology of more solar-like stars suggests that required convective mixing length values correlate with intrinsic stellar properties \citep[\teff, $\log g$, metallicity;][]{Bonaca2012}. 

Vectors shown in Figure~\ref{fig:knob_fiddle} were calculated for a set of models with a convective mixing length parameter \amlt\ = 1, as compared to the solar-calibrated value of \amlt\ = 1.88. This leads to models with shallower convective envelopes and a more extensive radiative interior, causing a reduction in model surface \teff s, an increase in predicted stellar radii, and a slight decrease in the luminosity. Note that the effects are mass dependent, with higher mass stars being relatively more affected than lower-mass stars. Table \ref{tab:knob_fiddle} illuminates this mass dependence, particularly the dependence of the rate of change of \teff\ with \amlt. The mass dependence is a consequence of the extent of the super-adiabatic layers in the atmosphere and the degree to which they are super-adiabatic. Lower mass stars are more dense and therefore largely undergo near-adiabatic convection with convective properties being fairly insensitive to the choice of input parameters, as compared to higher mass stars, at least within the context of mixing length theory.

Mass estimates are not overly sensitive to variations in the convective mixing length, owing to the fact that the convective mixing length does not affect the stellar luminosity at a significant level. At the high-mass end of our sample, reduction of the convective mixing length from solar \amlt\ = 1.88 to \amlt\ = 1 yields  models with \teff s that are 1.5\% cooler and radii about 1\% larger. The mixing length parameter would need to be further reduced to provide agreement at the lower end of the MS where stars are fully convective. Tests models with \amlt\ = 0.5 indicate that such a reduction largely affects model \teff s and luminosities, owing to the direct coupling between the stellar surface and the core. Mass estimates are affected at the 3\% level at 0.30 \ms, with \teff\ reductions of about 2\% and radius increases of around 2.5\%. We can therefore conclude that \amlt\ needs to be continually reduced towards lower masses, with \amlt\ $\sim$ 1 at 0.5 \ms\ and $\sim$ 0.5 at 0.3 \ms. These reductions would provide general agreement between model \teff\ and observations. However, model radii are still under-predicted by several  percent.

\subsubsection{Radiative Opacity}
We previously mentioned (Section \ref{sec.composition}) the effect that radiative opacities have on stellar structure. To isolate the specific effects of the radiative opacity, test models were computed with the Rossland mean opacity $\kappa$ increased by 50\%, such that $\kappa = 1.5\kappa_0$, where $\kappa_0$ is the Rosseland mean 
opacity in our standard models with a \citet{GS98} solar abundance distribution. Increasing the radiative opacity has the opposite effects as those described in Section~\ref{sec:mixing}. Evolution through the pre-MS of all the test models is unaffected by the radiative opacity, as the stars contract in convective equilibrium. Ignition of hydrogen burning and increasing temperatures precipitates the formation of a radiative core, after which the base of the convective envelope recedes toward the  stellar surface. Increasing the opacity causes the radiative core to form at an older age. Once the radiative core is established, the gas temperature and density decrease with increasing opacity. Stars continue to contract following this occurrence until the abundance of $^3$He is sufficient for the {\it p--p} chain to establish equilibrium. Since a higher opacity cools the core, the star will end up contracting more than a star with a lower opacity. The result is that increasing the opacity leads to cooler, more compact, and therefore less luminous stars. Below approximately 0.28 \ms, radiative opacities have no noticeable affect on stellar interiors as models predict the stars will remain entirely convective throughout their pre-MS and MS lifetime. For higher mass stars increasing the radiative opacity by 50\% would increases our mass estimates by about 10\%. 

However, the insensitivity of models to radiative opacities below the fully convective boundary means that errors in radiative opacities do not provide a robust solution to the disagreements noted in Section \ref{sec:model_results}. One shortcoming of the present investigation is that opacities were not modified in the optically thin regions above $\tau = 10$, as pre-computed model atmospheres provide the surface boundary condition. It is not clear to what degree opacity changes are required in the outer layers to impart significant structural changes to fully convective stars.

\section{Summary \& Discussion}\label{sec:discussion} 
To better characterize the physical properties of M dwarfs, we determined precise temperatures, luminosities, radii, and metallicities for 183 nearby M dwarfs with well-determined distances. We obtained spectra of each star spanning 0.3--2.4\um\ and took advantage of accurate spectrophotometric calibrations to derive \teff, \fbol, and [Fe/H], and hence $R_*$ using the Stefan-Boltzmann relation. We then derived empirical relations between \teff{} and [Fe/H] and radii and luminosities and compared our derived parameters to predictions from new predictions by the Dartmouth stellar evolution model.

Our method of measuring \teff\ was calibrated using stars with determinations from interferometry \citep{Boyajian2012, 2013ApJ...779..188M}, so consistency between this method of estimation and LBOI is expected, and indeed shown for the set of LBOI stars. Further, using masses derived from the empirical mass-$M_{K_S}$ relation of \citet{Delfosse2000} yields a mass-radius relation for our stars consistent within formal uncertainties for that determined from observations of LMEBs. 

Our determinations are largely model-independent. One concern is that LBOI-determined angular diameters and \teff\ require an estimate of limb darkening, which is typically based on from model spectra. These corrections are $\simeq$3\%. However, the error from this correction is likely much less than 0.1\%, which is a consequence of taking observations in the NIR where limb darkening effects are smaller. Further, \citet{2014Natur.505...69K} found that model limb darkening parameters are consistent with those derived from fitting high-S/N transits with HST within errors (5-15\%). 

More importantly, different interferometric observations of the same star have sometimes yielded different angular diameters \citep[e.g., ][]{2011A&A...526L...4B, Boyajian:2012fk}. \citet{2014MNRAS.439.2060C} found that disagreement between angular diameter measurements grows with decreasing diameter beyond expected growth in measurement uncertainties. This suggests the presence of systematic errors that become increasingly important with decreasing angular size. This may be due to differences in how calibrators stars are handled, which become important for stars near the resolution limit of the array. Additional observations are needed to better understand these differences.

Another concern involves systematic uncertainties in our estimates of [Fe/H], which were calibrated with spectroscopically-determined metallicities of solar-type (FGK) companions to a small number of M dwarfs. Different analyses have yielded systematically different metallicities for the same solar-type dwarfs \citep{2014AJ....148...54H}. This was corrected for in \citet{Mann2013a} and \citet{Mann2014} by adjusting metallicity values from a given reference using stars common to the reference source and to the SPOCS catalog \citep{2005ApJS..159..141V}. While this puts all metallicities on the same scale, it is not necessarily the true scale. 

Metallicity appears to have a minor but statistically significant effect on the $M_{K_S}$-$R_*$ relation, and a highly significant effect on the \teff-[Fe/H]-$R_*$ relation. If there is some systematic offset in the metallicity scale, our empirical relations that include [Fe/H] can still be used provided metallicities are adjusted to match our adopted scale. However, systematic offsets in our metallicity scale could more significantly impact our model analysis, since the models rely on absolute metallicities. If our metallicities are systematically too high, the shift to a correct metallicity scale will force the models toward hotter \teff, likely exaggerating the differences between model predictions and observations described in  Section~\ref{sec:model_results}. Conversely, a shift of the metallicity scale in the opposite direction would make models in better agreement with observations.

Pre-MS stars are systematically larger than their older counterparts, and their inclusion could affect our derived relations. However, M dwarfs need to be $<100$~Myr to be significantly inflated for a given $M_K$ magnitude. Based on the age distribution of M dwarfs in the Solar neighborhood \citep{Ansdell2015} there should be just 2-5 such young stars in our sample. Unsurprisingly, two of our stars (Gliese 896A and 896B) are purported to be members of young moving groups with ages $\lesssim$100\,Myr \citep{Zuckerman2013}. As expected, all three stars fall above our best-fit $M_{K_S}$-$R_*$ relation, by 4.4\% and 9.5\%, respectively. However, neither of these differences are significant (1.3 and 2.3$\sigma$), and removing them from our fits resulted in negligible changes. Even if we assume there are three more such young stars in our sample, this would not create any significant changes in our results. 

We used the parameters of this sample to construct empirical relations between more readily observed parameters (\teff\ [Fe/H], and $M_K$), and less accessible parameters, i.e., $R_*$ and $M_*$ (Equation~\ref{eqn:MKR}, Table~\ref{tab:fits}). Our $M_{K_S}$-$R_*$ relation has a scatter of only 2.9\%, and 2.7\% when [Fe/H] is included. The scatter and \rchisq\ in the absolute magnitude-$R_*$ relation is smallest when using $K_S$ compared to other filters, a consequence of the increasing role of [Fe/H] at bluer wavelengths. Unfortunately redder photometry is not available, as {\it WISE} photometry saturates on a sizable fraction of our stars, and our empirical spectra only go to 2.5\um, so we could not generate reliable empirical synthetic {\it WISE} magnitudes.

\begin{figure}[t]
    \centering
    \includegraphics[width=0.45\textwidth]{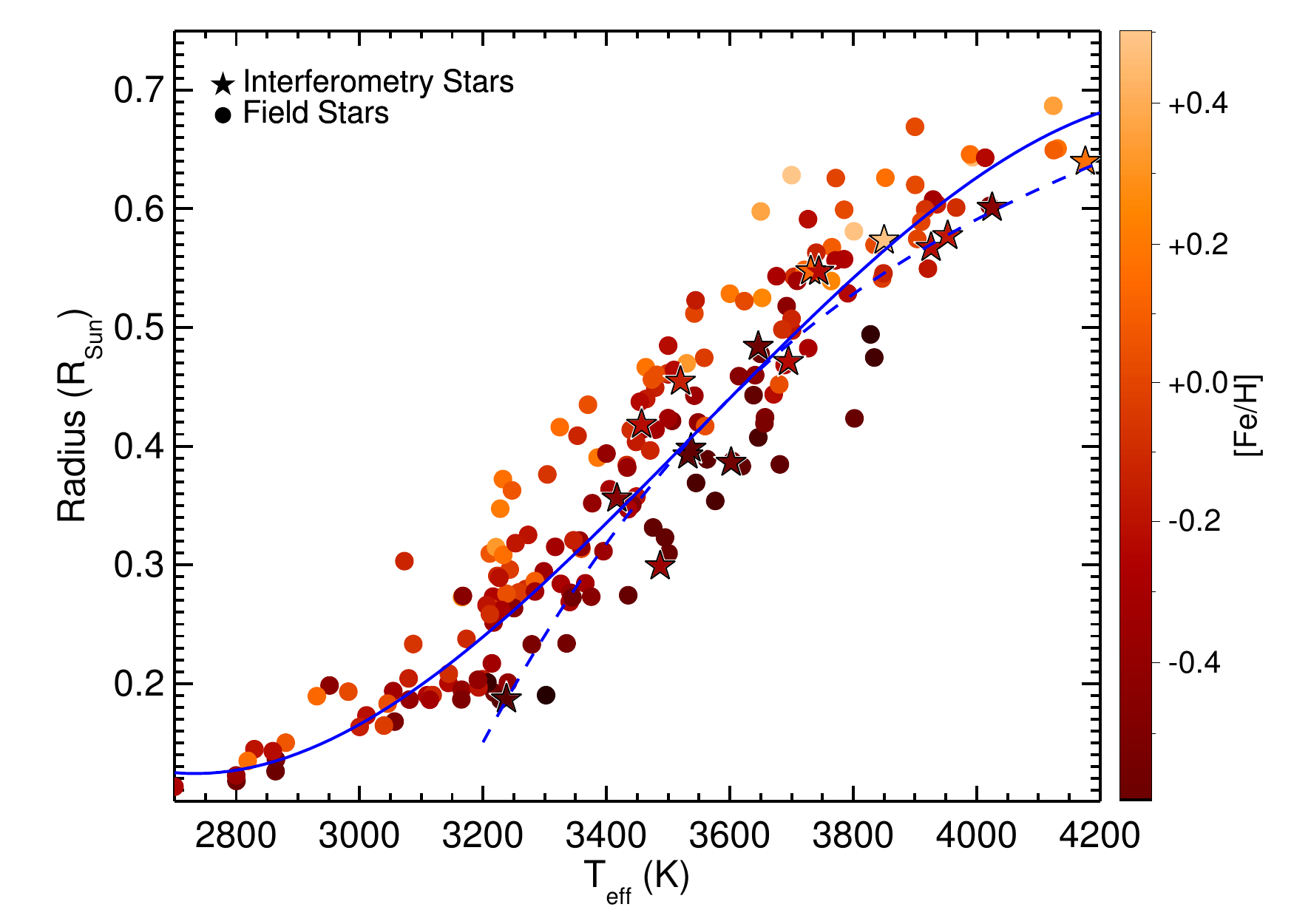} 
    \caption{Radius vs. \teff\ of our sample (circles) and the LBOI sample (5-point stars) used by \citet{2013ApJ...779..188M}. Gliese~725B is not shown because it was not included in the fits from \citet{2013ApJ...779..188M}. The best-fit relation from \citet{2013ApJ...779..188M} is shown as a blue dashed line, while the fit from this paper (Equation~\ref{eqn:MKR}) is shown as a blue solid line. Points are color-coded by metallicity.}
    \label{fig:nometal}
\end{figure}

Our derived \teff-$R_*$ relation can be used to predict stellar radii accurate to 13\%. When [Fe/H] is included the relation is accurate to 9\% in radius, or perhaps better, as suggested by the low \rchisq\ value. In this work, we found a significant correlation between [Fe/H] and the \teff-$R_*$ relation. The effect of [Fe/H] was not seen in previous empirical studies with smaller collections of of stars and sparser sampling in [Fe/H] \citep{Boyajian2012, 2013ApJ...779..188M}. These earlier findings did not necessarily exclude [Fe/H] dependence, but only showed that any correlation was not statistically significant. They were also hampered by a coincidence in the metallicities and temperatures of their sample. In Figure~\ref{fig:nometal} we show the \teff-$R_*$ distribution of stars from our sample along with those stars used in \citet{Boyajian2012} and \citet{2013ApJ...779..188M}. The LBOI (long-baseline optical interferometry) stars with \teff$<3700$\,K happen to be more metal poor. Thus much of the effect of metallically on the \teff-$R_*$ relation was masked by an (inaccurately) steeper dependence on \teff. 

The coverage and quality of our spectra enabled us to generate synthetic $VR_CI_CgrizJHK_S$ and {\it Gaia} $G$, $G_{RP}$, and $G_{BP}$ photometry for all stars in our sample, which we used to generate empirical relations between color and \teff\ (Table~\ref{tab:teffcolor}), and derive bolometric corrections for all relevant pass-bands (Table~\ref{tab:bcorr}). We find that \teff\ can be determined to an accuracy of 70-80~K and bolometric corrections to an accuracy of 2-3\% from color relations alone. More accurate \teff\ and BC values are possible if [Fe/H] is independently established, but even when [Fe/H] is not known, the [Fe/H] dependence can be mitigated by exploiting relations between $JHK_S$ colors and [Fe/H] for M dwarfs.
  
We compared our parameter values to predictions from the Dartmouth model grid using a MCMC method to obtain best-fit model parameters. We find that, while \fbol s are accurately predicted by the models (by design), there are still slight discrepancies that correlate significantly with inferred \teff\ (or mass) above 3500~K ($\sim$0.4 \ms). The broad consensus that models systematically under-estimate stellar radii and over-estimate stellar \teff\ among M dwarfs is supported by our data, which show average systematic offsets of 4.7\% and $-2.2$\%, respectively. These correspond to average deviations of $1.1\sigma$ and $1.2\sigma$ from the observations. While not all offsets between model predictions and the observations for individual stars are significant compared to the formal uncertainties, the offsets are systematic and hence significant when considering the full sample. This characterization of model errors is to similar that for LMEBs \citep{Feiden2012a,Spada2013} and stars in the LBOI sample \citep{Spada2013, Boyajian2012}. 

Although we uncover significant disagreements between model predictions and observations, we find no significant correlation between these offsets and stellar mass, metallicity, or magnetic activity. There have been no previous claims that modeling errors correspond with stellar mass, so it is not surprising that no correlation was uncovered. A correlation between model radius offsets and metallicity was suggested by previous investigations of single field stars \citep{Berger2006,LopezMorales2007}, but we find no evidence to support this claim. In a similar manner, we find model errors to be independent of observed H$\alpha$ activity measures, the ratio of coronal X-ray flux to \fbol, and both NUV and FUV fluxes. This is a somewhat surprising result given the mounting evidence in favor of magnetic fields and/or activity inflating stars in LMEBs \citep[e.g.,][]{Kraus2011, MacDonald2012, Feiden2012b, Feiden2013, Torres2014}. The stars in our sample appear to be systematically larger and cooler than model predictions, {\it independent of their level of magnetic activity.} If activity does inflate the radii of late-type stars, then the effect must be much weaker than previously suggested. Furthermore, since single stars and LMEBs are similarly inflated, magnetic activity is no longer necessary to explain LMEB inflation.

Previous studies have suggested a need to increase the He abundance to reconcile models and observations of LMEBs \citep[e.g., ][]{Paczynski1984,Metcalfe1996, Lastennet2003, Feiden2014}. Interestingly, we find the opposite, that He abundance must be reduced to provide better agreement with models. This is most likely a consequence of using luminosity, \teff, and radius as the observables, whereas studies of LMEBs use mass and radius. However, if we assume that the inflation we observe has the same origin as that observed in LMEBs---which is likely given that the discrepancies have almost identical magnitude and direction---this suggests that incorrect He abundances {\it cannot} be the origin. If He abundance were the cause, the requisite adjustments for LMEBs and our sample would be in agreement.

It would be interesting to apply our methodology to yet cooler stars (i.e., late M dwarfs and even L dwarfs), probing the physics at the stellar-substellar boundary. While previous studies have used similar methods on such ultracool dwarfs \citep[e.g.,][]{Dieterich2014}, parameters were either less precise or model-dependent. More importantly, it has only become recently possible to measure metallicities to spectral types as late as M9.5 \citep{Mann2014}, and there are no dwarfs with interferometric measurements beyond M6, making it difficult to test the \teff\ determinations empirically. Adaptive optics for CHARA \citep{Che2013}, and other instrumentation upgrades may enable measurements for the brightest and closest ultracool dwarfs.

The ESA {\it Gaia} mission, launched in 2013 December, will enormously expand the number of M dwarfs with precise trigonometric parallaxes \citep{2012Ap&SS.341...31D}. Our empirical and model-based relations can be applied to these stars, including those hosting planets expected to be discovered by the NASA K2 and TESS missions and the ESA PLATO mission. Currently {\it Kepler} M dwarf radii have been recently estimated using model fits \citep{Gaidos2013}, \teff-based relations from models \citep[e.g., ][]{Muirhead2014} or nearby stars \citep[e.g.,][]{Mann:2013vn, Newton2015}. But these methods are only good to $\simeq$10\%. With {\it Gaia} trigonometric parallaxes combined with the $M_{K_S}-R_*$ and $M_{K_S}-M_*$ relations from this paper will be possible to measure stellar radii accurate to 3\% and stellar masses to 2\% for M dwarfs, enabling far more precise determination of planet radii and densities (with radial velocity measurements). This will also provide accurate parameters for the entire set of targets, important when calculating planet occurrence rates or searching for planet-metallicity correlations.

\acknowledgements
Thanks to the anonymous referee for their fast and generous comments. Thanks to Keivan Stassun for his comments on activity diagnostics, to Michael Cushing for providing a pre-release version of SpexTools for uSpeX, and to Coryn Bailer-Jones for providing the spectral response profiles for {\it Gaia}. Additional thanks thanks to Adam Kraus, Chao-Ling Hung, and Aaron Rizzuto for useful discussions on the content of this manuscript. PPD Figures were made with the aid of the triangle.py package \citep{Foreman-Mackey:11020}. This research was supported by NASA grants NNX10AQ36G and NNX11AC33G to EG and NASA grants ADAP12-0172 and 14-XRP14 2-0147 to TB. 

SNIFS on the UH 2.2-m telescope is part of the Nearby Supernova Factory project, a scientific collaboration among the Centre de Recherche Astronomique de Lyon, Institut de Physique Nuclaire de Lyon, Laboratoire de Physique Nuclaire et des Hautes Energies, Lawrence Berkeley National Laboratory, Yale University, University of Bonn, Max Planck Institute for Astrophysics, Tsinghua Center for Astrophysics, and the Centre de Physique des Particules de Marseille. Based on data from the Infrared Telescope Facility, which is operated by the University of Hawaii under Cooperative Agreement no. NNX-08AE38A with the National Aeronautics and Space Administration, Science Mission Directorate, Planetary Astronomy Program. Some/all of the data presented in this paper were obtained from the Mikulski Archive for Space Telescopes (MAST). STScI is operated by the Association of Universities for Research in Astronomy, Inc., under NASA contract NAS5-26555. Support for MAST for non-HST data is provided by the NASA Office of Space Science via grant NNX09AF08G and by other grants and contracts. The CHARA Array is funded by the National Science Foundation through NSF grants AST-0606958 and AST-0908253 and by Georgia State University through the College of Arts and Sciences, as well as the W. M. Keck Foundation. This research made use of the SIMBAD and VIZIER Astronomical Databases, operated at CDS, Strasbourg, France (http://cdsweb.u-strasbg.fr/), and of NASA Astrophysics Data System, of the Jean-Marie Mariotti Center SearchCal service (http://www.jmmc.fr/searchcal), co-developed by FIZEAU and LAOG/IPAG. 

{\it Facilities:} \facility{IRTF:SpeX}, \facility{UH:2.2m (SNIFS), \facility{CHARA}}


\clearpage

\tabletypesize{\scriptsize}
\setlength{\tabcolsep}{0.02in} 
\begin{deluxetable*}{l l l | r r r r r r | r r r}
\tablewidth{0pt}
\tablecaption{Mass and Radius Relations}
\tablehead{
\colhead{Y} & \colhead{X} & \colhead{Eqn} & \colhead{a}  & \colhead{b} & \colhead{c} & \colhead{d} & \colhead{e} & \colhead{f} & \colhead{$\sigma$\tablenotemark{a} } & \colhead{\rchisq} \\
\colhead{} & \colhead{} & \colhead{\#} & \colhead{} & \colhead{} & \colhead{} & \colhead{} & \colhead{}  & \colhead{}  & \colhead{\%} & \colhead{} 
 }
\startdata
$R_*$ & $M_{K_S}$ & (\ref{eqn:MKR}) & $ 1.9515$ & $-0.3520$ & $0.01680$ & \nodata & \nodata & \nodata &   2.89 &    0.93\\
$R_*$ & $M_{K_S}$,[Fe/H] & (\ref{eqn:T_R2}) & $ 1.9305$ & $-0.3466$ & $0.01647$ & \nodata & \nodata & \phantom{-0}$0.04458$ &   2.70 &    0.88\\
$R_*$ & \teff/3500 & (\ref{eqn:MKR}) & $ 10.5440$ & $ -33.7546$ & $  35.1909$ & $-11.5928$\phantom{0} & \nodata & \nodata &  13.4 &    2.35\\
$R_*$ & \teff/3500,[Fe/H] & (\ref{eqn:T_R2}) & $16.7700$ & $-54.3210$ & $57.6627$ & $-19.6994$\phantom{0} & \nodata & $  0.4565$\phantom{0} &   9.3 &    1.10\\
\hline
$M_{\star}$\tablenotemark{c}  & $M_{\rm K_{S}}$& (\ref{eqn:mk_mass}) & $0.5858$ & $0.3872$ & $-0.1217$ & $0.0106$ & $-2.7262\times10^{-4}$ & \nodata & 1.8 & 0.37 
\enddata
\label{tab:fits}
\tablecomments{For the first, third, and fifth equation $Y=a+bX+c^2\ldots$, for the equations including [Fe/H] the right-hand side is multiplied by (1+f[Fe/H]).}
\tablenotetext{a}{For the first three relations $\sigma$ is given as the percent scatter in $R_*$, i.e., the standard deviation of $\frac{R_{*, \rm observed}-R_{*, \rm predicted}}{R_{*, \rm observed}}$. The last relation is quoted as the percent scatter in $M_{\star}$.}
\tablenotetext{c}{Semi-empirical relation derived using empirical $K_S$-band magnitudes and masses estimated from our model analysis. Coefficients are calculated using maximum likelihood and a MCMC method. See Section~\ref{sec:models} for details.}
\end{deluxetable*}

\tabletypesize{\scriptsize}
\setlength{\tabcolsep}{0.025in} 
\begin{deluxetable*}{l l | l l l l l l l | r r}
\tablewidth{0pt}
\tablecaption{\teff\ Relation Coefficients}
\tablehead{
\colhead{Y} & \colhead{X} & \colhead{a}  & \colhead{b} & \colhead{c} & \colhead{d} & \colhead{e}  & \colhead{f} & \colhead{g} & \colhead{$\sigma$\tablenotemark{a} } & \colhead{\rchisq} \\
\colhead{} & \colhead{} & \colhead{} & \colhead{} & \colhead{} & \colhead{}  & \colhead{}  &  \colhead{}  & \colhead{}  & \colhead{K} & \colhead{} 
 }
\startdata
\teff/3500 & $BP-RP$ & $3.245$ & $-2.4309$ & $1.043$ & $-0.2127$ & $0.01649$ & \nodata & \nodata & 52 & 0.88\\
\teff/3500 & $V-J$ & $2.840$ & $-1.3453$ & $0.3906$ & $-0.0546$ & $0.002913$ & \nodata & \nodata & 55 & 0.93\\
\teff/3500 & $V-Ic$ & $2.455$ & $-1.5701$ & $0.6891$ & $-0.1500$ & $0.01254$ & \nodata & \nodata & 53 & 0.94\\
\teff/3500 & $r-z$ & $1.547$ & $-0.7053$ & $0.3656$ & $-0.1008$ & $0.01046$ & \nodata & \nodata & 58 & 1.06\\
\teff/3500 & $r-J$ & $2.445$ & $-1.2578$ & $0.4340$ & $-0.0720$ & $0.004502$ & \nodata & \nodata & 58 & 1.04\\
\hline
\teff/3500 & $BP-RP$,[Fe/H] & $2.835$ & $-1.893$ & $0.7860$ & $-0.1594$ & $0.01243$ & $0.04417$ & \nodata & 45 & 0.60\\
\teff/3500 & $V-J$,[Fe/H] & $2.515$ & $-1.054$ & $0.2965$ & $-0.04150$ & $0.002245$ & $0.05262$ & \nodata & 42 & 0.53\\
\teff/3500 & $V-Ic$,[Fe/H] & $1.901$ & $-0.6564$ & $0.1471$ & $-0.01274$ & \nodata & $0.04697$ & \nodata & 48 & 0.67\\
\enddata
\label{tab:teffcolor}
\tablecomments{The first five formulae follow Equation~\ref{eqn:MKR}, the middle five follow Equation~\ref{eqn:teff2} ($f$ is the coefficient of the [Fe/H] term), and the last five follow Equation~\ref{eqn:teff3} ($f$ and $g$ are the coefficients for the $J-H$ and $(J-H)^2$ terms, respectively). Equations using $J-H$ as an additional variable are meant for when the metallicity is not known. }
\tablenotetext{a}{We report the scatter in the predicted - observed (from spectrum) \teff. Conservatively, these errors should be added (in quadrature) with our typical spectroscopic uncertainty (60~K).}
\end{deluxetable*}

\tabletypesize{\scriptsize}
\setlength{\tabcolsep}{0.025in} 
\begin{deluxetable*}{l l | l l l l l | l r}
\tablewidth{0pt}
\tablecaption{Bolometric Correction Formulae}
\tablehead{
\colhead{BC$_Y$} & \colhead{X} & \colhead{a} & \colhead{b} & \colhead{c} & \colhead{d} & \colhead{e} &  \colhead{$\sigma$} & \colhead{\rchisq} 
 }
\startdata
$V$ & $V-J$ & $0.5817$ & $-0.4168$ & $-0.08165$ & \phantom{$-$}$4.084\times10^{-3}$ & \nodata & 0.016 & 0.88\\
$Rc$ & $V-J$ & $2.127$ & $-1.059$ & \phantom{$-$}$0.1029$ & $-7.881\times10^{-3}$ & \nodata & 0.031 & 2.97\\
$Ic$ & $V-J$ & $0.4440$ & \phantom{$-$}$0.2331$ & $-0.05313$ & \nodata & \nodata & 0.037 & 2.47\\
\hline
$r$ & $r-J$ & $0.8958$ & $-0.5081$ & $-0.07387$ & \phantom{$-$}$3.999\times10^{-3}$ & \nodata & 0.016 & 0.56\\
$i$ & $r-J$ & $0.4431$ & $-0.06470$ & $-0.04038$ & \phantom{$-$}$2.798\times10^{-5}$ & \nodata & 0.031 & 2.86\\
$z$ & $r-J$ & $0.05373$ & \phantom{$-$}$0.2980$ & $-0.05001$ & \nodata & \nodata & 0.035 & 3.53\\
\hline
$Gaia$ & $BP-RP$ & $0.7384$ & $-0.7398$ & \phantom{$-$}$0.01340$ & \nodata & \nodata & 0.045 & 5.93\\
\hline
$J$ & $V-J$ & $0.8694$ & \phantom{$-$}$0.3667$ & $-0.02920$ & \nodata & \nodata & 0.016 & 0.90\\
\enddata
\label{tab:bcorr}
\tablecomments{All relations are of the form BC$_Y = a + bX + cX^2 +dX^3$ +e([Fe/H]), where $Y$ is a the filter listed above, and $X$ is the specified color.}
\end{deluxetable*}

\begin{deluxetable*}{l c c r r r c r r r c r r r c r r r}
    \renewcommand{\arraystretch}{1.2}
    \tablewidth{0.9\linewidth}
    \tablecaption{Influence of Standard Model Parameter Values on Four Representative Stellar Models}
    \tablehead{
        \colhead{} & \colhead{} 
        & & \multicolumn{3}{c}{Solar Composition\tablenotemark{a}} 
        & & \multicolumn{3}{c}{Helium Abundance\tablenotemark{b}} 
        & & \multicolumn{3}{c}{Mixing Length, $\alpha_{\rm MLT}$\tablenotemark{b}} 
        & & \multicolumn{3}{c}{Opacity $\kappa = 1.5 \kappa_0 $\tablenotemark{b}} \\
        \cline{4-6}\cline{8-10}\cline{12-14}\cline{16-18} \\
        \colhead{Star} & \colhead{class\tablenotemark{c}} 
        & & \colhead{\teff} & \colhead{$L/L_{\odot}$} & \colhead{$R/R_{\odot}$} 
        & & \colhead{\teff} & \colhead{$L/L_{\odot}$} & \colhead{$R/R_{\odot}$} 
        & & \colhead{\teff} & \colhead{$L/L_{\odot}$} & \colhead{$R/R_{\odot}$}  
        & & \colhead{\teff} & \colhead{$L/L_{\odot}$} & \colhead{$R/R_{\odot}$}
    }
    \startdata
    Gl~411 & PC--MP & & $0.0039$ & $-0.0089$ & $-0.0127$ & & $2630$ & $0.0988$ & $0.2160$ & & $13.07$ & $-0.0002$ & $-0.0044$ & & $-32.62$ & $-0.0019$  & $-0.0093$ \\
    Gl~699 & FC--MP & & $0.0054$ &  $0.0225$ & $-0.0001$ & & $1614$ & $0.0077$ & $0.0187$ & & $5.055$ &  $0.0000$ & $-0.0001$ & & \nodata  & \nodata & \nodata \\
    Gl~876 & FC--MR & & $-0.0583$ & $-0.2263$ & $0.0109$ & & $1590$ & $0.0355$ & $0.1377$ & & $49.88$ &  $0.0004$ & $-0.0046$ & & $-43.68$ & $ 0.0000$ &  $0.0090$ \\
\enddata
    \tablenotetext{a}{Total variation measured as a relative difference, $(x_{\rm GS98} - x_{\rm AGSS09})/x_{\rm GS98}$, 
    where $x$ is one of the stellar parameters.}
    \tablenotetext{b}{Quoted as a rate of change $\Delta x / \Delta q$, where $q$ is the parameter being changed.}
    \tablenotetext{c}{PC: partially convective; FC: fully convective; MP: metal-poor; MR: metal-rich.}
    \label{tab:knob_fiddle}
\end{deluxetable*}

\clearpage
\LongTables
\tabletypesize{\scriptsize}
\setlength{\tabcolsep}{0.025in}
\begin{deluxetable*}{l l | l l l l l l l l l l l l l l l}
\tablewidth{0pt}
\tablecaption{Parameters of All Stars in Our Sample}
\tablehead{
\colhead{LG11/LSPM Name} & \colhead{CNS3} & \colhead{R.A.} & \colhead{Dec.} & \colhead{\fbol} & \colhead{$\sigma_{\rm{F}_{\rm{bol}}}$} & \colhead{\teff} &\colhead{$\sigma_{\rm{T_{eff}}}$} &\colhead{$R_*$} &\colhead{$\sigma_{\rm{R_*}}$} &\colhead{$M_*$} &\colhead{$\sigma_{M_*}$} & \colhead{[Fe/H]} & \colhead{$\sigma_{\rm{[Fe/H]}}$} & \colhead{SpT} \\
\colhead{} & \colhead{} & \multicolumn{2}{c}{J2000}     &  \multicolumn{2}{c}{10$^{-8}$ erg\,s$^{-1}$\,cm$^{-2}$}        & \multicolumn{2}{c}{K} & \multicolumn{2}{c}{$\rm{R_{\odot}}$}& \multicolumn{2}{c}{$\rm{M_{\odot}}$} & \colhead{} & \colhead{} & \colhead{}
}
\startdata
PM I00115+5908 & \nodata & 00:11:31.8 & $+$59:08:39 & 0.04210 & 0.00091 & 2864 & 60 & 0.1361 & 0.0061 & 0.100 & 0.010 & $ -0.27$ &   0.08 & M6.1\\
PM I00118+2259 & \nodata & 00:11:53.0 & $+$22:59:04 & 0.1307 & 0.0017 & 3359 & 62 & 0.314 & 0.015 & 0.294 & 0.029 & $+  0.13$ &   0.08 & M3.5\\
PM I00183+4401 & Gl 15A & 00:18:22.9 & $+$44:01:22 & 5.669 & 0.045 & 3603 & 60 & 0.388 & 0.013 & 0.398 & 0.040 & $ -0.30$ &   0.08 & M1.4\\
PM I00184+4401 & Gl 15B & 00:18:25.8 & $+$44:01:38 & 0.891 & 0.010 & 3218 & 60 & 0.1923 & 0.0082 & 0.159 & 0.016 & $ -0.30$ &   0.08 & M4.1\\
PM I00219-3124 & GJ 1009 & 00:21:56.0 & $+$31:24:21 & 0.4175 & 0.0035 & 3600 & 60 & 0.528 & 0.028 & 0.551 & 0.055 & $+  0.27$ &   0.08 & M2.1\\
PM I01056+2829 & GJ 1029 & 01:05:37.6 & $+$28:29:33 & 0.06680 & 0.00091 & 2981 & 61 & 0.193 & 0.011 & 0.148 & 0.015 & $+  0.14$ &   0.08 & M5.3\\
PM I01076+2257E & Gl 53.1B & 01:07:38.5 & $+$22:57:20 & 0.0675 & 0.0012 & 3232 & 60 & 0.308 & 0.013 & 0.276 & 0.028 & $+  0.25$ &   0.03 & M4.0\\
PM I01125-1659 & Gl 54.1 & 01:12:30.5 & $+$16:59:56 & 0.5206 & 0.0082 & 3056 & 60 & 0.1680 & 0.0085 & 0.130 & 0.013 & $ -0.26$ &   0.08 & M4.9\\
\enddata
\label{tab:sample}
\end{deluxetable*}
\clearpage

\clearpage
\LongTables
\tabletypesize{\tiny}
\setlength{\tabcolsep}{0.02in}
\begin{deluxetable*}{l l | l l l l l l l l l l l l l l l l l l l l l l l l l l l l l l l l l l l }
\tablewidth{0pt}
\tablecaption{Synthetic Photometry of All Stars in Our Sample}
\tablehead{
\colhead{Name} & \colhead{$V$} & \colhead{$\sigma$} & \colhead{$R_C$} & \colhead{$\sigma$} & \colhead{$I_C$} & \colhead{$\sigma$} & \colhead{$g$} & \colhead{$\sigma$} & \colhead{$r$} & \colhead{$\sigma$} & \colhead{$i$} & \colhead{$\sigma$} & \colhead{$z$} & \colhead{$\sigma$} & \colhead{$J$} & \colhead{$\sigma$} & \colhead{$H$} & \colhead{$\sigma$} & \colhead{$K_S$} & \colhead{$\sigma$} & \colhead{Gaia $G$} & \colhead{$\sigma$} \\
}
\startdata
PM I00115+5908 & 15.611 &  0.040 & 13.926 &  0.012 & 11.805 &  0.020 & 16.325 &  0.024 & 14.953 &  0.011 & 12.703 &  0.013 & 11.594 &  0.014 &  9.957 &  0.018 &  9.405 &  0.017 &  9.075 &  0.020 & 13.827 &  0.014\\
PM I00118+2259 & 12.958 &  0.016 & 11.780 &  0.013 & 10.253 &  0.020 & 13.834 &  0.021 & 12.362 &  0.011 & 11.010 &  0.013 & 10.263 &  0.013 &  8.861 &  0.017 &  8.267 &  0.017 &  8.016 &  0.020 & 12.024 &  0.014\\
PM I00183+4401 &  8.119 &  0.016 &  7.116 &  0.012 &  5.991 &  0.020 &  8.973 &  0.020 &  7.505 &  0.011 &  6.638 &  0.012 &  6.148 &  0.013 &  4.896 &  0.017 &  4.303 &  0.017 &  4.039 &  0.020 &  7.523 &  0.014\\
PM I00184+4401 & 11.007 &  0.016 &  9.723 &  0.012 &  8.128 &  0.020 & 11.993 &  0.021 & 10.327 &  0.011 &  8.911 &  0.013 &  8.134 &  0.013 &  6.745 &  0.017 &  6.211 &  0.017 &  5.940 &  0.020 &  9.953 &  0.014\\
PM I00219-3124 & 11.175 &  0.016 & 10.142 &  0.013 &  8.910 &  0.020 & 11.956 &  0.021 & 10.577 &  0.011 &  9.583 &  0.013 &  9.036 &  0.013 &  7.713 &  0.017 &  7.025 &  0.017 &  6.753 &  0.020 & 10.500 &  0.014\\
PM I01056+2829 & 14.820 &  0.018 & 13.297 &  0.013 & 11.361 &  0.020 & 15.740 &  0.023 & 14.153 &  0.011 & 12.260 &  0.013 & 11.152 &  0.013 &  9.477 &  0.017 &  8.882 &  0.017 &  8.588 &  0.020 & 13.311 &  0.014\\
PM I01076+2257E & 14.054 &  0.016 & 12.770 &  0.012 & 11.095 &  0.020 & 14.899 &  0.021 & 13.438 &  0.011 & 11.904 &  0.013 & 11.038 &  0.013 &  9.538 &  0.017 &  8.935 &  0.017 &  8.662 &  0.020 & 12.937 &  0.014\\
PM I01125-1659 & 12.082 &  0.016 & 10.662 &  0.012 &  8.866 &  0.020 & 13.021 &  0.021 & 11.399 &  0.011 &  9.710 &  0.013 &  8.789 &  0.013 &  7.285 &  0.017 &  6.751 &  0.017 &  6.441 &  0.020 & 10.780 &  0.014\\
\enddata
\label{tab:photsample}
\end{deluxetable*}
\clearpage

\LongTables
\tabletypesize{\scriptsize}

\begin{deluxetable*}{l l c c c c r c r r c r c c r c }
\tablewidth{\linewidth}
\tablecaption{Inferred stellar evolution model parameters and associated errors.\tablenotemark{$\dagger$}}
\tablehead{
    LG11/LSPM & CNS3 & \colhead{Mass} & \colhead{$\sigma_M$} & \colhead{$\log({\rm age})$} & 
    \colhead{$\sigma_{\rm age}$} &
    \colhead{\fbol} & \colhead{$\sigma_{\rm{F}_{\rm{bol}}}$} & \colhead{$\delta$\fbol} &
    \colhead{\teff} & \colhead{$\sigma_{T}$} & \colhead{$\delta$\teff} & \colhead{Radius} & 
    \colhead{$\sigma_{R}$} & \colhead{$\delta R$} & \colhead{$\chi^2$}  \\
    Name & \colhead{} & \multicolumn{2}{c}{(\ms)} & \multicolumn{2}{c}{(yr)} &
    \multicolumn{2}{c}{($10^{-8}$erg s$^{-1}$cm$^{-2}$)} & & \multicolumn{2}{c}{(K)} & &
    \multicolumn{2}{c}{(R$_\odot$)} & & \colhead{}
}
\startdata
PM I00115+5908       &  \nodata     &  $ 0.107$  &  $ 0.009$  &  $ 9.46$  &  $0.59$  &  $ 0.04237$  &  $ 0.00138$  &  $ -0.00636$  &  $2921$  &  $32$  &  $-0.020$  &  $ 0.130$  &  $ 0.003$  &  $ 0.043$  &  $ 1.935$ \\ 
PM I00118+2259       &  \nodata     &  $ 0.312$  &  $ 0.016$  &  $ 9.43$  &  $0.60$  &  $ 0.13054$  &  $ 0.00271$  &  $  0.00130$  &  $3376$  &  $35$  &  $-0.005$  &  $ 0.309$  &  $ 0.012$  &  $ 0.016$  &  $ 0.197$ \\ 
PM I00183+4401       &  Gl 15A      &  $ 0.406$  &  $ 0.009$  &  $ 9.48$  &  $0.56$  &  $ 5.67125$  &  $ 0.06996$  &  $ -0.00043$  &  $3626$  &  $32$  &  $-0.006$  &  $ 0.382$  &  $ 0.007$  &  $ 0.015$  &  $ 0.345$ \\ 
PM I00184+4401       &  Gl 15B      &  $ 0.169$  &  $ 0.006$  &  $ 9.44$  &  $0.59$  &  $ 0.88968$  &  $ 0.01519$  &  $  0.00114$  &  $3250$  &  $32$  &  $-0.010$  &  $ 0.187$  &  $ 0.005$  &  $ 0.026$  &  $ 0.692$ \\ 
PM I00219-3124       &  GJ 1009     &  $ 0.538$  &  $ 0.019$  &  $ 9.46$  &  $0.52$  &  $ 0.41856$  &  $ 0.00521$  &  $ -0.00265$  &  $3657$  &  $45$  &  $-0.016$  &  $ 0.507$  &  $ 0.019$  &  $ 0.041$  &  $ 1.646$ \\ 
PM I01056+2829       &  GJ 1029     &  $ 0.153$  &  $ 0.009$  &  $ 9.33$  &  $0.62$  &  $ 0.06661$  &  $ 0.00136$  &  $  0.00272$  &  $3084$  &  $33$  &  $-0.034$  &  $ 0.178$  &  $ 0.007$  &  $ 0.078$  &  $ 4.961$ \\ 
PM I01125-1659       &  Gl 54.1     &  $ 0.136$  &  $ 0.005$  &  $ 9.40$  &  $0.62$  &  $ 0.51956$  &  $ 0.01213$  &  $  0.00202$  &  $3132$  &  $31$  &  $-0.025$  &  $ 0.159$  &  $ 0.006$  &  $ 0.055$  &  $ 2.817$ \\ 
PM I01186-0052S      &  Gl 56.3 B     &  $ 0.611$  &  $ 0.018$  &  $ 9.53$  &  $0.53$  &  $ 0.35883$  &  $ 0.00645$  &  $ -0.00262$  &  $4012$  &  $51$  &  $-0.019$  &  $ 0.579$  &  $ 0.015$  &  $ 0.042$  &  $ 2.593$ \\
\enddata
\tablenotetext{$\dagger$}{Relative errors quoted as $\delta X = (X_{\rm obs} - X_{\rm model})/X_{\rm obs}$}
\label{tab:model_results}
\end{deluxetable*}

\clearpage

\end{document}